\documentclass[aps,pre,superscriptaddress,showpacs,amsmath,amssymb,reprint,nofootinbib]{revtex4-1}
\usepackage{amsmath}
\usepackage{amsfonts}
\usepackage{amsthm}
\usepackage{graphics}
\usepackage{graphicx}
\usepackage{amssymb}
\usepackage{mathtools}
\usepackage[justification=justified]{caption,subcaption}
\usepackage{color}
\usepackage{url}
\usepackage[abs]{overpic}
\usepackage[usenames,dvipsnames]{xcolor}
\usepackage[normalem]{ulem}

\def\kmax{k_{\rm max}}

\DeclareMathOperator{\sgn}{sgn}

\graphicspath{{../}}

\begin{document}

\title{Poles, Shocks and Tygers: The time-reversible Burgers equation}

\author{Arunava Das}\thanks{Both authors contributed equally to this paper.}
\affiliation{Department of Physics, Indian Institute of Technology Kharagpur, Kharagpur - 721 302, India}%

\author{Pinaki Dutta$^*$}
\affiliation{Department of Physics, Indian Institute of Technology Kharagpur, Kharagpur - 721 302, India}%

\author{Vishwanath Shukla}
\email{research.vishwanath@gmail.com}
\affiliation{Department of Physics, Indian Institute of Technology Kharagpur, Kharagpur - 721 302, India}%

\date{\today}
\begin{abstract}
We construct a formally time-reversible, one-dimensional forced Burgers equation by imposing a global constraint of energy conservation, wherein the constant viscosity is modified to a fluctuating state-dependent dissipation coefficient. The new system exhibits dynamical properties which bear strong similarities with those observed for the Burgers equation and can be understood using the dynamics of the poles, shocks and truncation effects such as tygers. A complex interplay of these give rise to interesting statistical regimes ranging from hydrodynamic behavior to a completely thermalized warm phase. The end of the hydrodynamic regime is associated with the appearance of a shock in the solution and a continuous transition leading to a truncation-dependent state. Beyond this, the truncation effects such as tygers and the appearance of secondary discontinuity at the resonance point in the solution strongly influence the statistical properties. These disappear at the second transition, at which the global quantities exhibit a jump and attain values that are consistent with the establishment of a ‘quasi-equilibrium’ state characterized by energy equipartition among the Fourier modes. Our comparative analysis shows that the macroscopic statistical properties of the formally time-reversible system and the Burgers equation are equivalent in all the regimes, irrespective of the truncation effects, and this equivalence is not just limited to the hydrodynamic regime, thereby further strengthening the Gallavotti's equivalence conjecture. The properties of the system are further examined by inspecting the complex space singularities in the velocity field of the Burgers equation. Furthermore, an effective theory is proposed to describe the discontinuous transition.
\end{abstract}

\maketitle

\section{Introduction}
Equivalence of ensembles in equilibrium statistical physics is an important result. As a consequence, a given physical problem in equilibrium statistical physics can be studied using alternative ensembles (under appropriate thermodynamic limits), some of which can either be especially suited for the ease of calculation or for illustrating certain aspects of the problem. Such a luxury is often not present while dealing with general nonequilibrium systems; however, a few examples do exist where some progress has been made, for example, the construction of microcanonical and canonical ensembles for paths of Markov processes to describe the statistical properties of nonequilibrium systems driven in steady states and the examination of conditions for the equivalence of nonequilibrium ensembles~\cite{PRLChetriteHugo2013}. Moreover, the nature of interaction between the constituents of a system under consideration can further complicate the situation, both for equilibrium and out-of-equilibrium general many-body systems~\cite{Vroylandt2019,Touchette_2011,CAMPA200957}.

Turbulence described by driven-dissipative hydrodynamic equations is a paradigmatic example of a nonequilibrium phenomena. Hence, the availability of nonequilibrium ensembles and their equivalence may significantly further theoretical understanding of statistical properties of turbulence. Such an approach was initiated in~\cite{gallavotti1995dynamical}, where the idea of \textit{dynamical ensembles} was introduced for the nonequilibrium stationary states of a two-dimensional sheared fluid. This was further extended to turbulence in~\cite{gallavotti1996equivalence} by constructing a time-reversal invariant version of the Navier-Stokes equations (NSE), hereafter, called the time-reversible NSE (RNS). The ensembles associated with nonequilibrium stationary states of the NSE and RNS were conjectured to be equivalent in the limit of  an infinite Reynolds number (\textit{equivalence conjecture})~\cite{gallavotti1996equivalence}. This conjecture was later weakened to include cases with finite Reynolds number~\cite{Gallavotti2018quantandconj2,Gallavotti_2019conj2,gallavotti2020nonequilibrium,Gallavotti2021conj2}.

Following~\cite{gallavotti1996equivalence} a time-reversible variant of the NSE is constructed by imposing a global constraint on the system so certain prescribed macroscopic observables, such as energy, enstrophy, etc. remain conserved in time however this fundamentally modifies the dissipation mechanism; the dissipation terms now act as thermostatting terms. A priori it is not certain that such a formulation will reproduce the statistical properties of turbulence in detail.

Since the \textit{equivalence conjecture} was put forward, several studies have examined and verified its validity both in reduced models~\cite{biferale1998time,DePietro2018,Biferale_2018} and direct numerical simulations (DNSs) of fluid turbulence~\cite{shukla2019phase,jaccod2021const,margazoglou2022nonequilibrium}. Time-reversible shell model studies provided early insight into the problem, ranging from the statistical characterization of its dynamical regimes~\cite{biferale1998time} to the influence of different global constraints on the nature of equivalence~\cite{Biferale_2018}. Moreover, in~\cite{DePietro2018} the reversible shell model was used to disentangle and explore the contribution to time irreversibility of the nonequilibrium energy cascades from the explicit time-reversal symmetry breaking due to the viscous term. Recent DNS studies of three-dimensional (3D) RNS further confirm many of these findings and support the conjecture~\cite{shukla2019phase,jaccod2021const,margazoglou2022nonequilibrium}. For example, in~\cite{shukla2019phase} the constraint of a conservation of total energy was imposed and a continuous transition from `warm' to  `hydrodynamic' states was observed, when an appropriate control parameter was varied. At the critical point of this transition, the reversible system exhibited turbulence (hydrodynamic regime) whose statistics overlapped with the original 3D NSE. Recently, in~\cite{dubrullelog2023} the same transition was found in RNS defined on logarithmic lattices. In contrast, the total enstrophy was constrained to be conserved in~\cite{margazoglou2022nonequilibrium} and~\cite{jaccod2021const}, and the equivalence conjecture was verified.

The validity of the equivalence conjecture has also been studied for 2D turbulence~\cite{seshasayanan2021equivalence,rondoni1999fluctuations,GALLAVOTTI2004338,gallavotti2023reversibility,Gallavotti_2019conj2}, where the equivalence has been suggested. In~\cite{seshasayanan2021equivalence} two simultaneous global constraints were imposed by conserving the energy and enstrophy for the dual cascade setup in 2D; it was found that 2D RNS reproduces very well the global observables and is able to capture the essential features of the dual-cascade system such as energy spectra and fluxes, which match with their NSE counterparts. Moreover, the third-order longitudinal velocity structure function obtained using the RNS system exhibits scaling behavior similar to what is seen for the 2D NSE. The conjecture has also been verified for the Lorenz model, a simplified model of atmospheric turbulence~\cite{Gallavotti2014}.

The one dimensional Burgers equation (BE) provides a simple model of a nonlinear system driven out of equilibrium; hence, it has often been employed to gain insights into general nonlinear fluid phenomena~\cite{BEC20071}. Because of its numerical and analytical tractability, the BE has served as a veritable testing ground for studies of turbulence~\cite{kida1979asymptotic,Boritchev2014,parisi1995,chakraborty_frisch_ray_2010,schakrabortynelkin2012}. However, despite its apparent similarity to the NSE, the BE is solvable using the famous Cole-Hopf transformation~\cite{colebe,Hopf1950ThePD}, which maps it into a linear heat equation. Turbulence studies in BE generally involve adding intermittency externally either by stochastic forcing or stochastic initial conditions~\cite{Shih1985,verma2022univfunctions,eqstatesburgersverma2022,cartes2022galerkin,multscalindartifactpandit2005,murray2018energy,She1992,vergassola1994burgers,De2023,beturbjayprakash2000,Menon2007,Menon2010,Menon2012}.

Galerkin truncated 3D Euler equation, in the presence of energy conservation, admits statistically stationary solutions called absolute equilibria with Gaussian statistics and equipartition of energy among all Fourier modes, the latter results in an energy spectrum with $k^2$ scaling~\cite{lee1952some,kraichnan1967inertial,KRAICHNAN1989160,cicowlas,bos2006,verma2020euler}. Generally, these appear as a final stage of a relaxation dynamics that ensues for non-equilibrium initial conditions in conservative hydrodynamical equations. Moreover, such a relaxation dynamics may involve long-lasting transients which behave just as those of high Reynolds-number viscous flow with $k^{-5/3}$ inertial range and are accompanied by partial thermalization, wherein the $k^2$ spectrum is present at high wave number modes and acts like a micro-world or sink, mimicking the behavior of molecular viscosity in pumping out the energy of larger-scale modes~\cite{cicowlas,krstulovic2008two,krstulovic2009helicaltherm}. Also, the recent study of the relation between the absolute equilibrium state of the spectrally truncated Euler equations to the forced and dissipated flows of the spectrally truncated NSE in~\cite{alexakis_brachet_2020} offers an interesting insight into the nature of energy fluxes in quasi-equilibrium flows. For a one-dimensional (1D) inviscid, Galerkin-truncated BE, such a thermalized spectrum with scaling $k^0\sim \text{const.}$ was observed in~\cite{majda2000remarkable} and a Hamiltonian description was formulated in~\cite{abramovmajda2003}. Thermalization has also been investigated in the truncated Gross-Pitaevski equation for the 3D~\cite{krstulovic2011gpetherm,krstulovic2011thermbottleneck} and 2D cases~\cite{Shukla_2013, pandit2017overview}.

In~\cite{ray2011resonance}, the mechanism of thermalization in the inviscid BE and 2D Euler equations was identified and appropriately named after W. Blake's famous poem as \textit{tygers}. It is well known that solutions of the Burgers equation develop singularities in finite time, and when such singularities condense, they manifest as shocks or discontinuities~\cite{bessis1984pole,senouf1997dynamics,Senouf19971490,VANDENHEUVEL2023133686}. Most interestingly, the tygers are born from the combined effect of Galerkin truncation and pole-condensation and a loss of analyticity. The quantitative analysis of the dynamics of tygers and the onset of thermalization was discussed in~\cite{di2018dynamics,venkataraman2017onset}. Such tygers have also been encountered in the 3D Euler equation~\cite{tyger3deulersray2023}. From a more practical point of view, the studies of Galerkin truncated equations are critical since any pseudospectral numerical procedure necessitates a wave number cut-off at a particular $\kmax$. Note that spectral methods are the most accurate methods of integrating PDEs and are crucial in studies of complex systems~\cite{canuto1993spectral}.

Here, we examine the equivalence of nonequilibrium ensembles for the forced-dissipated 1D BE. We construct a formally time-reversible Burgers equation (RB) by imposing a conservation of total energy, as a consequence the viscosity becomes state dependent and can fluctuate in time. To identify different regimes of the RB, we use the dimensionless number $R=f_0\ell_f/E_0$, introduced in~\cite{shukla2019phase}, which serves as measure of the ratio of kinetic energy injected by the forcing, of amplitude $f_0$, at the length scale $\ell_f$ to the total kinetic energy $E_0$. We not only characterize the statistical features of different regimes of the RB system in between the truncation independent hydrodynamic regimes ($R\to \infty$) and the truncation dominated warm regimes ($R \to 0$), but closely examine and explain the observed dynamical behavior. The latter is made possible due to the analytical solvability of the BE, while determining any such comparable results for the 3D NSE is an extremely non-trivial task.

We find that the hydrodynamical regime persists from $R\to \infty$ to a critical point $R \overset{>}{\to} R_{c_1}$ at which the system undergoes a bifurcation akin to a continuous transition, but immediately beyond this critical point the dynamical and statistical properties are strongly influenced by the truncation effects. For example, we witness the emergence of a dynamical regime characterized by the presence of the resonance phenomenon called tygers. Quite surprisingly, a further decrease of the control parameter leads to yet another transition, but a discontinuous one at $R_{c_2}\ll R_{c_1}$, where the global quantities exhibit jump and attain values associated with the presence of a quasi-equilibrium state.

The comparative analysis of the BE and RB shows that the \textit{equivalence conjecture} holds true, not just for $R \overset{>}{\to}R_{c_1}$ in the hydrodynamic regime, but for all the values of $R$, irrespective of the truncation effects.

The remainder of this paper is organized as follows. In Sec.~\ref{sec:trb} we introduce the time-reversible formulation of the 1D BE and describe the numerical setup that we use in this paper. Section~\ref{sec:masymp} contains a discussion of the stationary solution of the forced BE obtained using the method of matched asymptotic expansion. In Sec.~\ref{sec:results} we present our results and discussion, followed by concluding remarks in Sec.~\ref{sec:conclusions}.

\section{The time-reversible 1D Burgers equation}
\label{sec:trb}

\subsection{Governing equation}

The 1D forced viscous BE is given by
\begin{equation}\label{eqn:BE}
	\frac{\partial u}{\partial t} + u\frac{\partial u}{\partial x} =\nu \frac{\partial^{2}u}{\partial x^{2}} + f,
\end{equation}
where $u = u(x,t)$ is the velocity field, $\nu$ is the kinematic viscosity and $f= f(x)$ is the forcing term. The presence of the viscous term breaks the invariance of the above equation under the transformation:
\begin{center}
	$\mathcal{T} : t\rightarrow -t ;\hspace{2mm} u\rightarrow -u.$
\end{center}
The macroscopic dynamics described by the BE is irreversible. However, if the \textit{equivalence conjecture} extends to the Burgers system, then the invariance under $\mathcal{T}$ can be restored by the imposition of a suitable global constraint on the system. We choose the latter to be the conservation of the total energy, defined by
\begin{equation}
E\equiv\langle u(x,t)^2/2 \rangle = \frac{1}{2\pi} \int_{0}^{2\pi} (u^2/2) dx,
\end{equation}
where $\langle \cdot\rangle$ denotes spatial averaging. This constraint, together with the energy balance equation
\begin{equation}
	\frac{1}{2} \frac{\partial}{\partial t} \langle u^{2}\rangle =- \boldsymbol{\nu_{r}}\Bigl\langle \Bigl(\frac{\partial u}{\partial x}\Bigr)^2\Bigr\rangle  +\langle f u\rangle  = 0,
\end{equation}
allows us to determine the modified dissipation coefficient as
\begin{equation}\label{eq:visrev}
	\nu_r[u] = \frac{\langle fu \rangle}{\langle (\partial u/\partial x)^2\rangle}=\frac{\epsilon}{\Omega},
\end{equation}
where  $\epsilon$ is the energy injection rate
\begin{equation}
	\epsilon \equiv \langle fu \rangle  = \frac{1}{2\pi} \int_{0}^{2\pi} (fu) dx
\end{equation}
and $\Omega$ is the strain-rate-squared average
\begin{equation}
\Omega \equiv \biggr \langle \biggr ( \frac{\partial u}{\partial x} \biggr )^2 \biggr \rangle  = \frac{1}{2\pi} \int_{0}^{2\pi} (u_{x})^{2} dx.
\end{equation}

Equation~\eqref{eq:visrev} shows that the viscosity, $\nu_{r}$, is altered in a fundamental way. It now depends on the state of the system ($u(x,t)$) and can fluctuate in time. Moreover, it is easy to check that the dissipation term  $\nu_r \partial^2 u/ \partial x^2$ is now invariant under the transformation $\mathcal{T}$. The subscript $r$ in $\nu_r$ indicates that the new system is ``formally'' time-reversible and is governed by the equation
\begin{equation}\label{eq:RB}
	\frac{\partial u}{\partial t} + u\frac{\partial u}{\partial x} =\boldsymbol\nu_{r} \frac{\partial^{2}u}{\partial x^{2}} +f.
\end{equation}
Hereafter, we refer to the above equation as the RB.

\subsection{Numerical setup}

To compare the statistical properties of the BE and RB systems in various non-equilibrium steady states, we use a highly accurate pseudo-spectral method to numerically integrate these equations over a periodic interval of length $L=2\pi$. The Fourier series expansion of
$u(x,t)$ is given by:
\begin{equation}
u(x,t)=\sum_{k=-\infty}^{\infty} u(k,t)e^{ikx},
\end{equation}
where $k$ is the wave number. For the numerical implementation the Fourier series is truncated to have a finite number of collocation points $N_c$ and we use the $2/3$-dealiasing rule, so the maximum wave number is $\kmax = N_c/3$. For time stepping we use the standard RK4 scheme with sufficiently small time steps, $dt\in[10^{-5},10^{-7}]$, in order to ensure a good energy conservation in our DNSs; see Table~\ref{tab:enconsvserr} in Appendix~\ref{app:errconsE} for more details.

\subsubsection{Initial conditions and forcing}
The BE and RB systems are numerically integrated starting from the following initial velocity field, unless otherwise stated,
\[u(x,0)=u_0\sin(x),\]
where $u_0$ is the amplitude of the velocity profile. Thus, for the RB system the initial energy $E_0=u_0^2/4$ remains conserved and is fixed at a desired value by an appropriate choice of $u_0$.

We use a spatially periodic forcing given by \[f(x,0)=f_0\sin(x),\] where $f_0$ is the forcing amplitude and acts at the large length scale, $\ell_f=2\pi$.

\subsubsection{Simulation protocols and the non-dimensional control parameter $R$}

To carry out a systematic investigation of the RB system, we introduce a non-dimensional control parameter
\begin{equation}
R = \frac{\ell_f f_0}{E_0}.
\end{equation}
We can regard $R$ as the ratio of the energy injected into the system, $\ell f_0$, at the forcing length scale $\ell_f$ to the conserved initial energy $E_0$ of the system~\cite{shukla2019phase}.

For the statistical and dynamical characterization of the RB system, we vary $R$ for a set of DNSs according to the following two protocols:
\begin{enumerate}
	\item \textit{Protocol E}: The initial energy $E_0$ is varied, while the forcing amplitude $f_0$ is held fixed.
	\item \textit{Protocol F}: The forcing amplitude $f_0$ is varied, while $E_0$ is kept fixed.
\end{enumerate}
It is useful to consider the following two asymptotic cases: $R \to 0$ and $R \to \infty$. For the protocol $E$,  $R \to 0$ and $R \to \infty$ correspond to $E_0 \to \infty$ and $E_0 \to 0$, respectively. Similarly, for the protocol $F$, $R \to 0$ and $R \to \infty$ correspond to $f_0 \to 0$ and $f_0 \to \infty$, respectively.

We assert that $R \to \infty$ corresponds to a ``hydrodynamic'' regime of the RB system, wherein the modes near the truncation wave number $\kmax$ do not influence the dynamics. However, the limit $R \to 0$ corresponds to a ``warm'' phase of the RB system, where the truncation effects significantly influence the dynamics leading to partial thermalization at large wave numbers, which progressively spreads to small wave numbers as $R$ decreases. The thermalized scales exhibit $k^0$ scaling which is associated with the equipartition of energy.

\subsubsection{Simulation parameters}

We keep the value of $f_0$ fixed at $1.0$ for the protocol $E$, whereas we use $E_0=0.25$ for the protocol $F$. In Table~\ref{tab:dnsparam} we give the additional details of our DNS runs.

\begin{table}[h!]
	\centering
	\begin{tabular}{|c|c|c|c|c|c|}
		\hline
		SET & $N_c$ & $k_{max}$ & $f_0$ & $E_0$ & $R$\\
		\hline\hline

		$E_{1024}$ & 1024 & 341 & 1 & 0.15625-56.25 & 402.12-0.11\\
		\hline
		$F_{1024}$ & 1024 & 341 & 1-0.004 & 0.25 & 402.12-0.11\\
		\hline
		$E_{2048}$ & 2048 & 682 & 1 & 0.15625-56.25 & 402.12-0.11\\
		\hline
		$F_{2048}$ & 2048 & 682 & 1-0.004 & 0.25 & 402.12-0.11\\
		\hline
		$E_{4096}$ & 4096 & 1365 & 1 & 0.975-56.25 & 6.44-0.11\\
		\hline
		$E_{8192}$ & 8192 & 2730 & 1 & 0.975-30.25 & 8-0.21\\

		\hline

	\end{tabular}
	\caption{List of parameters used in the paper. The names of the sets are chosen as $E_{[N_c]}$ or $F_{[N_c]}$ to indicate the protocol being followed. $N_c$ is the number of collocation points.}\label{tab:dnsparam}
\end{table}

\section{Stationary state solution of BE and its corresponding Poles}
\label{sec:masymp}

We use the method of matched asymptotic expansion to obtain a stationary solution of the forced BE~\cite{banerjee2019fractional}. This will be used in the later sections to justify and understand several observations made on the basis of our DNS runs.

For a stationary solution, we set $\partial_{t}u = 0$ in the BE~\eqref{eqn:BE} and solve the following resulting differential equation
\begin{equation}\label{eq:stationaryBE}
	u\frac{du_{}}{dx} =\nu \frac{d^{2}u}{dx^{2}} + f_0 \sin x.
\end{equation}
To proceed further, we formulate an outer solution and an inner solution of the above equation. The complete solution is then constructed by matching the outer and the inner solutions.

The outer solution corresponds to the large scales, where the viscosity is not effective. Hence, we set $\nu=0$ to obtain the following equation
\begin{equation}
	\frac{d }{ dx }(u_{outer}^2) = 2f_0\sin x,
\end{equation}
whose solution is given by
\begin{equation}\label{sssBEouter}
	u_{outer}(x) = 2 \sgn(x-\pi)\sqrt{f_0} \sin(x/2).
\end{equation}
This solution implies the existence of a shock at $x=\pi$ for the chosen forcing.

The inner solution corresponds to the inner structure of the shock at $x=\pi$ governed by the viscosity. Therefore, we scale the ODE~\eqref{eq:stationaryBE} by $X=(x-\pi)/\nu$ and retain only the leading order terms in $\nu$ to obtain
\begin{equation}
	\frac{d}{dX} u_{inner}^2 = 2\frac{d^{2}}{dX^{2}} u_{inner}.
\end{equation}

The above equation must be solved with appropriate boundary conditions in order to ensure a match between the smallest scales of the outer solution (inner limit of the outer solution) and the largest scales of the outer solution (outer limit of the inner solution). To ensure this, we choose the boundary conditions given by \[u_{inner}(\pm \infty)=u_{outer}(\pi \pm),\] which yields \[u_{inner}(\pm \infty)=\mp 2 \sqrt{f_0}.\] Thence, we have
\begin{equation}\label{eq:sssBEinner}
	u_{inner}(X) = -2 \sqrt{f_0} \tanh(\sqrt{f_0}X).
\end{equation}

Equations~\eqref{sssBEouter} and~\eqref{eq:sssBEinner} allow us to write the final matched solution, valid over the entire domain, as
\begin{equation}\nonumber
	u=u_{inner}+u_{outer}-u_{overlap},
\end{equation}
where  $u_{overlap}$ is the outer limit of the inner solution (or correspondingly the inner limit of the outer solution). Hence, the stationary solution of the BE~\eqref{eq:stationaryBE} is given by
\begin{equation} \label{eq:sssBEmatched}
	\begin{split}
		u &=2\sgn(x-\pi)\sqrt{f_0}(\sin(x/2)-1) \\
		& \quad +2\sqrt{f_0}\tanh(\sqrt{f_0}(\pi-x)/\nu).
	\end{split}
\end{equation}
It is worthwhile to recall that we have assumed the viscosity to be a perturbation, therefore, the solution above begins to hold better as $\nu$ gets smaller and smaller. Also, our choice of forcing type plays an important role and significantly affects the dynamical behavior of the RB system. We explain this in more details in Sec.~\ref{sec:cpt}.

\paragraph{Poles and singularities} It is well known that the inviscid BE has branch point singularities which condense on the real axis in finite time; this condensation can be associated to a Curie-Weiss transition~\cite{bessis1990complex,Choquard2004}. Moreover, this condensation is responsible for the observed  finite time blow-up and the formation of a shock. In the case of a viscous BE, we have an infinite set of poles which march along the imaginary axis (for a paradigmatic initial condition) and do not condense on the real axis~\cite{senouf1997dynamics}. We refer to the Appendix~\ref{app:poleexpn} for a brief account of pole expansion solutions of the BE. \footnote{Although our choice of the initial condition is different from the paradigmatic initial condition, the features of the solutions remain intact.} For an unforced viscous BE, with an initial condition $u_0 = \sin (x)$, poles appear in the complex plane away from the x-axis along the line $\Re(x)=\pi$, instead of being along the imaginary axis.

We can use the inner solution Eq.~\eqref{eq:sssBEinner} to show that the poles of the solution lie at locations where $\cosh(\sqrt{f_0}X)=0$. Thus, we have
\begin{equation}\label{eq:sssBEpoles}
	x = \pi + \frac{i \nu \pi}{\sqrt{f_0}}( 2n \pm \frac{1}{2}), \quad \text{\space $n \in \mathbb{Z}$}.
\end{equation}
Note that as $\nu \to 0$, the poles condense at $x=\pi$ on the real-axis and lead to a development of a shock or a discontinuity in the velocity field.

\section{Results and Discussion}
\label{sec:results}

In this section we present our results and discussion based on the analytical estimations and numerical simulations of the BE and RB systems.

\subsection{Statistical regimes of the RB}
\label{sec:regimesRB}

\begin{figure*}
	\begin{subfigure}{.497\textwidth}
		\centering
		\includegraphics[width=\textwidth]{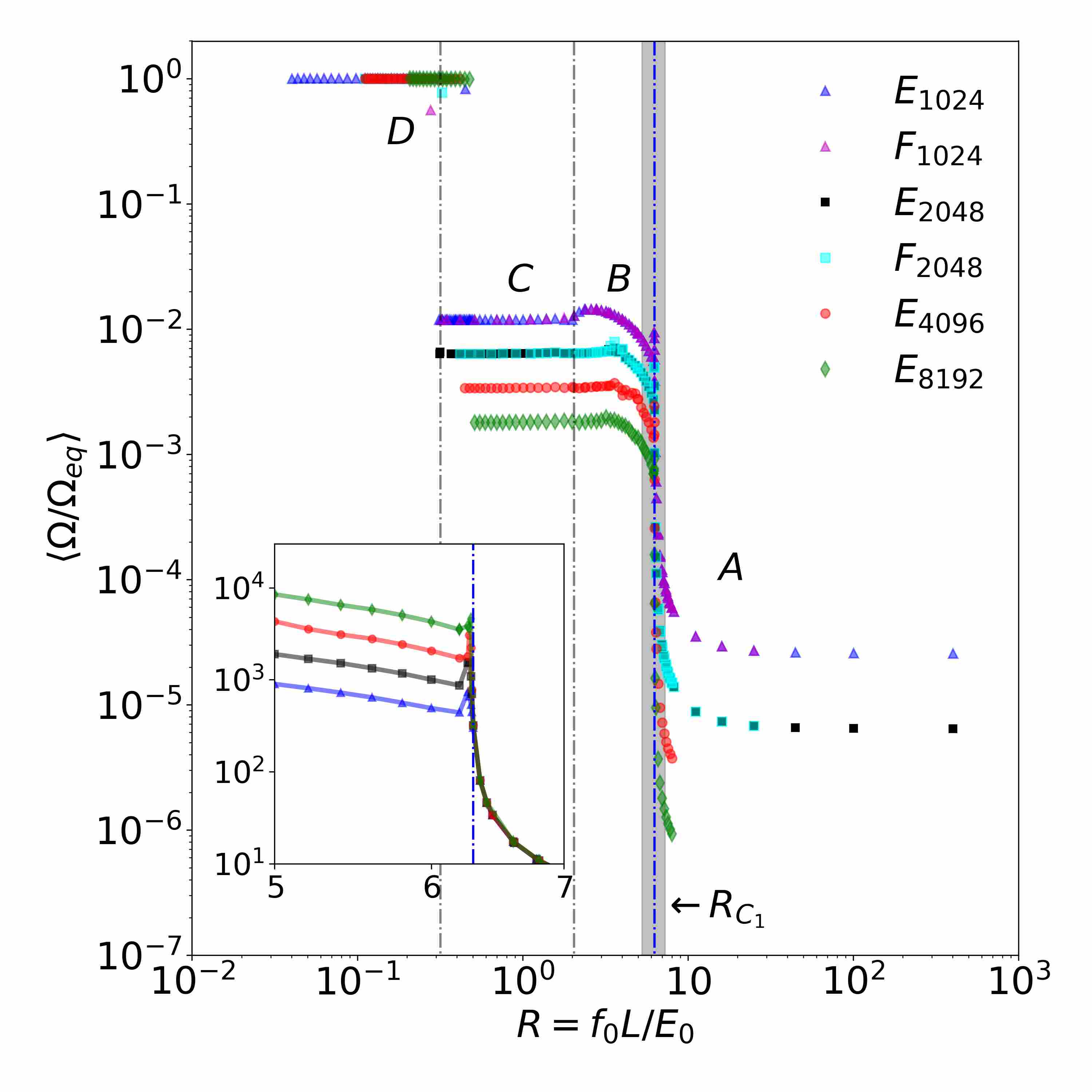}
		\put(-205,230){(a)}
		\put(-188,78){$\langle \Omega \rangle$}
		\put(-168,54){$R$}
		\phantomsubcaption\label{fig:ensavg}
	\end{subfigure}
	\begin{subfigure}{.497\textwidth}
		\centering
		\includegraphics[width=\textwidth]{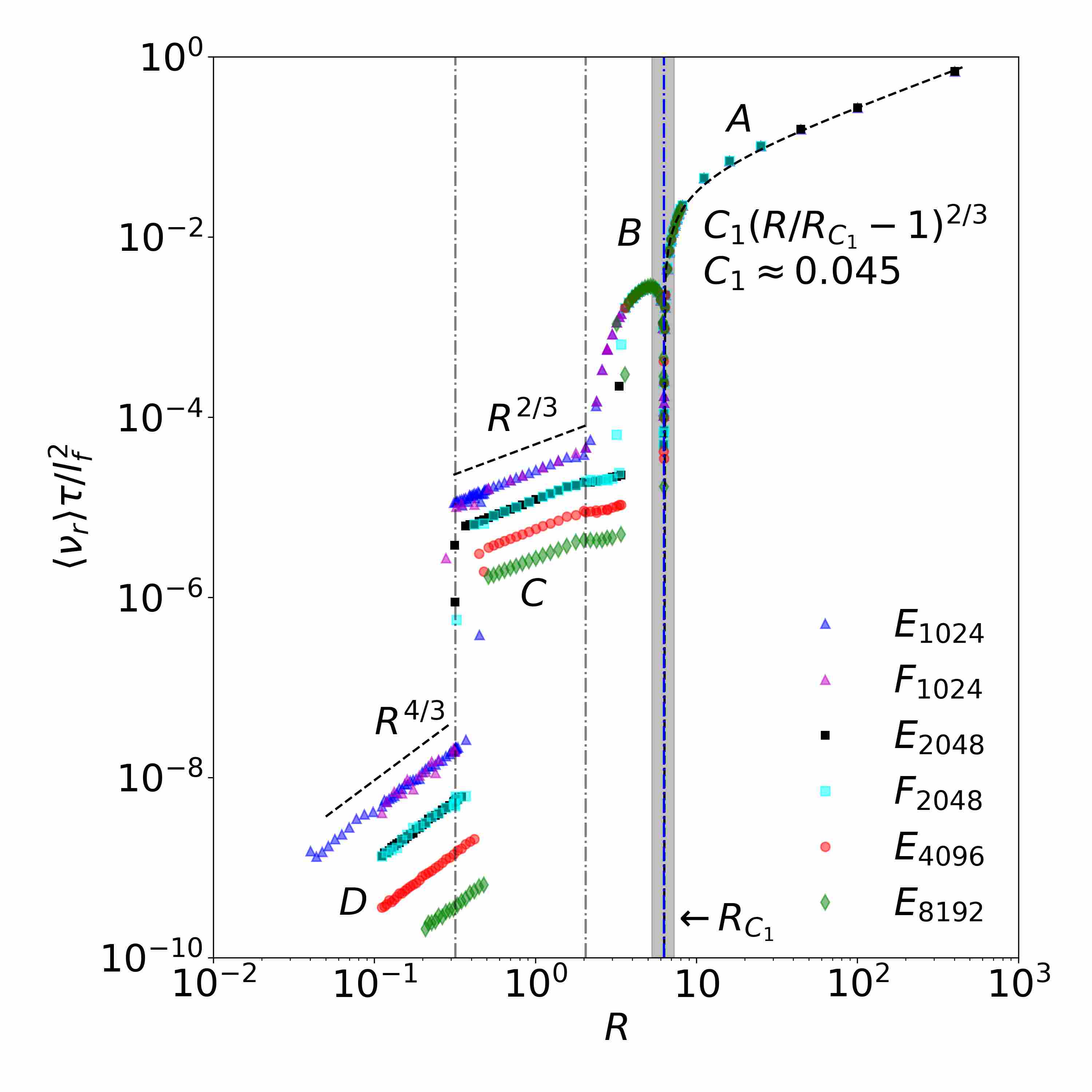}
		\put(-200,230){(b)}
		\phantomsubcaption\label{fig:nuavg}
	\end{subfigure}
	\caption{Statistical regimes of RB. (a) Time-averaged $\langle \Omega/\Omega_{eq} \rangle$ vs $R$, where $\Omega_{eq}=(2/3)E_0\kmax^2$. The inset shows the plots of un-normalized $\langle \Omega \rangle$ vs $R$ exhibiting a perfect collapse for $R>R_{c_1}$ and bifurcations to a Galerkin truncation wave number, $\kmax$, dependent state at the critical point $R_{c_1}$. (b) Time-averaged $\nu_r(\tau/\ell^2_f)$ vs $R$. $\tau=(\ell^2_f/\langle \epsilon \rangle)^{1/3}$, where $\langle \epsilon \rangle$ is the energy injection rate, $\ell_f$ is the forcing length-scale and $E_0$ is initial total energy. The dashed-dot lines indicate the boundaries between different regimes of RB for $E_{1024}$. The blue dashed-dot line at the critical point $R_{c_1}$ represents the continuous transition $\mathcal{T}_{AB}$ from regime A to B. The second grey dashed-dot line at $R_{\star}$ represents the transition $\mathcal{T}_{BC}$ from regime B to C, while the third grey dashed-dot line at $R_{c_{2}}$ represents the discontinuous transition $\mathcal{T}_{CD}$ from regime C to D. The black dashed lines in (b) indicate different scaling behavior of $\nu_r$.}
	\label{fig:globalensnur}
\end{figure*}

The time averaged behavior of the global quantities $\langle \Omega \rangle$ and $\langle \nu_r \rangle$ reveals four distinct regimes of the RB system, when the control parameter $R$ is varied. We have identified and named them (quite uninterestingly) as A, B, C, and D on Figs.~\ref{fig:globalensnur} (a) and (b). On these plots, the demarcation between different regimes are indicated by vertical lines. Also, hereafter, we indicate the transition from regime A to B as $\mathcal{T}_{AB}$; similarly, $\mathcal{T}_{BC}$ and $\mathcal{T}_{CD}$ represent the transitions from regime B to C and C to D, respectively. Moreover, the first and the last transitions are very distinct and clearly evident on these plots as: (i) $\mathcal{T}_{AB}$, a continuous transition from regime A to B; (ii) $\mathcal{T}_{CD}$, a clear jump from regime C to D. These two transitions are associated with a large increase in values of $\langle \Omega \rangle$ of nearly two orders of magnitude. \footnote{Here, we have normalized the value of $\Omega$ by $\Omega_{\rm eq}=2/3\,E_0\kmax^2$, see Appendix~\ref{app:abseq}. $\nu_r$ has been normalized by an unit constructed using $\ell_f$ and a time scale derived from the injection energy $\epsilon$, namely $\tau=(l_f^2/\langle \epsilon \rangle)^{1/3}$.} \footnote{Our choice of $\Omega_{eq}$ makes good sense as is clear from Fig.~\ref{fig:ensavg}, where the values of $\Omega/\Omega_{eq}\simeq1$ in regime D, indicating a thermalized state. Note that a transition to thermalized state is a well known result for Galerkin truncated equations of fluids~\cite{alexakis_brachet_2020}. Such a phenomenon of thermalization has been studied in great detail in the inviscid BE~\cite{ray2011resonance,di2018dynamics}, where a resonance phenomenon has been observed which is a precursor to a completely thermalized state. These have been termed as tygers. A brief account of tygers for the inviscid and forced-inviscid BE are provided in Appendix~\ref{app:tygersiBE} and {Appendix~\ref{app:tygersfiBE}}.}

Figure~\ref{fig:globalensnur}(b) shows that for $R > 6.3$ the values of $\nu_r$ do not depend on $\kmax$. This is also true for the un-normalized $\langle \Omega \rangle$ in the regime A, see the inset of Fig.~\ref{fig:globalensnur}(a). Since $\langle \Omega \rangle$ and $\langle \nu_r \rangle$  are resolution independent, the RB system is in a hydrodynamic regime. The values of $\nu_r$ can be shown to follow an empirical relation
\begin{equation}
	\langle \nu_r \rangle = C_1 \, (R/R_{c_1} - 1)^{2/3},
\end{equation}
where $C_1 \approx 0.045$ and $R_{c_1} \approx6.3$ is the critical point for the transition from regime A to B, indicated by the blue dash-dotted line in Fig.~\ref{fig:globalensnur}(a).

At the critical point $R = R_{c_1}$ the truncation wave number $\kmax$ dependence starts to becomes evident. Both $\langle \Omega \rangle$ and $\langle \nu_r \rangle$ exhibit bifurcations to a resolution dependent state. The inset of Fig.~\ref{fig:globalensnur}(a) demonstrates this for the un-normalized $\langle \Omega \rangle$. Moreover, as remarked earlier, at $R_{c_1}$ the bifurcation in $\langle \Omega/\Omega_{\rm eq} \rangle$ is accompanied by a sharp, but continuous, increase in its value, resembling a `cliff'. In contrast, $\langle \nu_r \rangle$ exhibits a very sharp drop, followed by an extremely quick recovery, which is almost akin to a `gorge', as $R$ is made to vary across $R_{c_1}$.

As $R$ decreases further below $R_{c_1}$, we enter the regime B, wherein the behavior of the strain-squared and viscosity is resolution dependent. Below $R_{c_1}$, $\langle \Omega/\Omega_{\rm eq} \rangle$ gradually rises and eventually plateaus to a constant at a resolution dependent value of $R=R_\star$. Note that the saturation value itself depends on the resolution. For example, there is a marked difference in the behavior of runs $E_{1024}$ and $F_{1024}$ in comparison with runs at higher resolutions. For the lowest resolution runs, values of $\langle \nu_r \rangle \tau/\ell^2_f$ fall gradually, whereas for the higher resolution runs a sudden drop-off is observed, which is followed by a passage through a shallow local minima. Post this, the system is in regime C. We assert that this behavior in regime B and the higher values of $\langle \Omega/\Omega_{\rm eq} \rangle$, are a result of truncation effects, namely tygers causing oscillatory behavior in the time-series of global quantities.

In regime C, the plateau in $\langle \Omega/\Omega_{\rm eq} \rangle$ that was attained at the end of regime B at $R_{\star}$ is maintained, whereas $\langle \nu_r \rangle \tau/\ell^2_f \sim R^{2/3}$. Recall that $\Omega_{\rm eq} \sim (2/3)\, E_0 \,\kmax^2$ and $E_0 \sim R^{-1}$; hence, $\langle \Omega \rangle \sim R^{-1}$. Moreover, later we show that $\langle \epsilon \rangle \sim R^{-1/2}$.

Interestingly, the regime C ends in a critical point at $R=R_{c_2}$, at which we observe a discontinuous jump to the final thermalized regime. Note that $R_{c_2}$ depends on resolution with thermalization appearing much earlier for the set of runs at higher resolutions. For $E_{1024}$ and $F_{1024}$ the critical point $R_{c_2} \simeq 0.3$. Upon entering regime D, $\langle \Omega/\Omega_{\rm eq} \rangle$ jumps by two orders of magnitude and then stays fixed at $\langle \Omega/\Omega_{\rm eq} \rangle = 1$, whereas for the viscosity $\langle \nu_r \rangle \tau/\ell^2_f$ we witness a change of scaling from $\sim R^{2/3}$ to $\sim R^{4/3}$. This can be accounted for by looking at the behavior of the injection energy $\epsilon$, which is responsible for the change in the scaling of $\nu_r$, even though $\Omega$ continues to exhibit $R^{-1}$ scaling. The scaling behavior of the injection energy is given in Sec.~\ref{sec:dpt}.

Below we describe the microscopic dynamics.

\subsubsection{Regime A}

\begin{figure*}
\centering
	\resizebox{\textwidth}{!}{%
\includegraphics[scale=0.2]{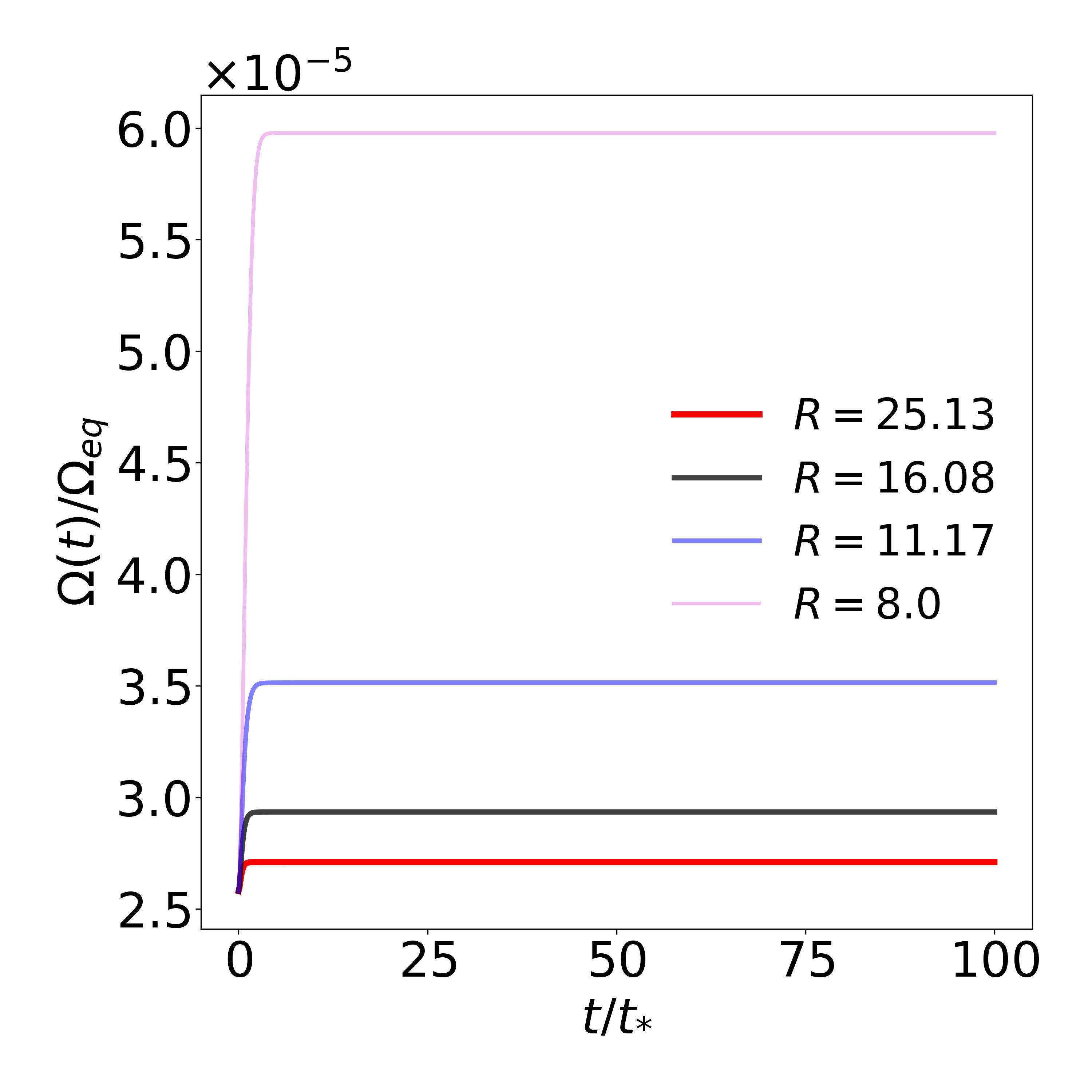}
\put(-25,105){(a)}
\includegraphics[scale=0.2]{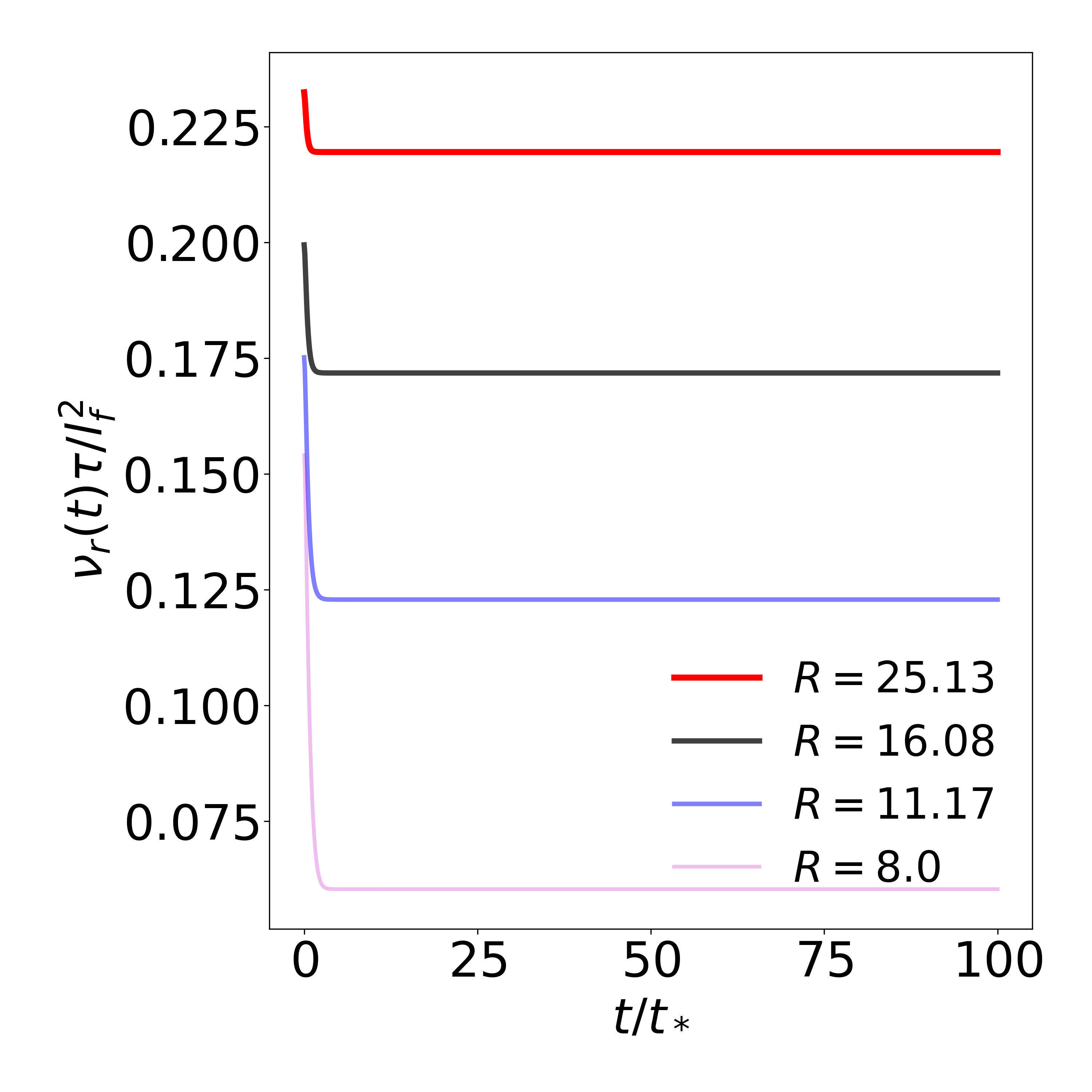}
\put(-25,105){(b)}
\includegraphics[scale=0.2]{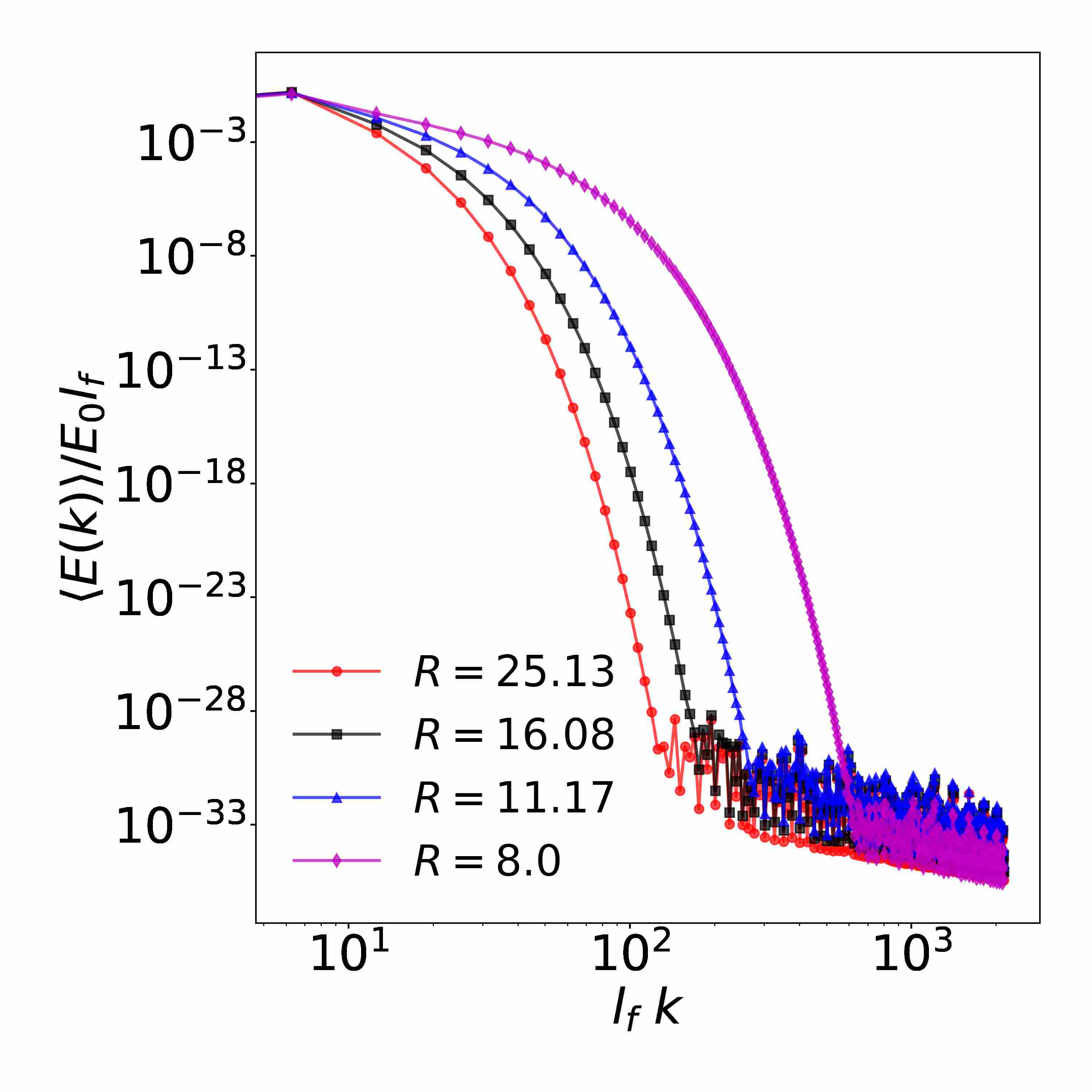}
\put(-25,105){(c)}
}
\caption{Regime A. Time series of: (a)$\Omega/\Omega_{eq}$ and (b) $\nu_r$. (c) Time-averaged energy spectrum $\langle E(k) \rangle$ decays exponentially at large wave numbers; the tail of the spectrum rises as $R$ decreases. $\Omega_{eq}=(2/3)E_0\kmax^2$, $\tau=(\ell^2_f/\langle \epsilon \rangle)^{1/3}$, where $\langle \epsilon \rangle$ is the energy injection rate, $\ell_f$ is the forcing length-scale, and $E_0$ is the initial total energy. $t_*=1$ is the time of shock formation in the inviscid BE for $u(x,0)=\sin x$.
}
\label{fig:regimeA}
\end{figure*}

As remarked earlier, regime A is the hydrodynamic regime, wherein the RB dynamics does not feel the truncation wave number $\kmax$, as the small scales are well resolved. In the limit $R \to \infty$, the effective scale-by-scale dissipation mechanism due to the reversible viscosity is strong enough to kill the nonlinear transfer of energy, \textit{i.e.}, the energy injected at the forcing length scale $\ell_f$ is quickly damped away in its immediate vicinity. Figure 2 (a) and (b) show that at finite, but large $R$, the $\Omega$ and $\nu_r$ time-series quickly go to a frozen-state, after short initial transients have subsided. The corresponding energy spectra show a quick exponential fall-off, thereby suggesting that the excitation of motion associated with large wave numbers is strongly suppressed, see Fig.~\ref{fig:regimeA} (c). However, as $R$ is decreased towards $R_{c_1}$, the $\Omega$ and $\nu_r$ time-series begin to exhibit a comparatively richer behavior, which is also reflected on the energy spectra in terms of excitation of larger number of Fourier modes. Note that when $R$ is decreased the values of $\langle \Omega \rangle$ gradually increases, whereas $\langle \nu_r \rangle$ goes down.

\subsubsection{Regime B}
\label{sec:regimeB}

\begin{figure}
	\centering
	\resizebox{\columnwidth}{!}{%
	\includegraphics[scale=0.2]{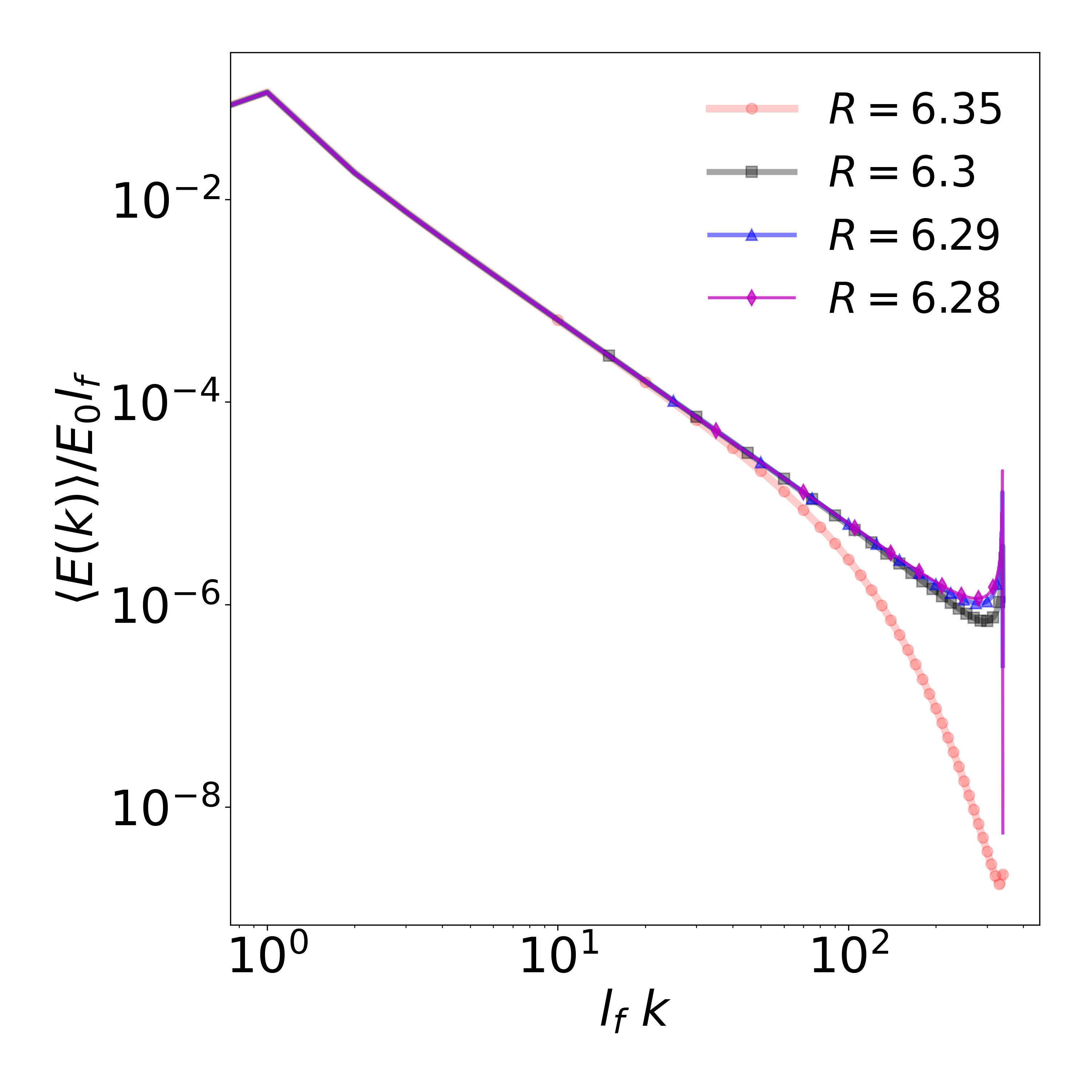}
}
	\caption{Plots of time-averaged $E(k)$ for $R_{c_1} < R < 8$ for $E_{1024}$. The behavior of the spectrum at high wave numbers is drastically modified as $R \overset{>}{\to}R_{c_1}$.}
	\label{fig:regimeABcross}
\end{figure}

\begin{figure*}
		\centering
	\resizebox{\textwidth}{!}{%
		\includegraphics[scale=0.2]{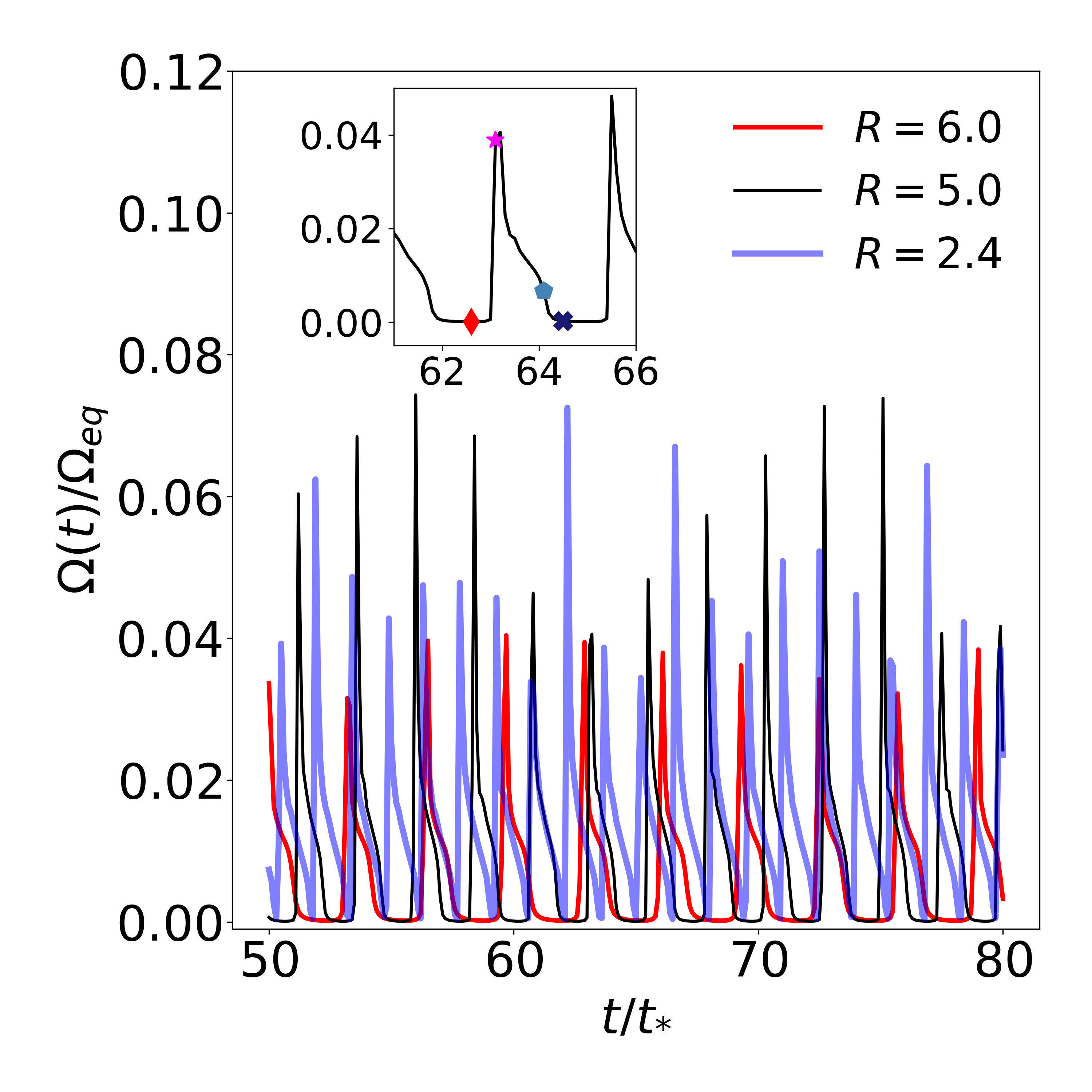}
		\put(-20,90){(a)}
		\includegraphics[scale=0.2]{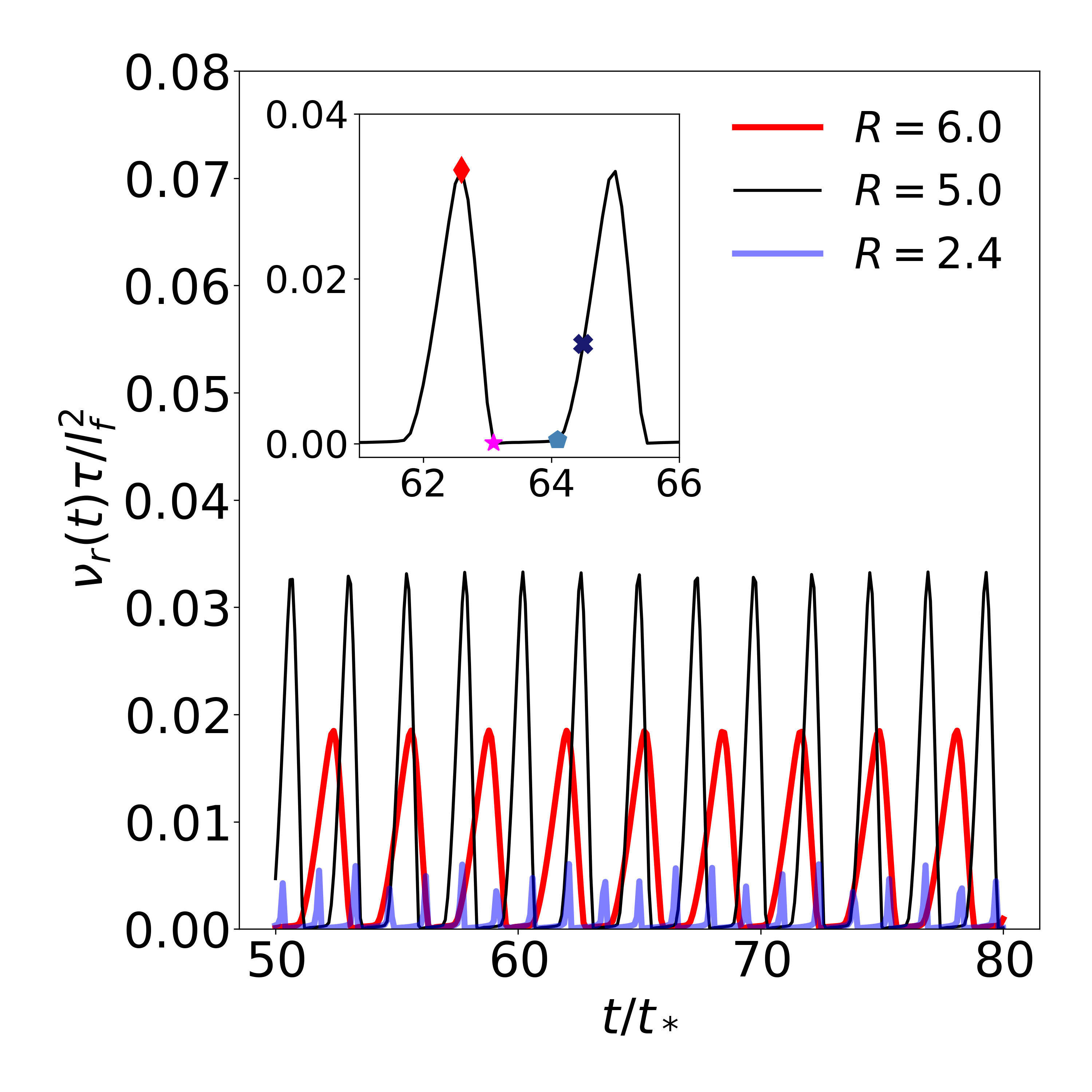}
		\put(-20,90){(b)}
		\includegraphics[scale=0.2]{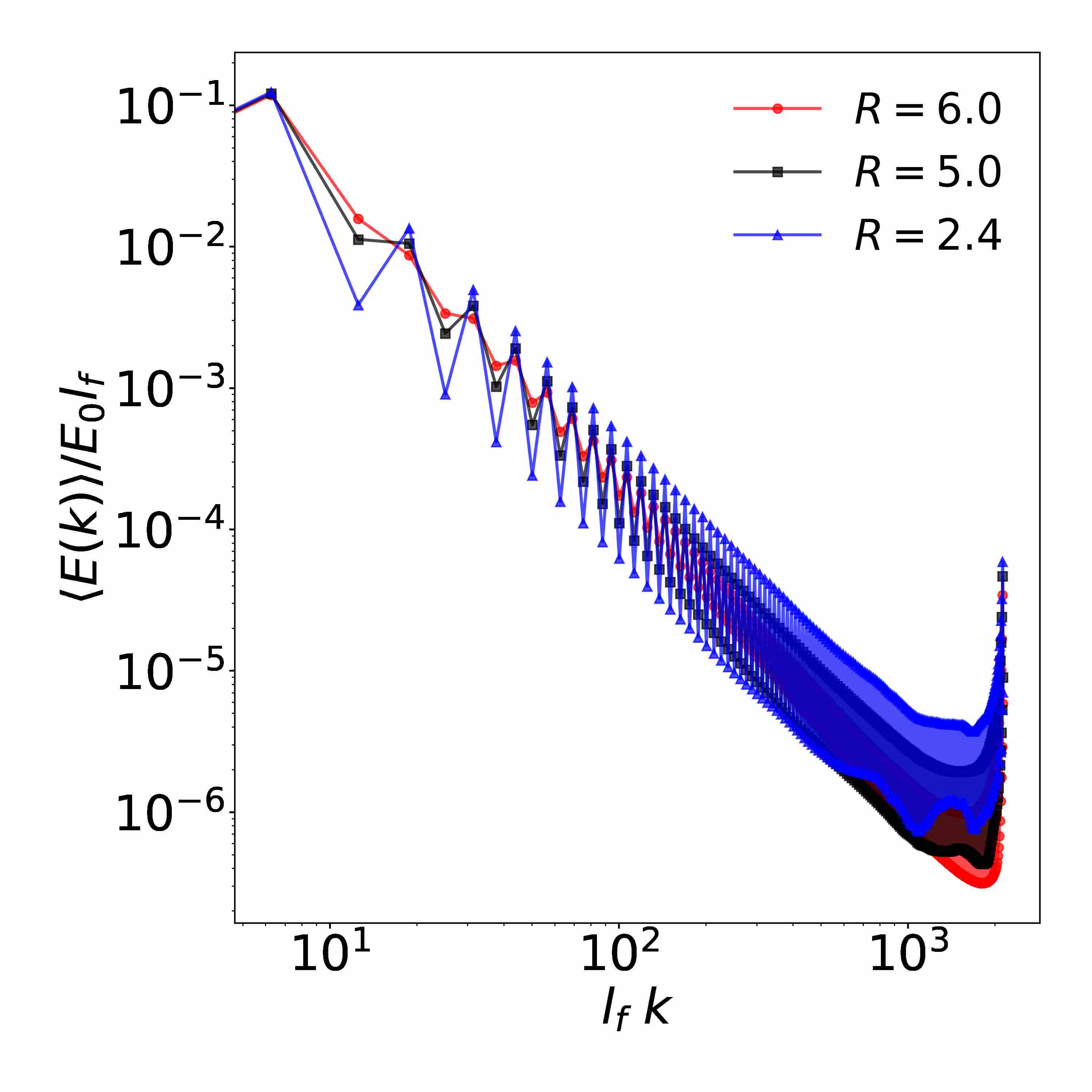}
		\put(-20,90){(c)}
}
	\caption{Regime B for $E_{1024}$. Time series of: (a)$\Omega/\Omega_{eq}$ and (b) $\nu_r$. Insets show two cycles of oscillation over the duration $t/t_* \approx 62$ to $66$. (c) Time-averaged energy spectrum exhibits $E(k) \sim k^{-2}$ behavior, with oscillations superimposed over it.  Plots are shown for $R=6.0$ (red line with filled circles), $R=5.0$ (black line with filled square) and $R=2.4$ (blue line with filled upward triangles). $\Omega_{eq}=(2/3)E_0\kmax^2$, $\tau=(\ell^2_f/\langle \epsilon \rangle)^{1/3}$, where $\langle \epsilon \rangle$ is the energy injection rate, $\ell_f$ is the forcing length-scale, and $E_0$ is the initial total energy. $t_*=1$ is the time of shock formation in the inviscid BE for $u(x,0)=\sin x$.}
	\label{fig:regimeB}
\end{figure*}

As $R$ crosses $R_{c_1}$, the truncation effects start to creep-in, see Fig.~\ref{fig:regimeABcross}. First, the exponential decay of the energy spectra is gradually lost and the scaling region $E(k)\sim k^{-2}$ gets extended till $\kmax$, followed by a piling-up of energy near $\kmax$. These truncation effects cause a bifurcation (transition) to occur at $R_{c_1}$, which is reflected in the global quantities, e.g., $\langle \Omega \rangle$ and $\langle \nu_r \rangle$. To understand this better, we look at the microscopics of the RB system for $R_{\star} < R < R_{c_1}$.

Figure~\ref{fig:regimeB} (a) and (b) show that in regime B, spatially averaged values, $\Omega$ and $\nu_r$ exhibit an oscillatory behavior. We have shown $\Omega(t)$ and $\nu_r(t)$ time series for $R=6$, $5$, and $2.4$. The amplitude of these oscillations in $\nu_r$ depends on $R$, it increases as $R$ decreases below $R_{c_1}$, reaches a maximum and then again starts to decrease till $R_{\star}$. It is largest near $R\approx 5$, see Fig.~\ref{fig:regimeB} (b). Moreover, it is interesting to note that (time-averaged) energy spectra also exhibit an oscillatory behavior, but in wave numbers ($k$). It is likely that the appearance of \textit{sustained} oscillations in this regime is associated with the existence of certain spatial structures in the velocity field.

\begin{figure*}
	\centering
	\resizebox{\textwidth}{!}{%
	\includegraphics[scale=0.2]{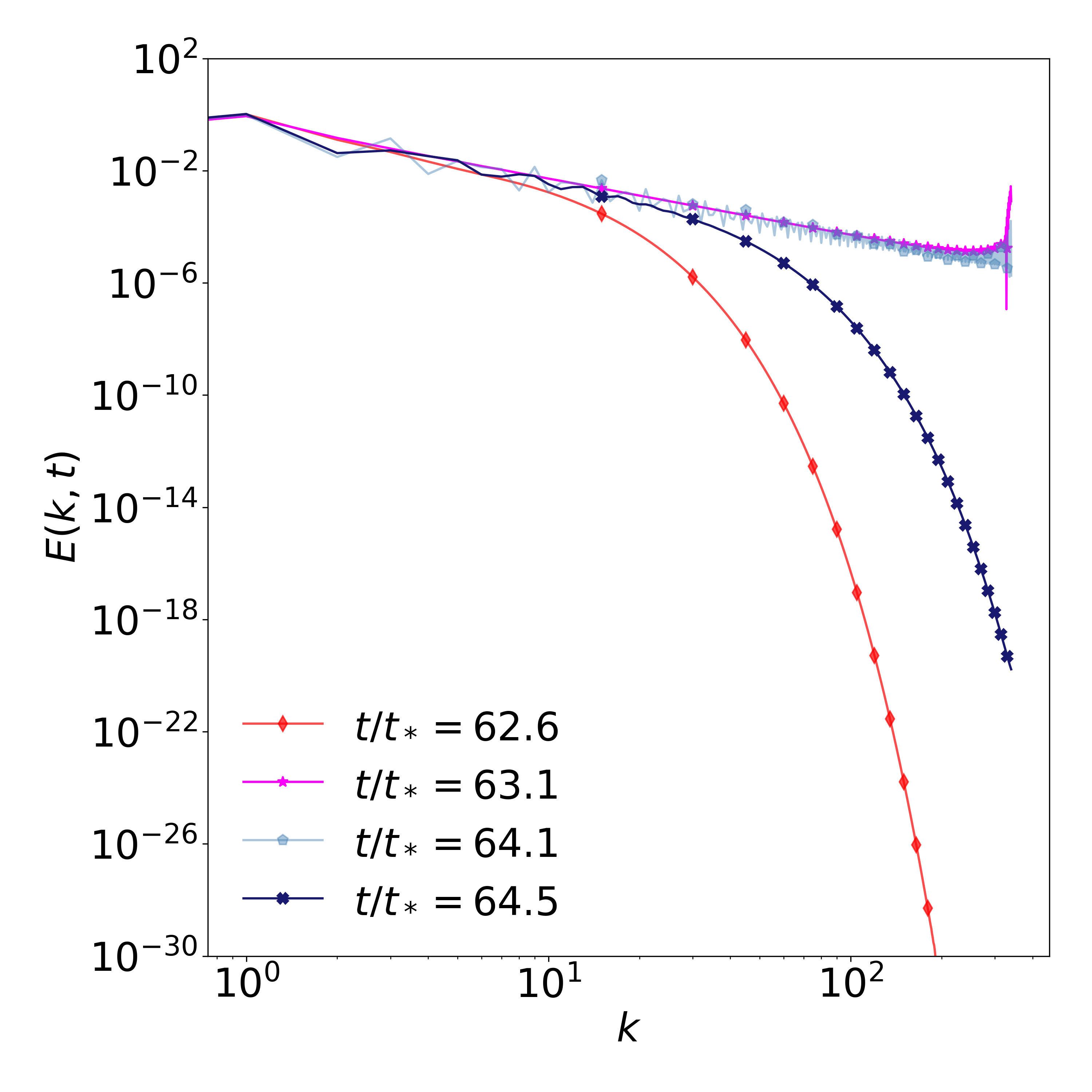}
	\put(-105,110){(a)}
	\includegraphics[scale=0.2]{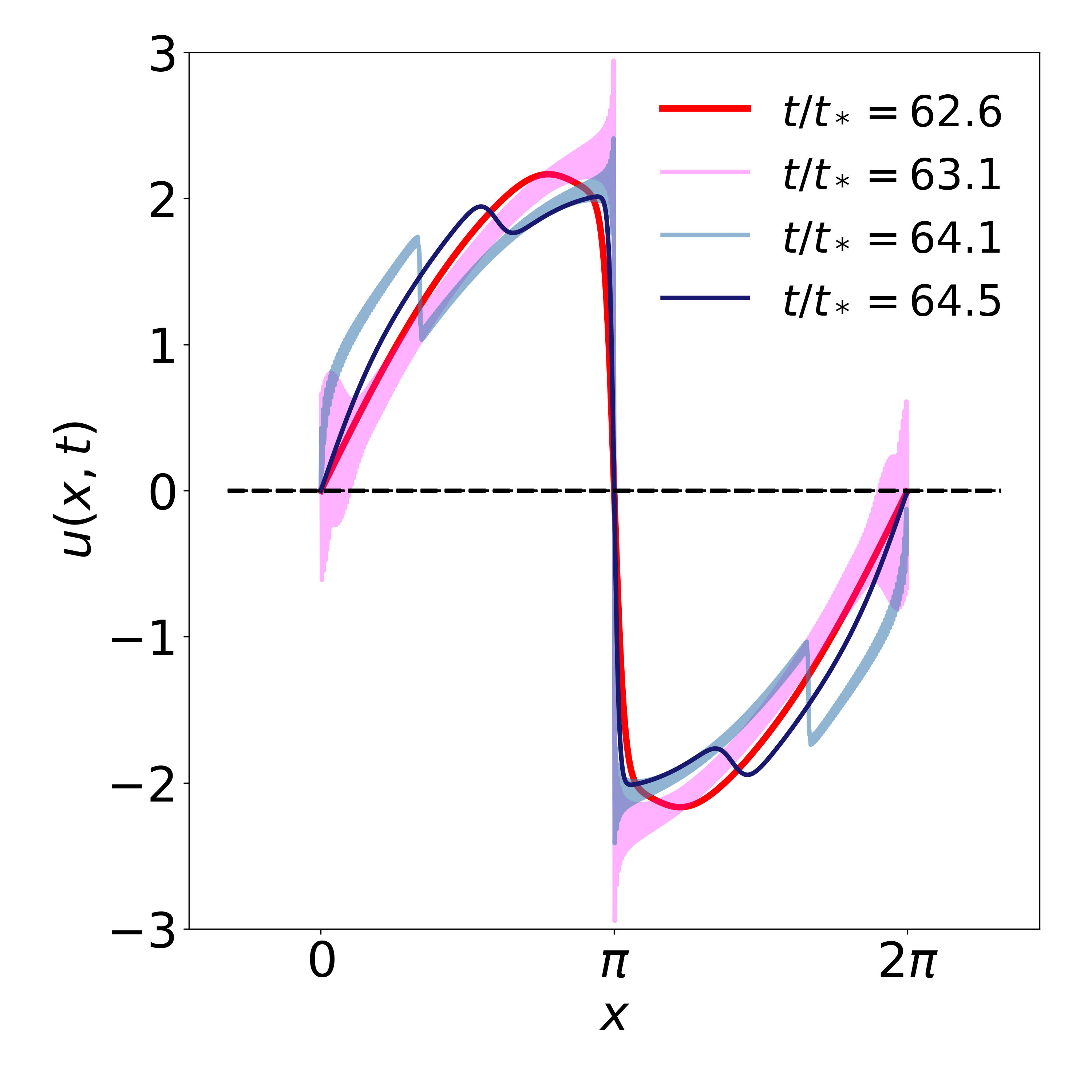}
	\put(-112,110){(b)}
	\includegraphics[scale=0.2]{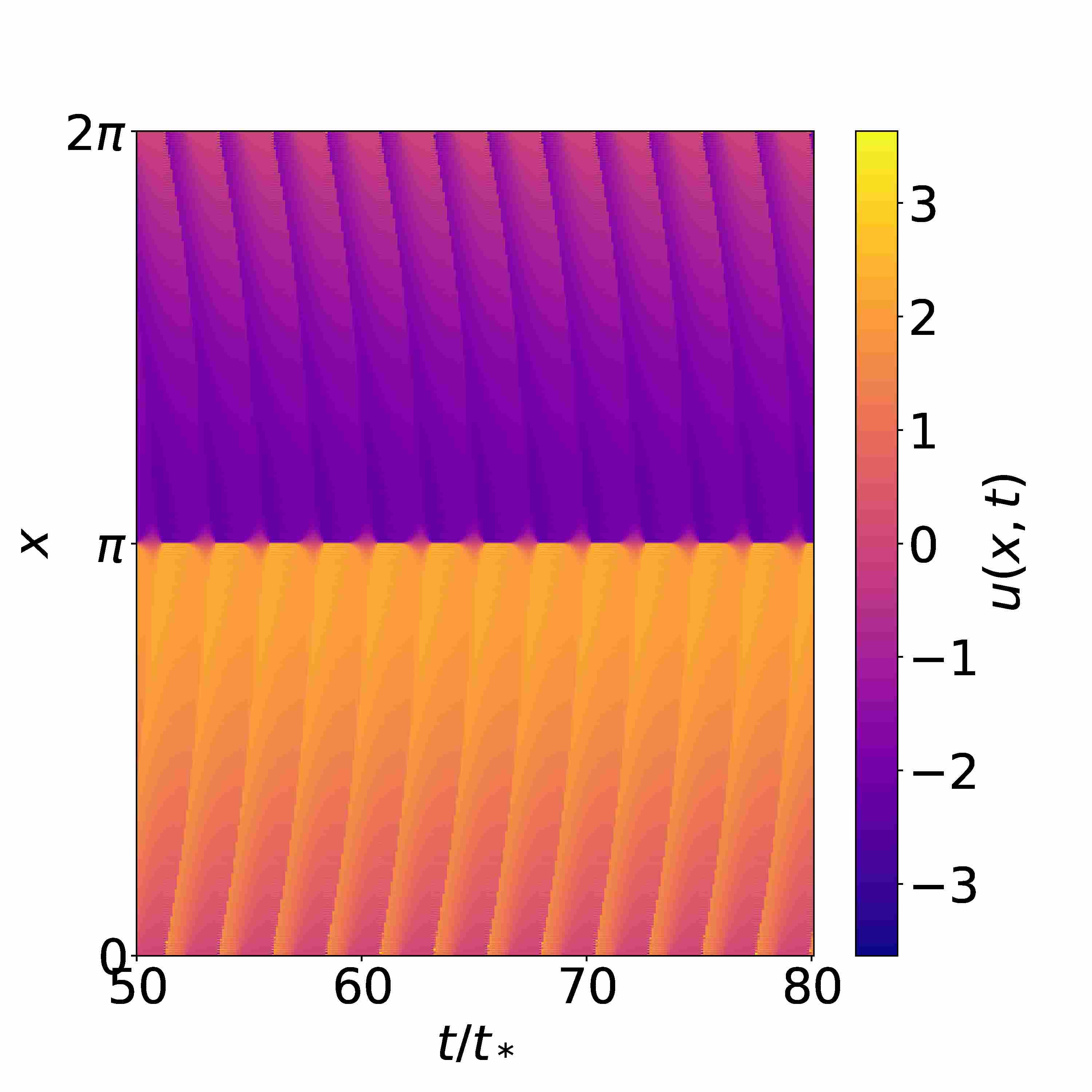}
	\put(-135,135){(c)}
}
	\caption{Signature of tygers in regime B for $E_{1024}$ at $R=5$. Plots show: (a) energy spectra $E(k)$ and (b) velocity profiles at different instants of time indicated in the insets of Fig.~\ref{fig:regimeB}. The spectrum transitions from having an exponential decay at high wave numbers at $t/t_*=62.6$ (dark shaded red line with filled diamonds) to an almost straight line at $t/t_*=63.1$ (lighter magenta line with filled asterisks). The corresponding velocity profiles exhibit a shock at $x=\pi$ (red line), which is followed by the appearance of tygers (magenta line). As time progresses, the tygers gradually become thinner (light blue line) and have greatly dissipated at the final step (dark blue line); consistent with this, the spectrum has also regained its exponential decay at high wave numbers (dark blue line with cross). (c) Spatio-temporal plot describing the above repetitive events. $E_0$ is the initial total energy and $t_*=1$ is the time of shock formation in the inviscid BE for $u(x,0)=\sin x$}
	\label{fig:traceB}
\end{figure*}

To understand this better, we show a part of the time-series, two cycles of oscillations, for $R=5$ in the insets of Fig.~\ref{fig:regimeB} (a) and (b). We select four instants of time $t/t_*=62.6$ (red diamond), $63.1$ (magenta star), $64.1$ (sky-blue pentagon), and $64.5$ (dark-blue cross) to capture the important aspects of the dynamics and show the energy spectra and the velocity fields in Fig.~\ref{fig:traceB} (a) and (b), respectively. Note that $t_*=1$ is the time of shock formation in the inviscid BE for the $u(x,0)=\sin x$ initial condition, see Appendix~\ref{app:tygersiBE} for more details.

At the instants at which $\Omega$ is minimum ($t/t_*=62.6$) or low ($64.5$), the corresponding energy spectra decay exponentially at large $k$. However, at the instants of large $\Omega$,
the energy spectra rather show a power-law scaling till $\kmax$ with oscillations superimposed on it. In fact, as we trace a full cycle of $\Omega(t)$ for $R=5$ starting at $t/t_*=62.6$, the energy spectrum also cycles through a state having a strongly exponentially decaying tail to a truncation dominated $E(k)\propto k^{-\alpha(t)}$ and then gradually back to a truncation effects free state, e.g., $E(k)$ at $t/t_*=64.5$ (dark-blue solid line in Fig.~\ref{fig:traceB} (a)).

This behavior of the global quantities and the energy spectra is easily explained by examining the rich dynamical behavior of the velocity fields depicted in Fig.~\ref{fig:traceB} (b). At around $t/t_* \approx 62.6$, the velocity field (red solid line) develops a steep-gradient, leading to the formation of a pre-shock singularity at $x=\pi$; the corresponding energy spectrum has an exponentially decaying tail. Slightly later, at $t/t_* \approx 63.1$, the velocity field almost develops a discontinuity at $x=\pi$ and a tyger appears at the origin (domain has a periodic boundary condition) with a bulge of truncation waves. At this instant of time, truncation waves are present over the entire domain though the bulge at the origin is much larger. The $\Omega(t)$  ($\nu_r(t)$) is close to its maximum (minimum) value for the cycle; the corresponding energy spectrum exhibits a power-law $k^{-\alpha(t)}$ region till $\kmax$, along with the presence of oscillations. Subsequently, at $t/t_*=64.1$ the amplitude of the truncation waves is decreased though the tyger is still present, see the velocity profile indicated by a light sky-blue solid line in Fig.~\ref{fig:traceB} (b). Finally, at $t/t_* \approx 64.5$ the truncation waves are completely damped out; the $\Omega(t)$ decreases to a much lower value, equivalently, $\nu_r(t)$ rises to a moderately high value; and $E(k)$ regains the exponentially decaying tail. For the duration of the RB runs, this cycle persists, see Fig.~\ref{fig:traceB} (c) for the spatio-temporal evolution. We illustrate the full dynamics in Appendix~\ref{app:tygersBEregB}.

In summary, the oscillatory behavior of the global quantities is associated with appearance, disappearance and reappearance of tygers in the velocity field.

A crucial thing that one must not ignore is that we are dealing with a forced viscous system and in effect the tygers are bound to be different from their inviscid counterpart. This is clear if one considers the velocity profile at the instant of tyger birth (the light pink line in of Fig.~\ref{fig:traceB})(b). While the tyger in the inviscid BE only appears as a bulge at the origin and a slight thickening of the velocity profile (see Fig.~\ref{fig:tyg-vel} for inviscid BE profile), in the RB this bulge is present along with oscillations at the ends of the shock straddling it near $x=\pi$. The thickening of the velocity profile is also much greater in RB. The oscillations straddling the shock have previously been seen in the fractional-hyperviscous BE~\cite{banerjee2019fractional}. Since we are dealing with truncation effects, which are the result of singularities approaching the real axis, such oscillations should be expected when tyger births occur. Thus, their appearance in the RB is not altogether unreasonable. The tyger bulge at the origin is due to the chosen initial condition, while the extensive oscillations along the rest of the velocity profile and the ones along the shock are due to the forcing. The singularities, in fact, are critical as they dictate this cycle of tyger birth and death. In a later section, we explain this cycle in detail after the equivalence of the RB and BE is established. But in brief, the tygers form as a result of the pre-shock singularity coming within one Galerkin wavelength of the real axis at $x=\pi$ as $\nu_r$ decreases. The oscillations straddling the shock are due to the interaction of the forcing with these singularities. This appearance of truncation effects coupled with the thermostat of the RB, causes an increase in the viscosity which subsequently regularizes the truncation effects and the cycle is set in motion.

\begin{figure*}
	\centering
	\resizebox{\textwidth}{!}{%
	\includegraphics[scale=0.2]{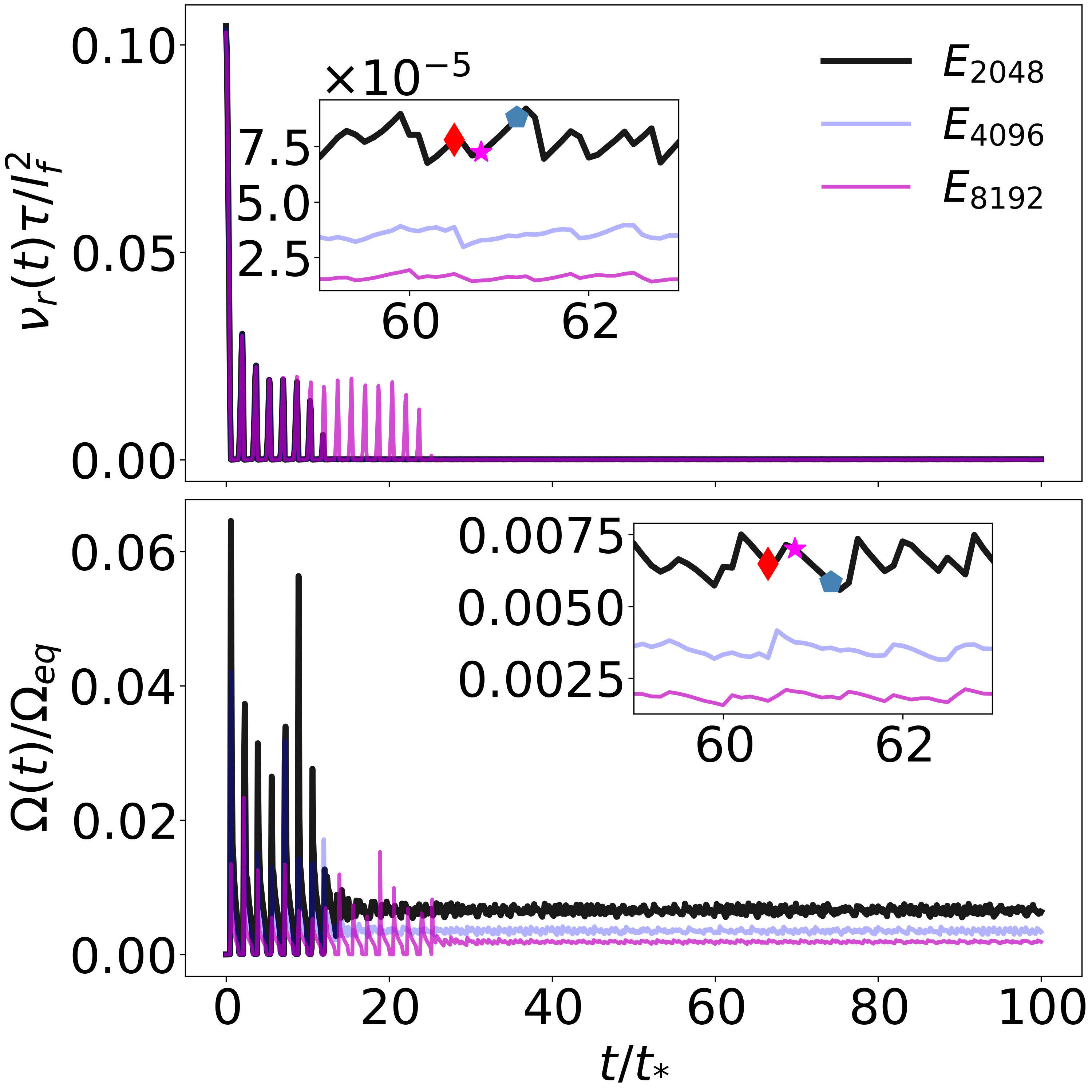}
	\put(-18,35){(a1)}
	\put(-18,95){(a2)}
	\includegraphics[scale=0.2]{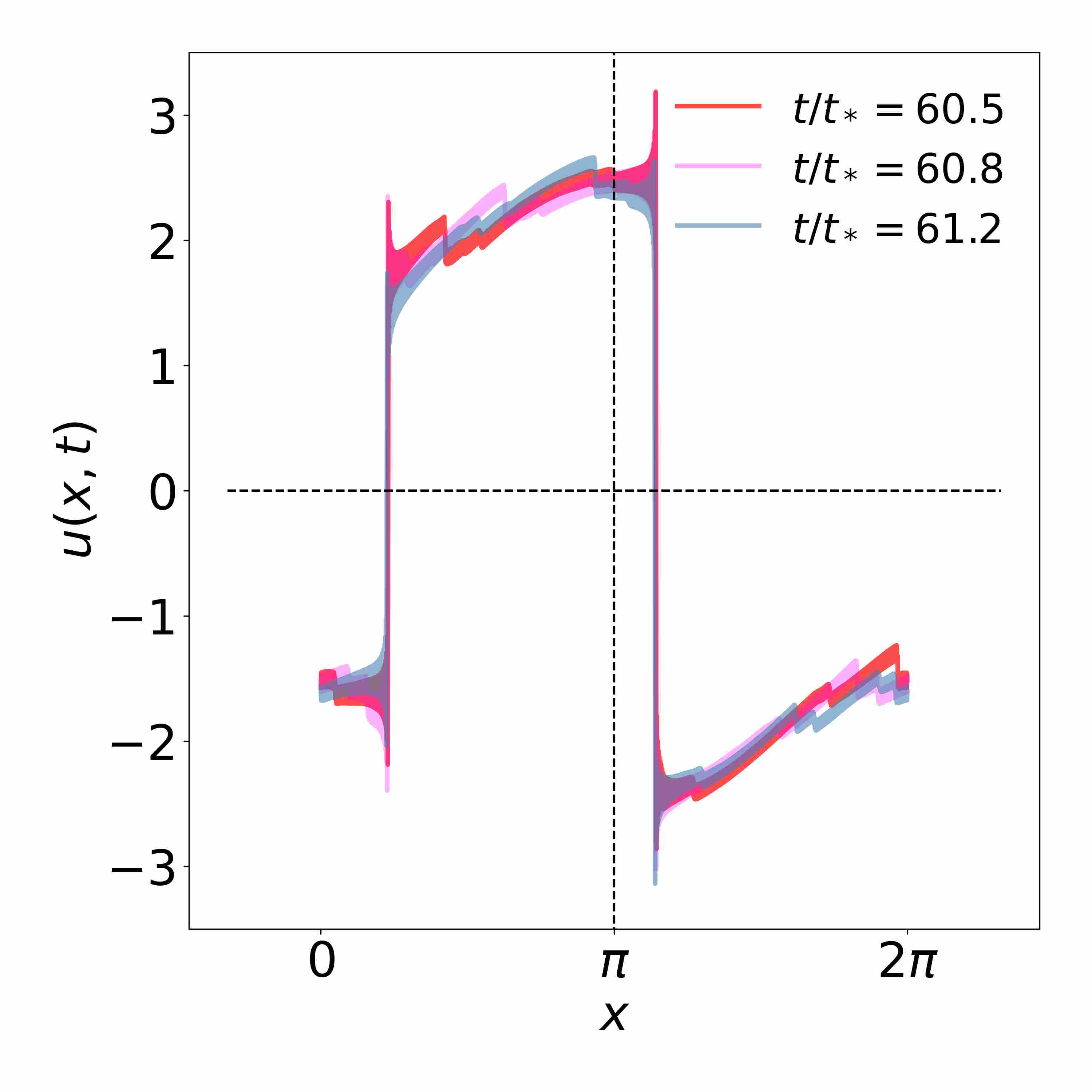}
	\put(-25,35){(b)}
	\includegraphics[scale=0.2]{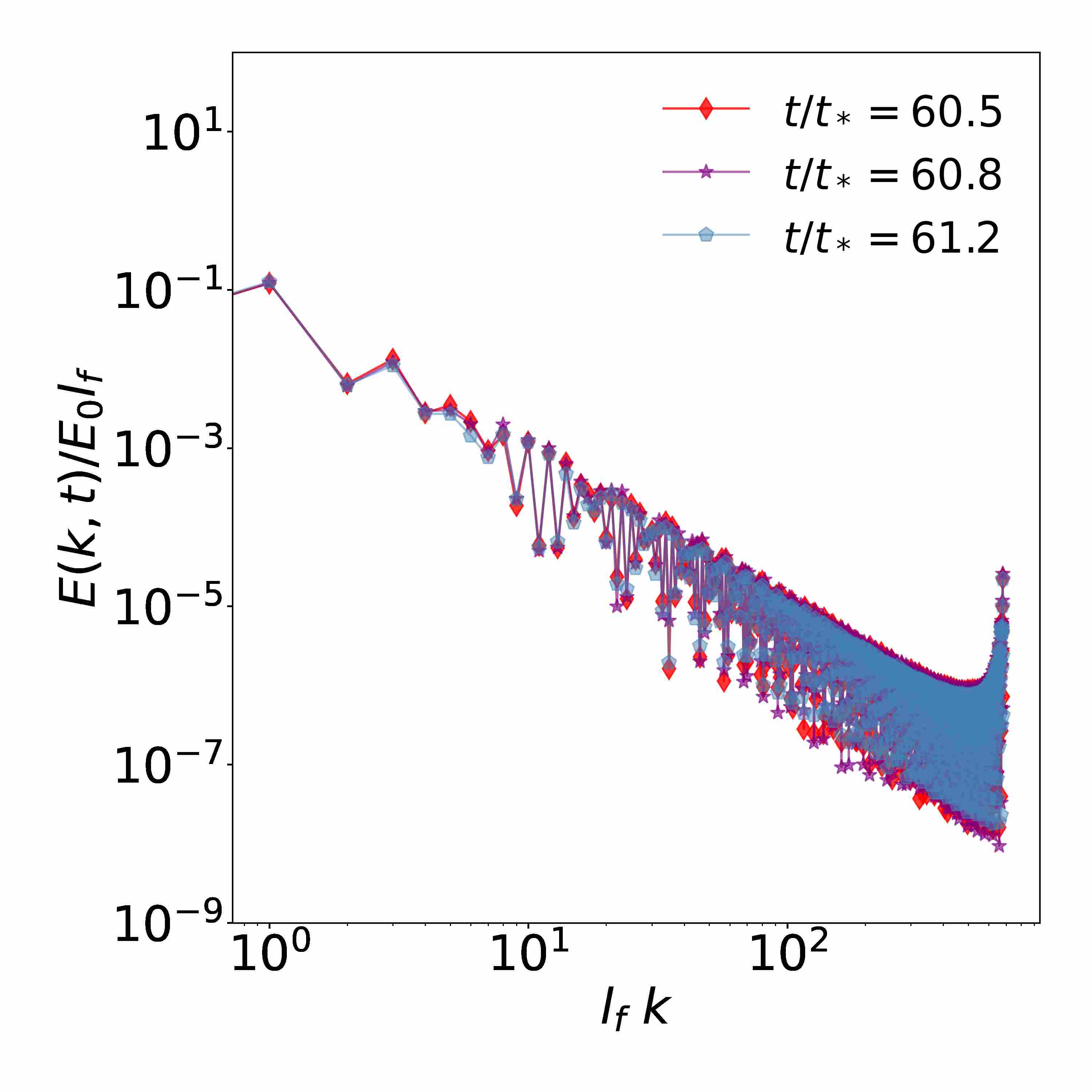}
	\put(-45,35){(c)}
}
	\caption{Regime B for $N_c=2048$ for $R=3$ for protocol $E_{2048}$. Time series of: (a1)$\Omega/\Omega_{eq}$ and (a2) $\nu_r$. Insets show zoomed-in part of the time series for the duration $t/t_*\approx 61$ to $62$.  (b) Velocity field, $u(x,t)$, at different instant of time, exhibiting a clear secondary discontinuity, (big jump in velocity near $x=0$) arising from the breakdown of tyger bulge at the resonance point, development of mini-shock like secondary structures between the secondary discontinuity and the original shock. (c) Energy spectra corresponding to time instants in (a). Note that once a secondary discontinuity has formed, it along with original shock at $x=\pi$ become mobile. Moreover, the minishocks also move in their respective domains. Plots are shown for $t/t_*=60.5$ (dark shaded red line with filled diamonds), $t/t_*=60.8$ (lighter magenta line with filled asterisks) and $t/t_*=61.2$ (lightest sky-blue line with filled pentagons). $\Omega_{eq}=(2/3)E_0\kmax^2$, $\tau=(\ell^2_f/\langle \epsilon \rangle)^{1/3}$, where $\langle \epsilon \rangle$ is the energy injection rate, $\ell_f$ is the forcing length-scale, and $E_0$ is the initial total energy. $t_*=1$ is the time of shock formation in the inviscid BE for $u(x,0)=\sin x$.}
	\label{fig:traceB2}
\end{figure*}

What remains to be explained is the discrepancy between the behavior of the lowest resolution case  $E_{1024}$ ($F_{1024}$) and the runs at higher resolutions, see regime B in Fig.~\ref{fig:globalensnur} (b). We know that in the inviscid BE, tygers initially appear as symmetric bulges and eventually break apart into two asymmetric wings of white noise, which are the inklings of thermalization (c.f. Appendix~\ref{app:tygersiBE}). Such a breakdown also occurs in the RB system. In fact, this breakdown leads to the formation of a small discontinuity that appears in the region between the shock and the tyger (c.f. Fig.~\ref{fig:traceB} (b)), which we dub as `mini-shock'.

For the larger resolutions ($N_c \geq 2048$) and $2 \lesssim R \lesssim 4$, the asymmetric breakdown is surprisingly associated with a movement of the shock and the tyger bulge at the origin, see Fig.~\ref{fig:traceB2}. The tygers give way to a secondary discontinuity smaller than the shock, which forms due to the breakdown, and moves away from $x=0$. In between the shock and the erstwhile tyger are oscillations along which the smaller discontinuities of minishocks travel (c.f. Fig.~\ref{fig:traceB2} (b)). The formation of such a structure is quite surprising and unexpected. This structure is also present in regime C, but with some differences. For the lowest resolution case $E_{1024}$, the breakdown appears, but does not cause the motion of the shock and tyger, and thus, the fall-off in the values of $\nu_r$ is continuous. This can be understood because at such a small resolution the tygers have fewer truncation waves, and similarly, the breakdown has fewer white noise disturbances. Hence, it leads to a persistence of tygers at the lower resolution indefinitely (at least within the finite run-time of our simulations). At the higher resolution, the formation of the structure is associated with a sudden collapse in the time-series of the global quantities  (c.f. Fig.~\ref{fig:traceB2} (a)). This collapse is also responsible for the drop-off in the viscosity values in Fig.~\ref{fig:globalensnur} (b).

\subsubsection{Regime C}
\label{sec:regimeC}

\begin{figure*}
	\centering
	\resizebox{\textwidth}{!}{
	\includegraphics[scale=0.2]{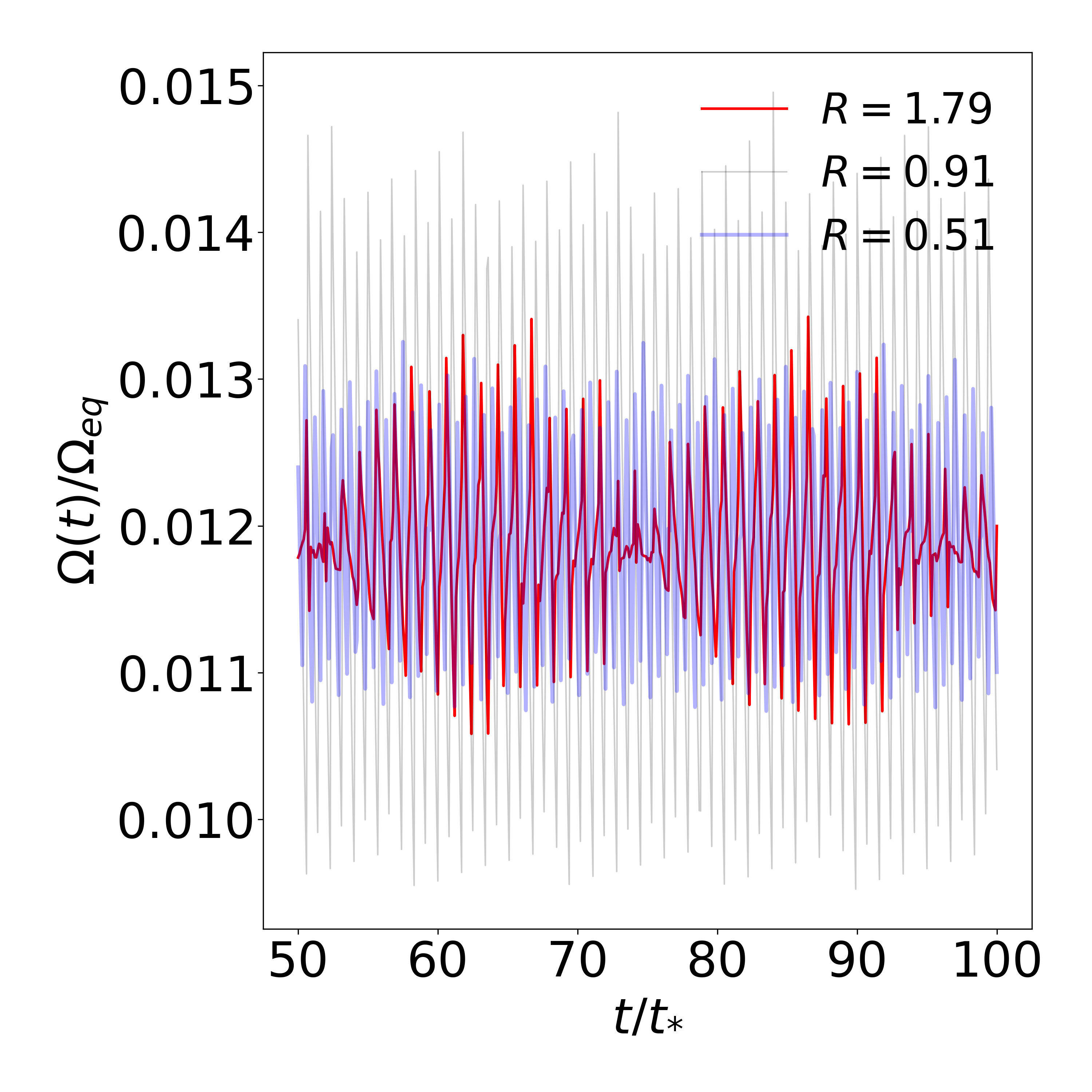}
	\put(-105,125){(a)}
	\includegraphics[scale=0.2]{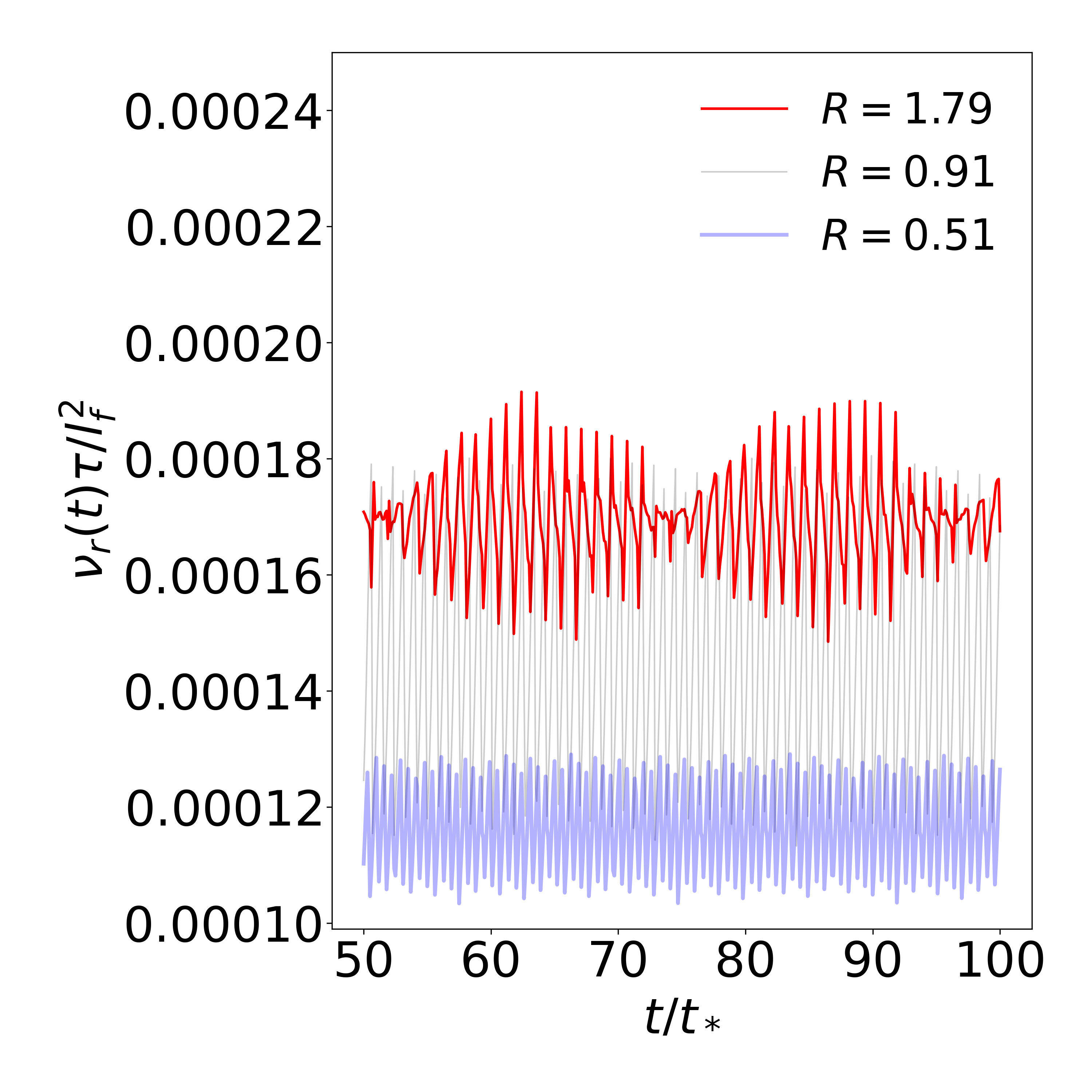}
	\put(-95,125){(b)}
	\includegraphics[scale=0.2]{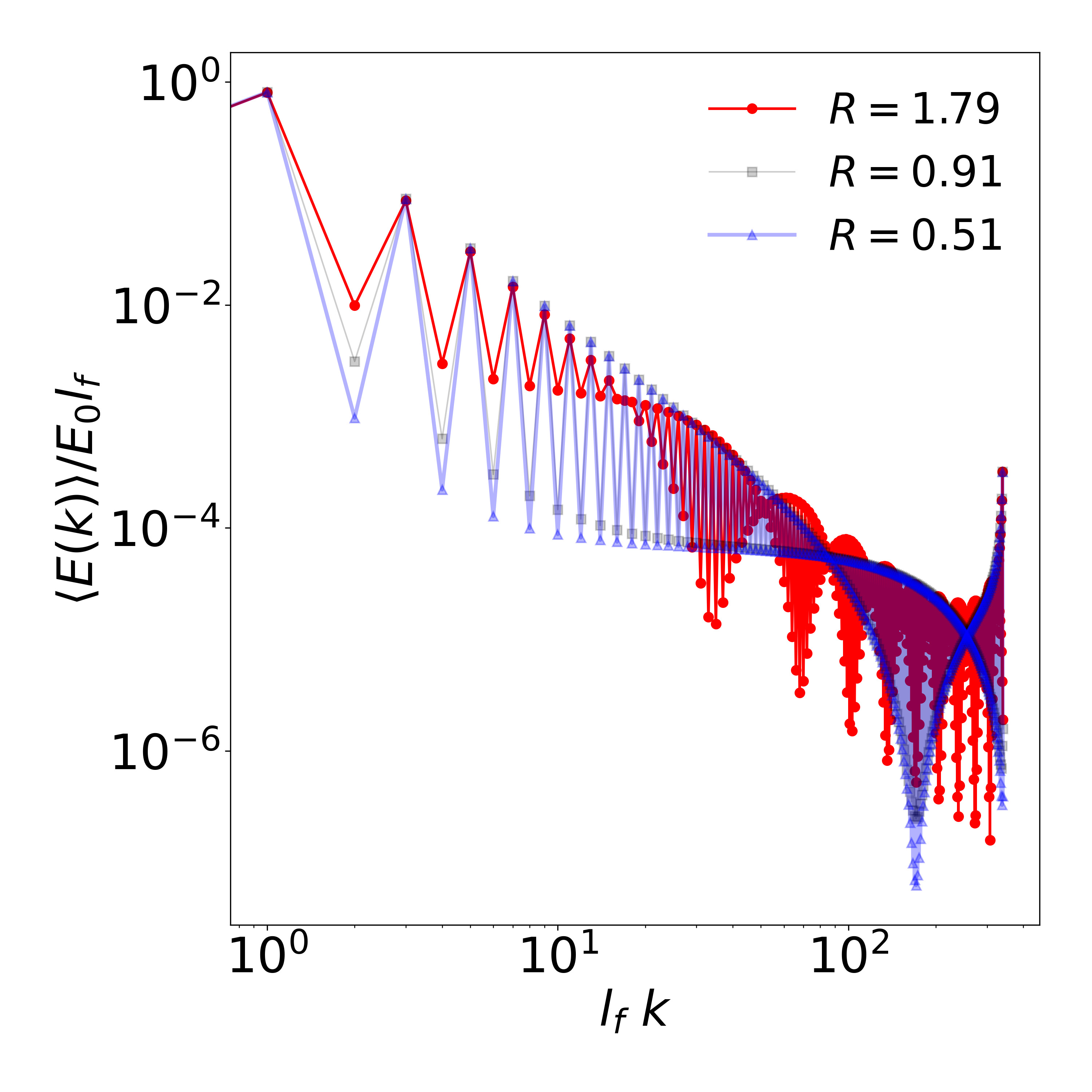}
	\put(-105,125){(c)}
	}
	\caption{Regime C. Time series of: (a)$\Omega/\Omega_{eq}$ and (b) $\nu_r$ exhibit very high frequency oscillations with amplitude decreasing as $R$. (c) Time-averaged energy spectrum, $E(k) \sim k^{-2}$ behavior, with oscillations superimposed over it. At smaller values of $R$, the spectrum begins to develop modulations, as a possible consequence of at least two poles appearing on the real axis. Plots are shown for $R=1.79$ (dark shaded red line with filled circles), $R=0.91$ (lightest black line with filled square) and $R=0.51$ (blue line with filled upward triangles). $\Omega_{eq}=(2/3)E_0\kmax^2$, $\tau=(\ell^2_f/\langle \epsilon \rangle)^{1/3}$, where $\langle \epsilon \rangle$ is the energy injection rate, $\ell_f$ is the forcing length-scale, and $E_0$ is the initial total energy. $t_*=1$ is the time of shock formation in the inviscid BE for $u(x,0)=\sin x$.}\label{fig:microC}
\end{figure*}

\begin{figure*}
	\centering
	\resizebox{\textwidth}{!}{
	\includegraphics[scale=0.25]{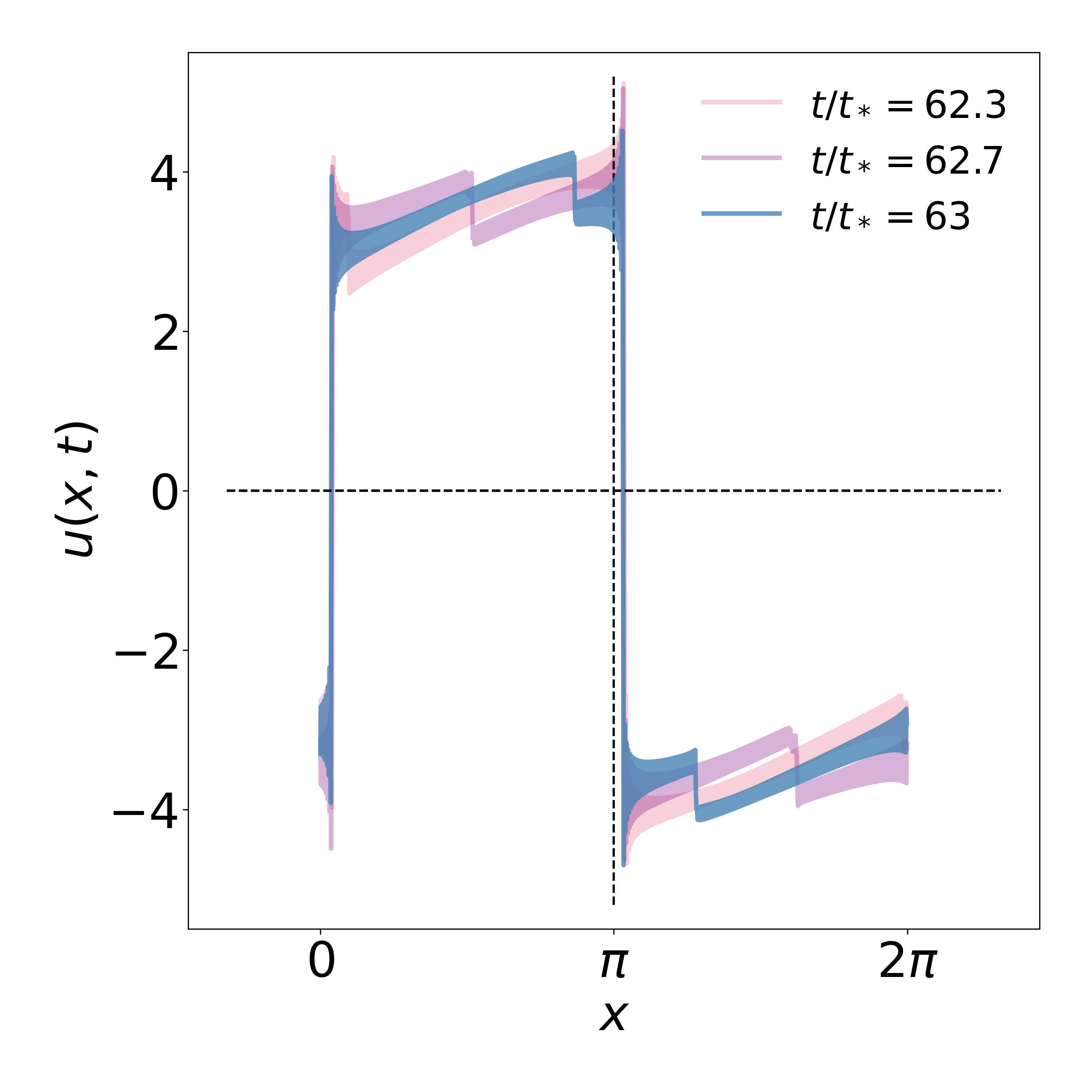}
	\put(-145,160){(a)}
	\includegraphics[scale=0.25]{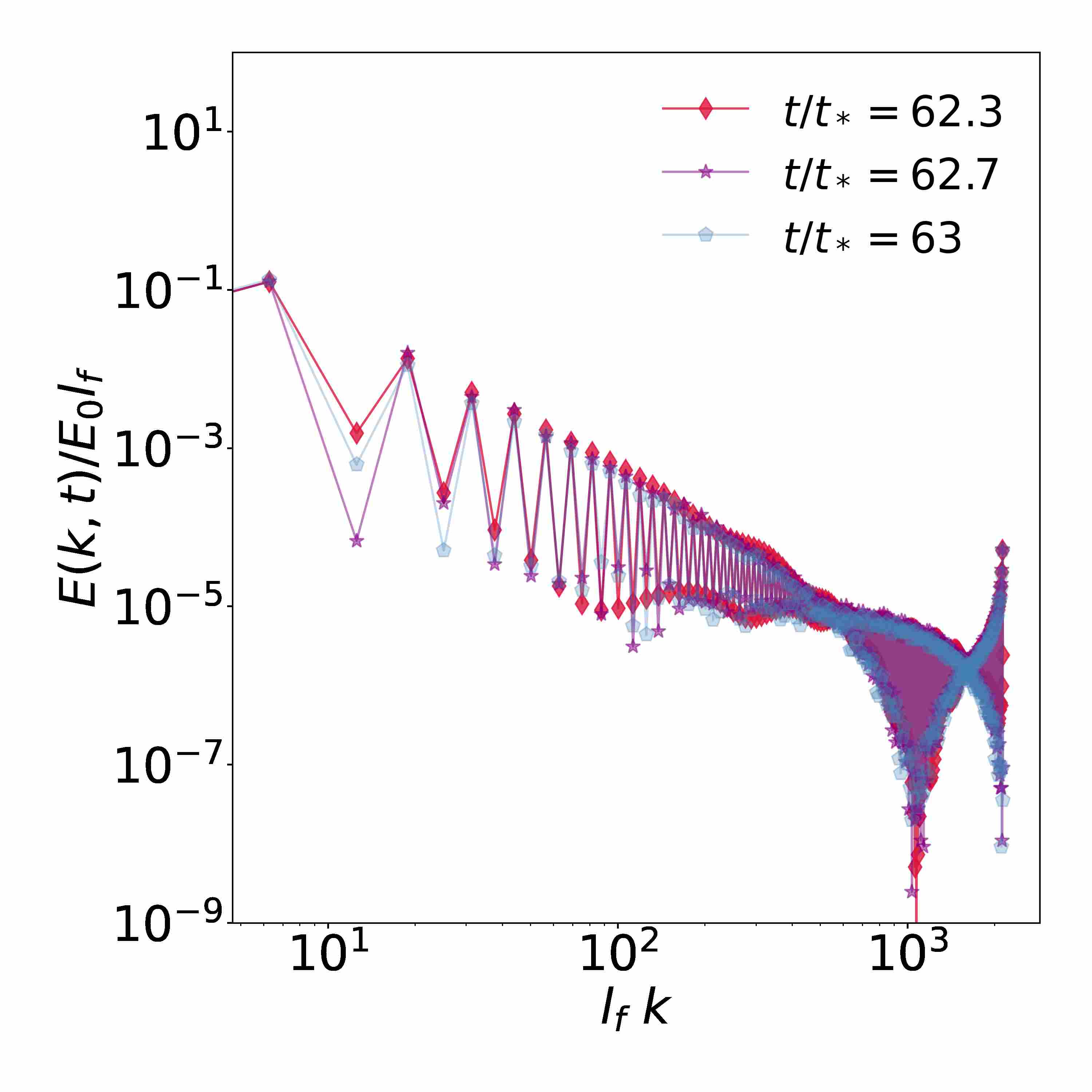}
	\put(-135,160){(b)}}
	\caption{Secondary discontinuities in regime C, $R=1.005$. (a) Velocity field, $u(x,t)$, at different instants of time, exhibiting a secondary discontinuity, (big jump in velocity near $x=0$) arising from the breakdown of tyger bulge at the resonance point, development of mini-shock like additional secondary structures between the secondary discontinuity and the original shock. (b) Energy spectra corresponding to time instants in (a). Note that once a secondary discontinuity has formed, it along with original shock at $x=\pi$ becomes mobile. Moreover, the mini-shocks also move in their respective domains. $t_*=1$ is the time of shock formation in the inviscid BE for $u(x,0)=\sin x$.}
	\label{fig:RC_minishock}
\end{figure*}

Figure~\ref{fig:microC} (a) and (b) show that $\Omega(t)$ and $\nu(t)$, respectively, exhibit high frequency oscillations; the amplitude of these oscillations goes down with $R$. More intriguingly, Fig.~\ref{fig:microC} (c) shows that the fast oscillations on the energy spectra develop low frequency modulations. It seems that the appearance of the modulations is associated with the shift of the shock slightly away from $\pi$, see Fig.~\ref {fig:RC_minishock}. The motion of the shock also triggers the formation of a structure that partitions the system in two halves, with one end lifted around $x=0$ (Fig.~\ref {fig:RC_minishock} (a)). Such structures were witnessed in regime B, however at resolutions larger than $1024$ though with less prominent modulations. The birth of this structures in regime C is exactly identical to its formation in regime B (due to the breakdown of a tyger).

Similarly, a smaller secondary  discontinuity arises at the resonance point, \textit{i.e.}, the origin, with mini-shock traveling in the partition between the shock and the tyger discontinuity (c.f. Fig.~\ref {fig:RC_minishock} (a)). The difference with the structure at higher resolutions in regime B is that in this regime the secondary discontinuity is taller and its height decreases with $R$. The structure persists for all the values of $R$ in regime C and disappears only when complete thermalization is attained. Corresponding to the velocity fields, Fig.~\ref {fig:RC_minishock} (b) shows the spectrum at the same time-instants; a signature modulation is quite evident in the spectrum. Later in Sec.~\ref{section:pole} we will see that the appearance of such a modulation is inevitably linked to the structure in the velocity field and the singularities of the solution.

This structure appears to prevent the complete thermalization by halting the spread of the white noise born from the tygers. This can be attributed to the fact that in regime C the system is inherently still a forced system and thus is in a state driven away from equilibrium. Although the viscosity values are small, the effect of forcing remains and the system is akin to a forced inviscid BE halted (or held) at a steady state by the non-zero viscosity. The viscosity is also responsible for the formation of the mini-shocks which are formed due to the regularization of the white noise arising from the tygers. It is shown in Appendix~\ref{app:tygersiBE} that in the forced inviscid BE, the pile-up of energy happens at two locations: the shock discontinuity at $x=\pi$ and the discontinuity created by the tyger at $x=0$. Such a pile-up forms impenetrable walls preventing the complete thermalization. The spectrum remains stuck in a partially thermalized state; this explains the behavior of the states in regime C.

\subsubsection{Regime D}
\label{sec:regimeD}

\begin{figure*}
	\centering
	\resizebox{\textwidth}{!}{
	\includegraphics[scale=0.2]{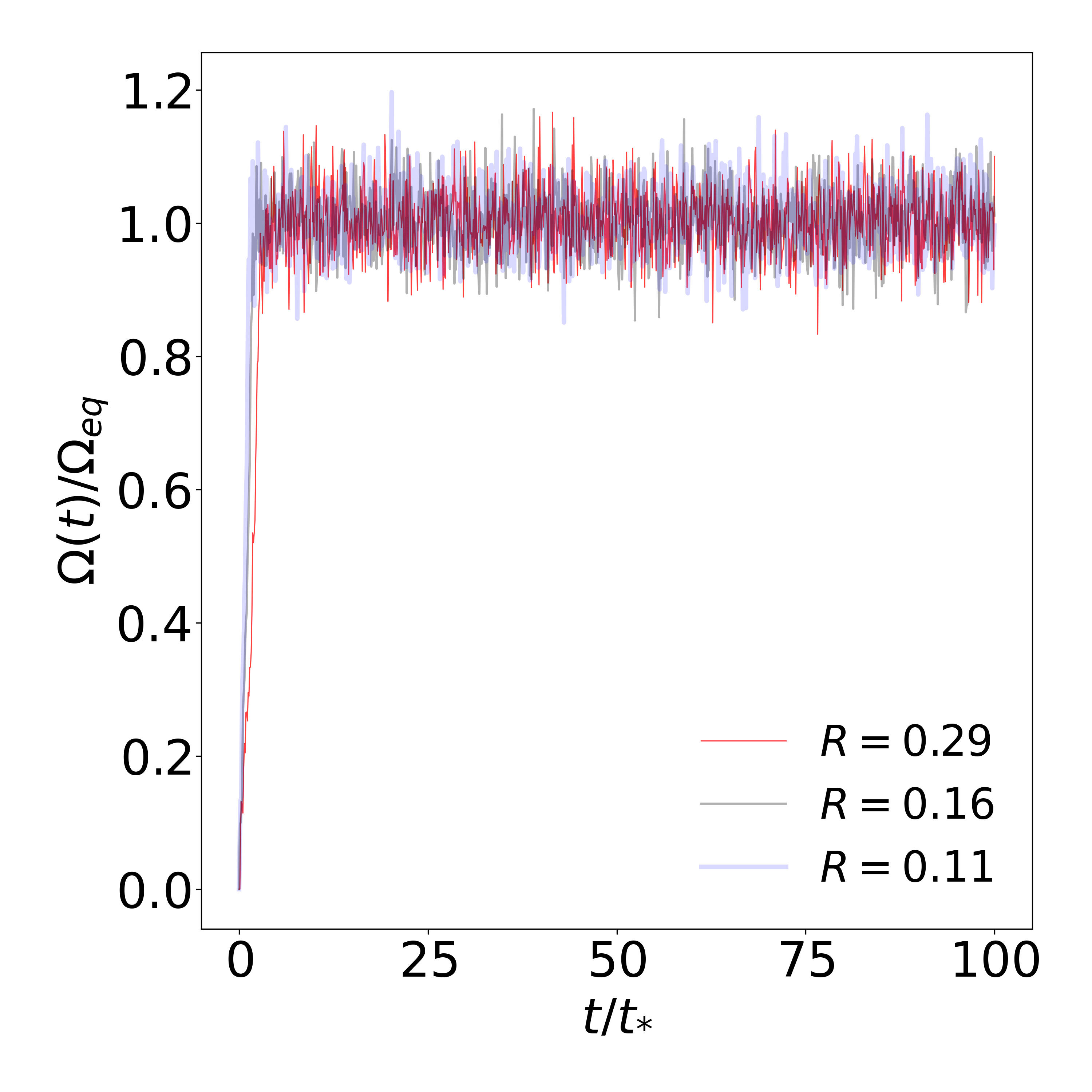}
	\put(-25,90){(a)}
	\includegraphics[scale=0.2]{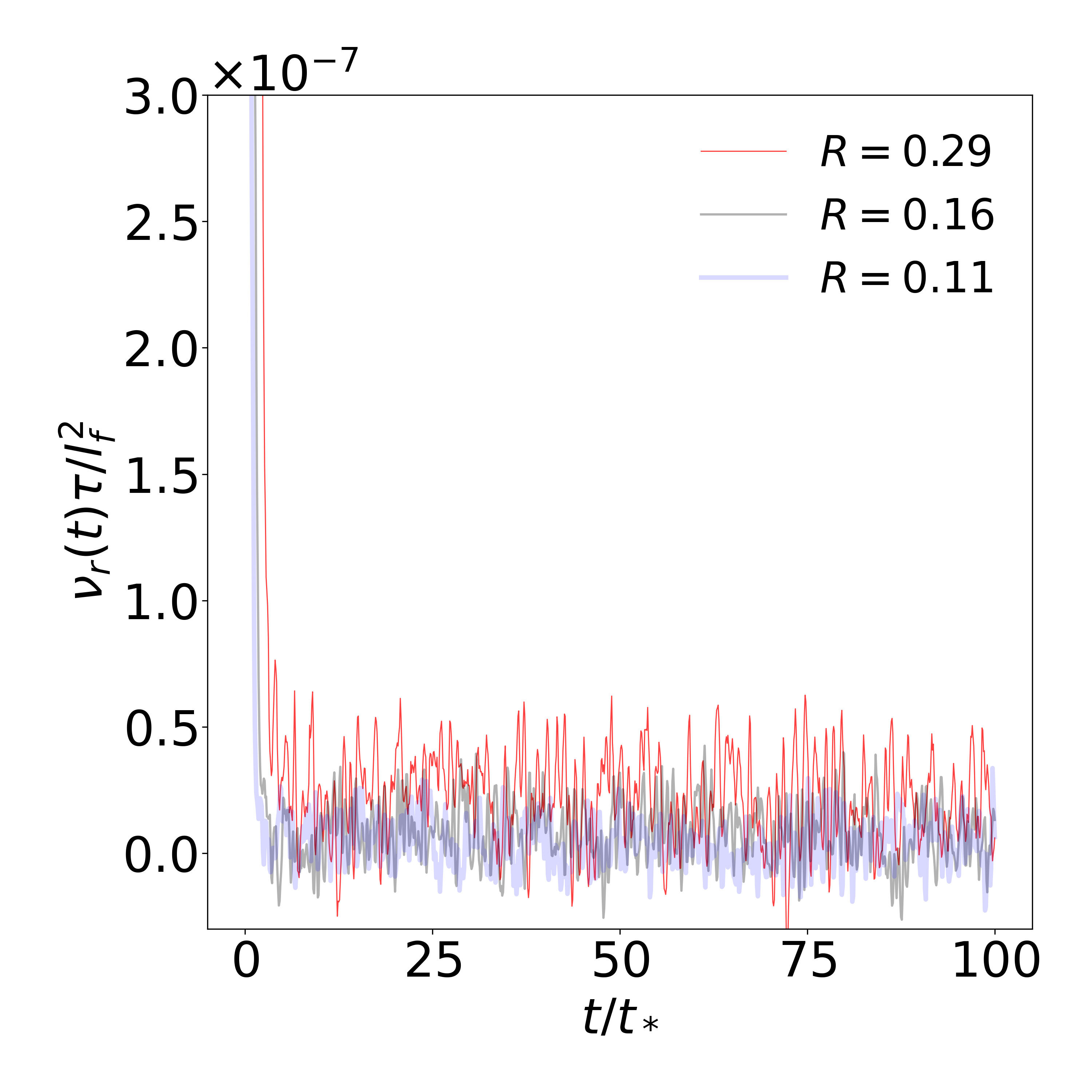}
	\put(-25,90){(b)}
	\includegraphics[scale=0.2]{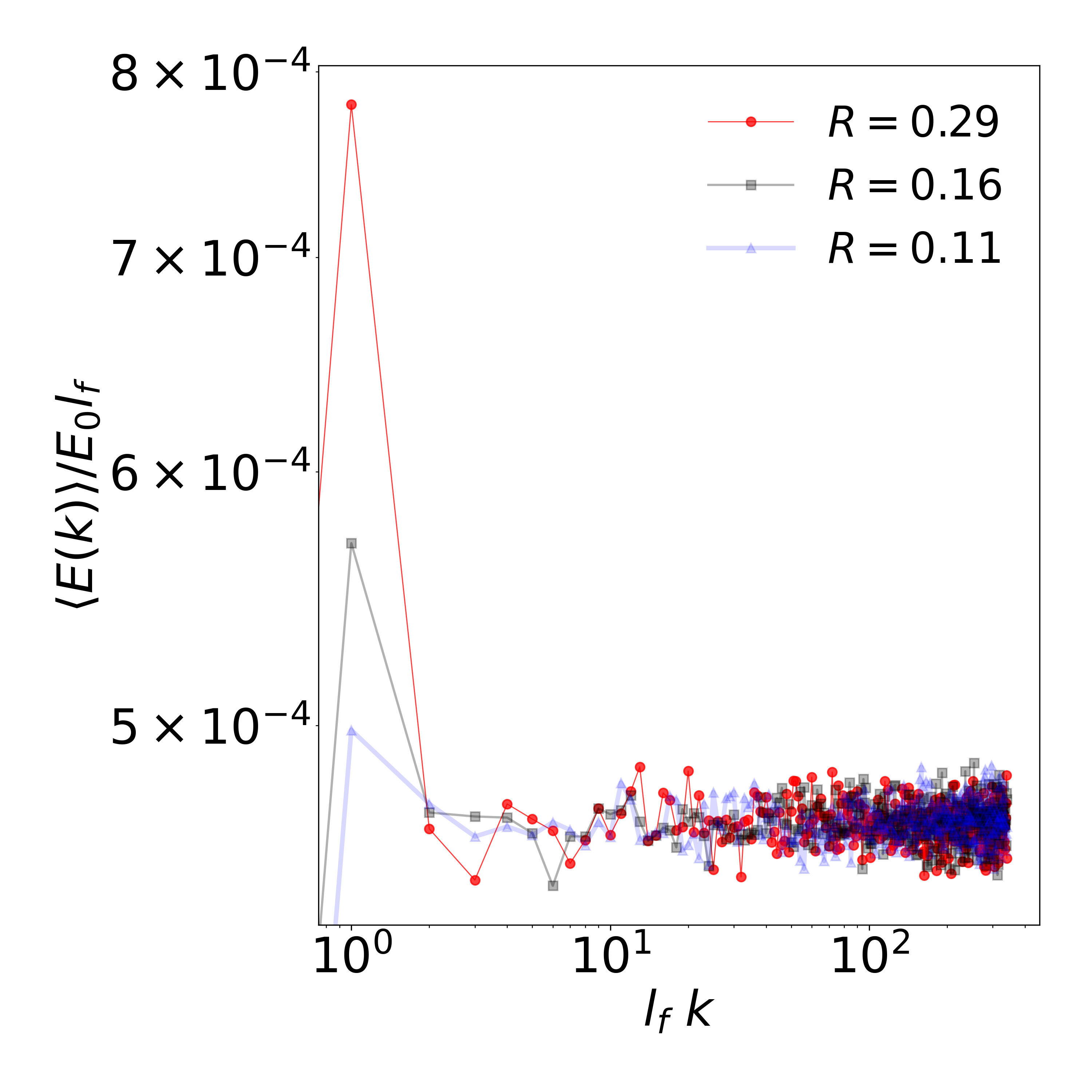}
	\put(-25,90){(c)}}
	\caption{Regime D. Time series of: (a)$\Omega/\Omega_{eq}$ and (b) $\nu_r$ exhibit random fluctuations. (c) Time-averaged energy spectrum, $E(k)\sim k^0$, indicates an equipartition of energy among the Fourier modes. Plots are shown for $R=0.29$ (red line with filled circles), $R=0.16$ (black line with filled square) and $R=0.11$ (blue line with filled upward triangles). $\Omega_{eq}=(2/3)E_0\kmax^2$, $\tau=(\ell^2_f/\langle \epsilon \rangle)^{1/3}$, where $\langle \epsilon \rangle$ is the energy injection rate, $\ell_f$ is the forcing length-scale, and $E_0$ is the initial total energy. $t_*=1$ is the time of shock formation in the inviscid BE for $u(x,0)=\sin x$.}
	\label{fig:regimeD}
\end{figure*}

Figure~\ref{fig:regimeD} (a) shows that in regime D, $\Omega(t)/\Omega_{eq}$, after passing through initial transients, and exhibits random fluctuations around $1$. Therefore, $\Omega \approx \Omega_{eq}$, where $\Omega_{eq}=2/3 E_0 \kmax^2$ is the absolute equilibrium value. Similarly, $\nu_r$ fluctuates around values close to zero for different $R$. Surprisingly, the fluctuations of the normalized viscosity lead to events wherein $\nu_r <0$.  The energy spectrum scales as $k^0$, except at very small wave numbers. Thus, this regime is in `quasi-equilibrium' and associated with equipartition of energy (in Fourier space). Here, the solutions of the RB are basically the thermal solutions of the inviscid BE.

\subsection{Equivalence of RB and BE}
\label{sec:equivalence}

\begin{figure*}
	\centering
	\resizebox{\textwidth}{!}{
	\includegraphics[scale=0.2]{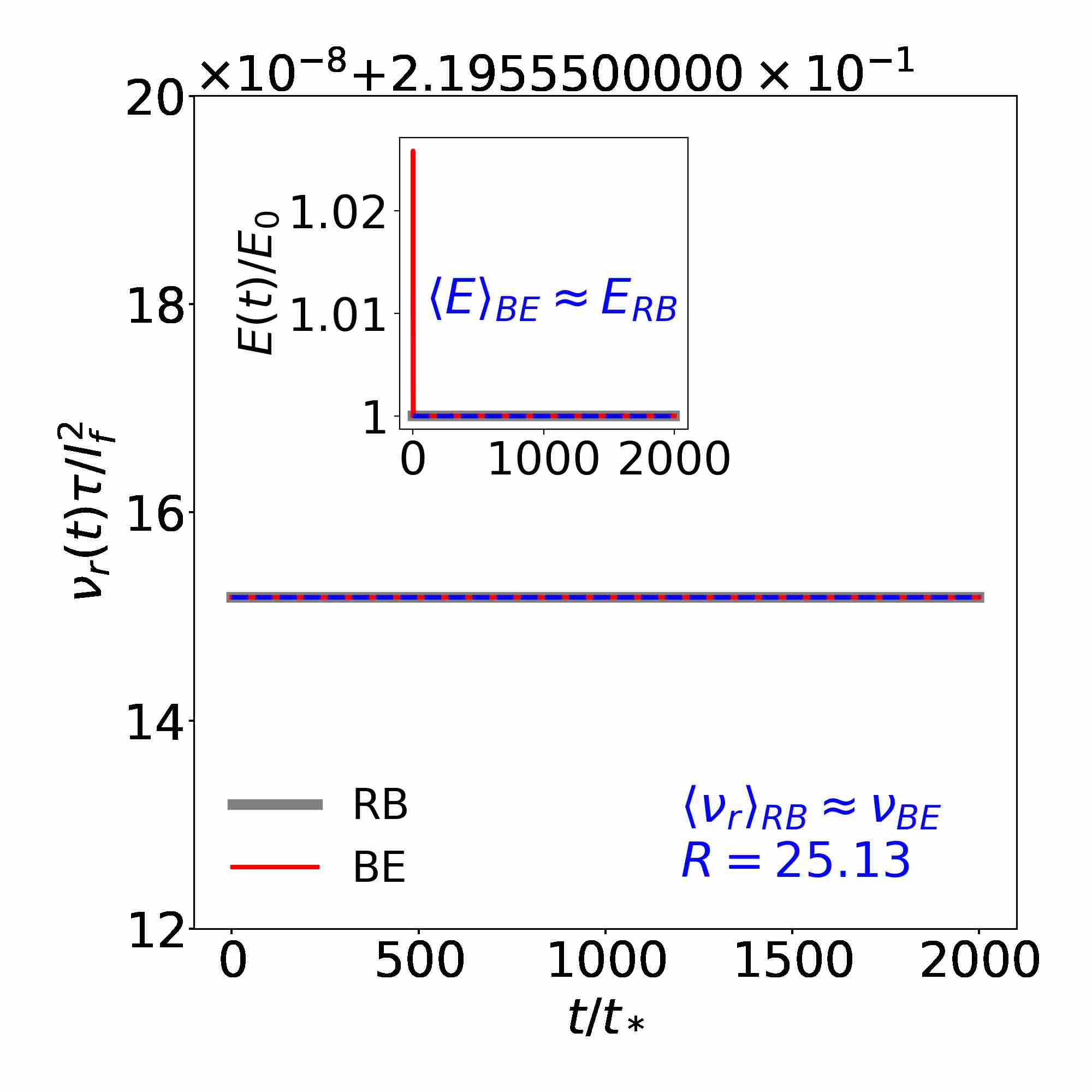}
	\put(-30,115){(a)}
	\includegraphics[scale=0.2]{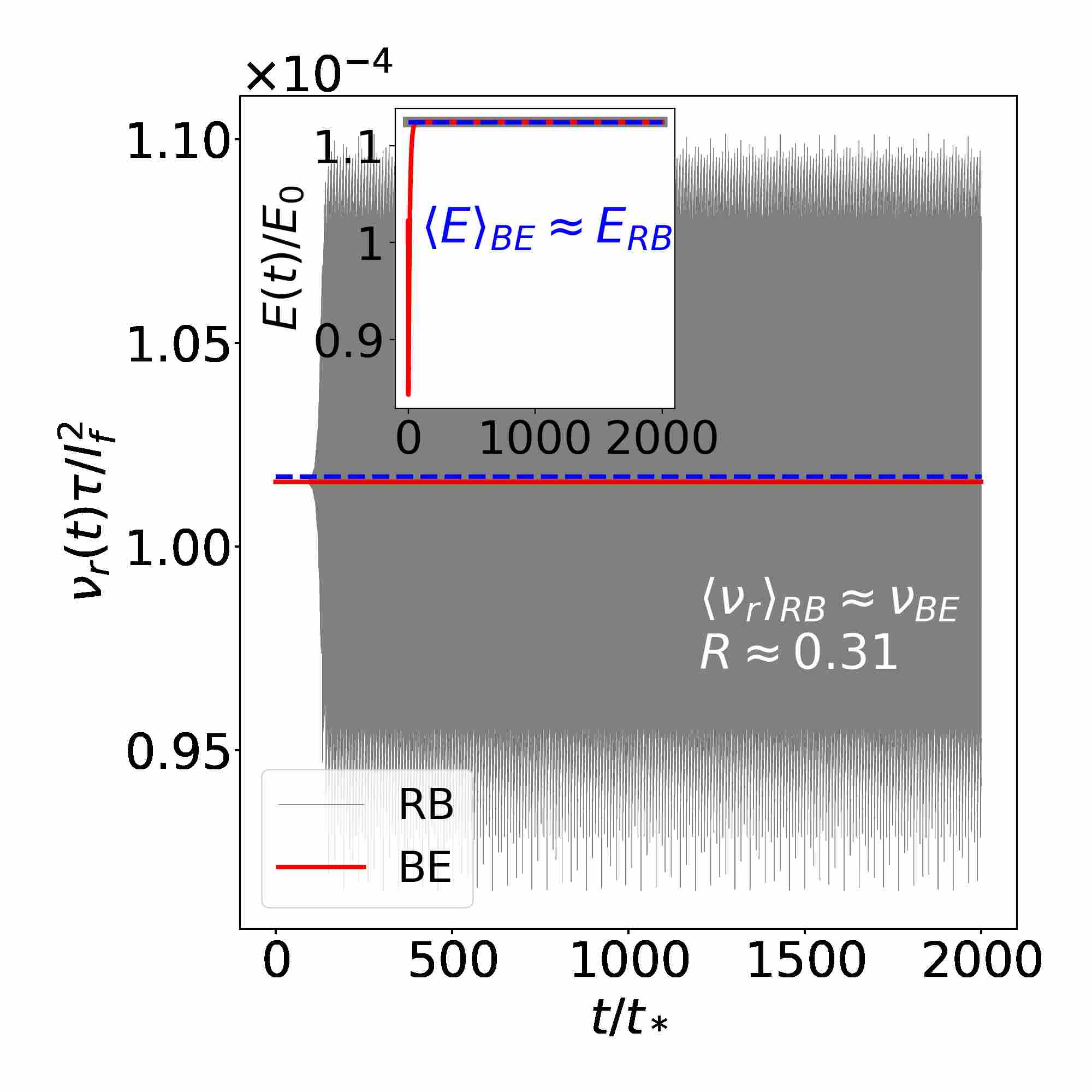}
	\put(-30,115){(b)}
	\includegraphics[scale=0.2]{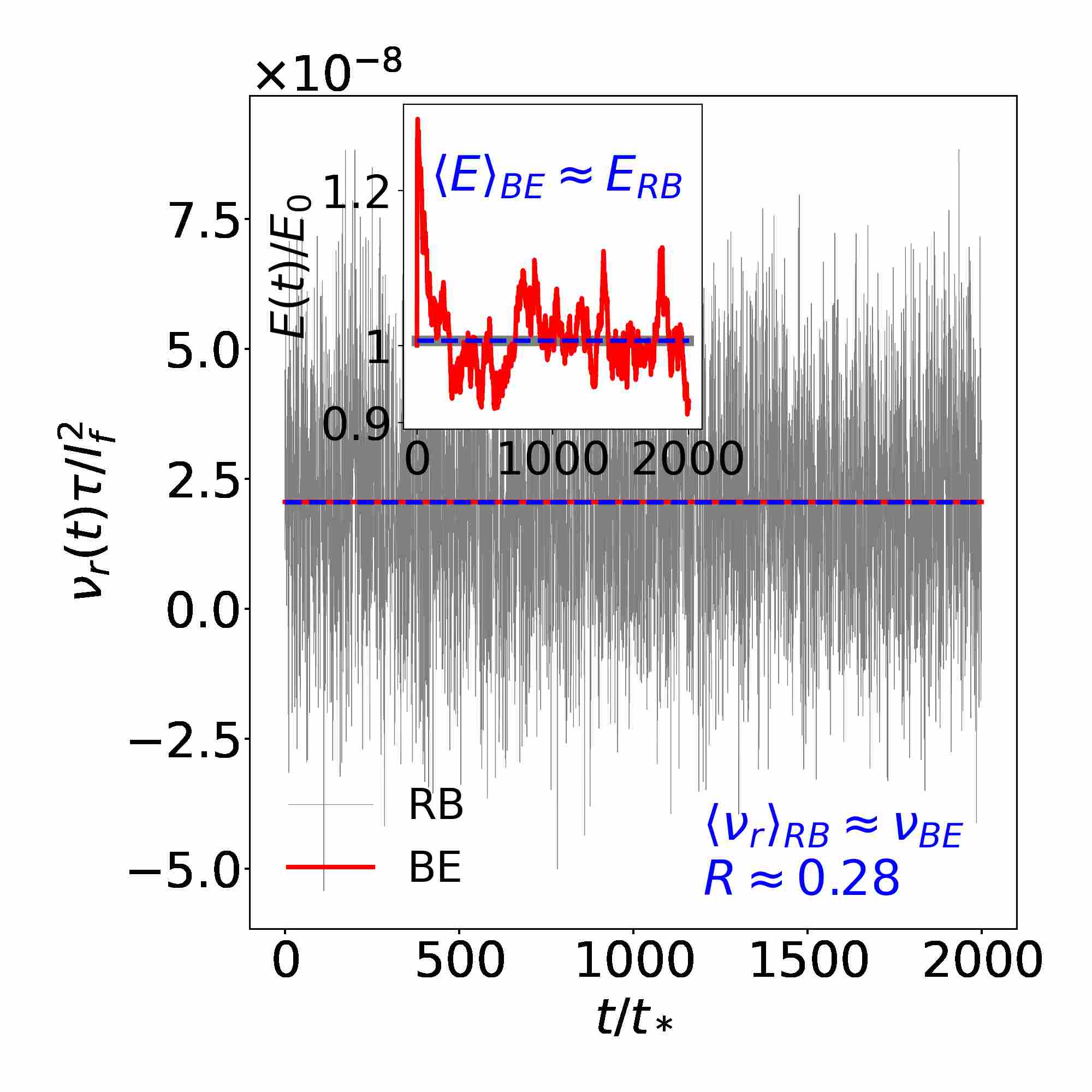}
	\put(-30,115){(c)}}
	\caption{Equivalence of RB and BE in different regimes. Plots of $\nu\tau/\ell^2_f$ vs $t/t_*$ for: (a) Regime A, $R = 25.13$. (b) Regime C, $R = 0.31$, and (d) Regime D, $R = 0.28$. The insets show energy time series $E(t)/E_0$ vs $t/t_*$, where $E_0$ is the initial total energy. Thin red and thick grey lines represent the BE and RB, respectively. Blue dashed lines indicate the average of a fluctuating quantity: ${\langle \nu_{r} \rangle}_{RB}$ in the main plots and ${\langle E \rangle}_{BE}$ in the insets. $\tau=(\ell^2_f/\langle \epsilon \rangle)^{1/3}$, where $\langle \epsilon \rangle$ is the energy injection rate and $\ell_f$ is the forcing length-scale. $t_*$ is the time of shock formation in the inviscid BE for $u(x,0)=\sin x$. Error in $\langle \nu_r \rangle$ relative to $\nu_{BE}$ is bounded by $0.2\%$.}
	\label{fig:equivnu}
\end{figure*}

\begin{figure*}
	\centering
	\resizebox{\textwidth}{!}{
	\includegraphics[scale=0.2]{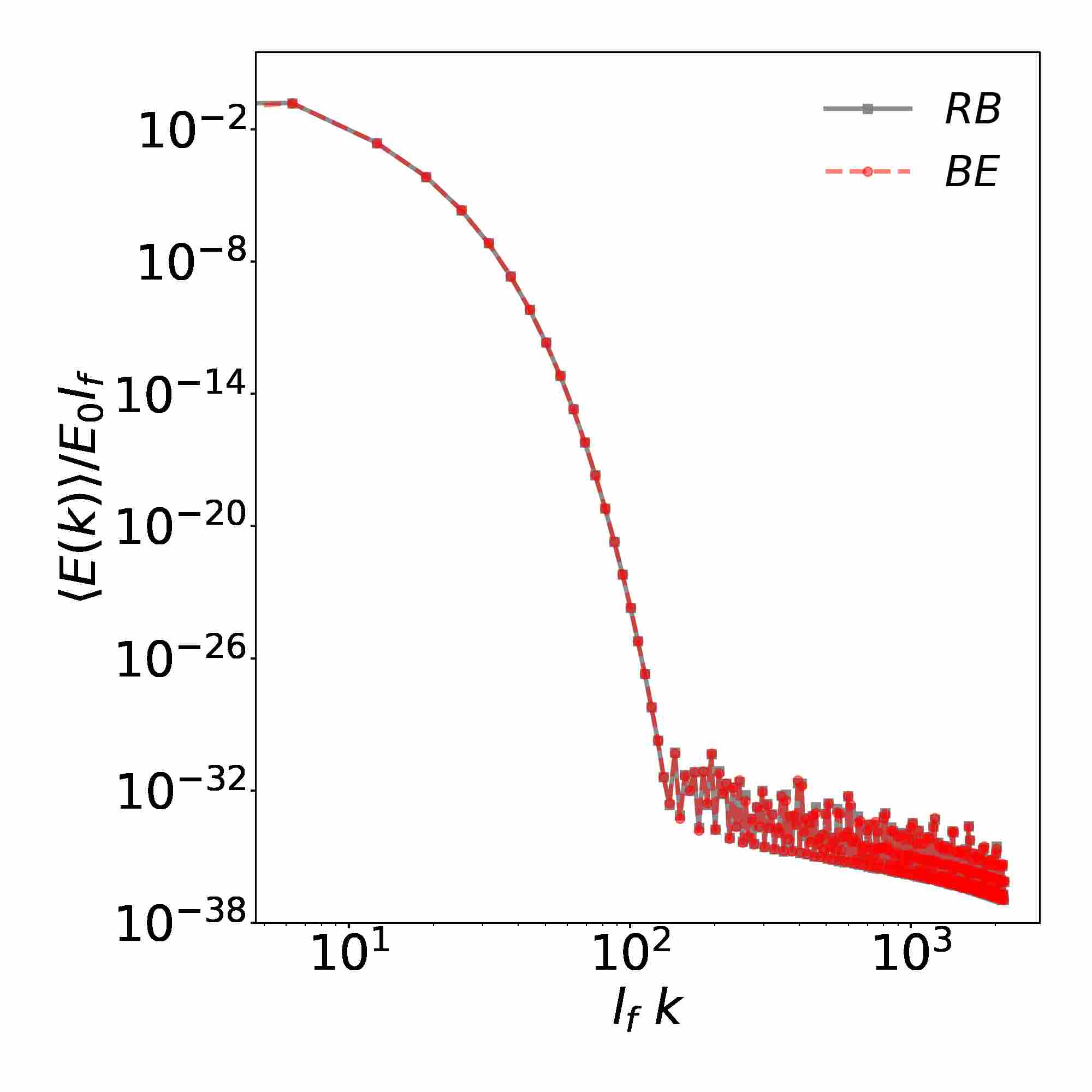}
	\put(-25,105){(a)}
	\includegraphics[scale=0.2]{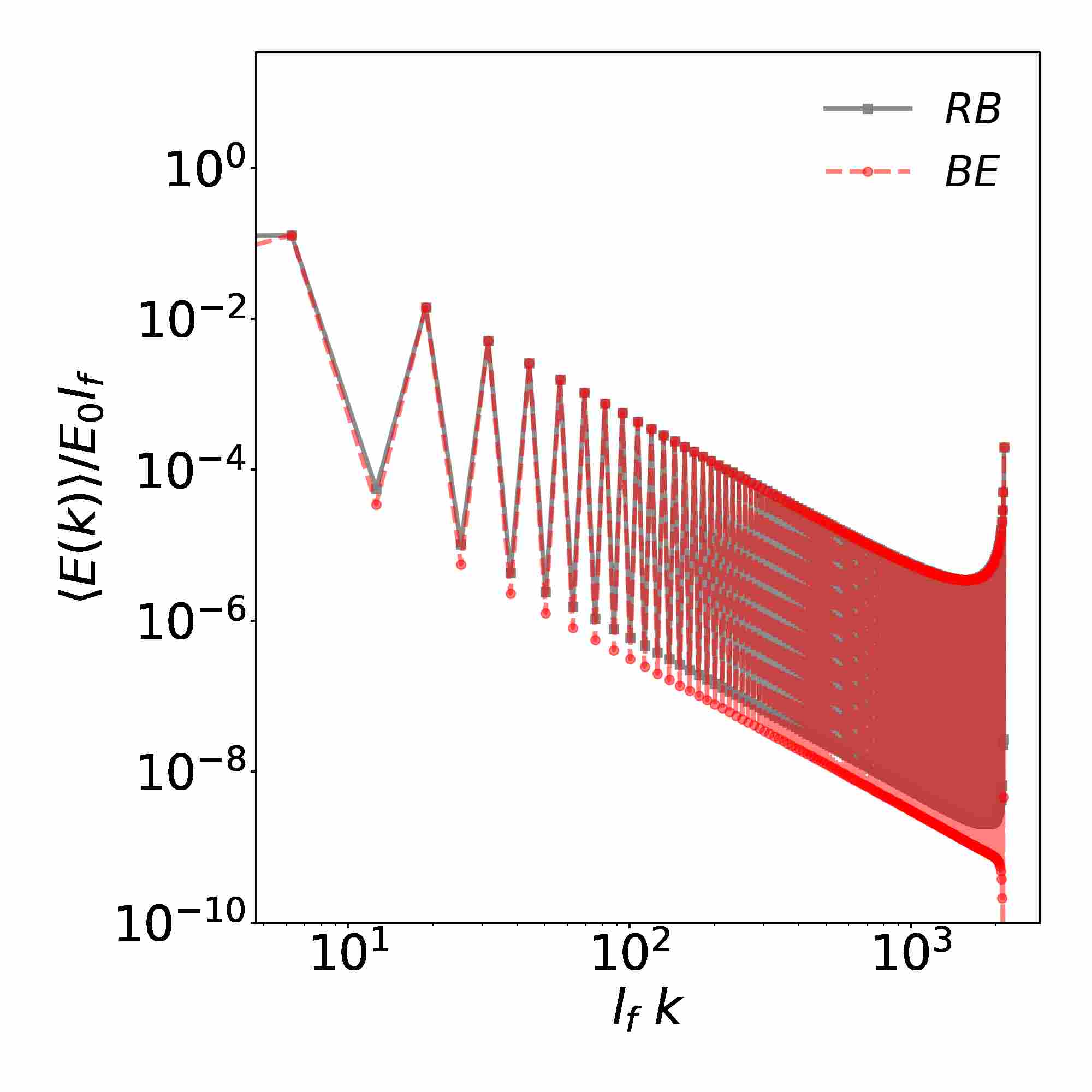}
	\put(-25,105){(b)}
	\includegraphics[scale=0.2]{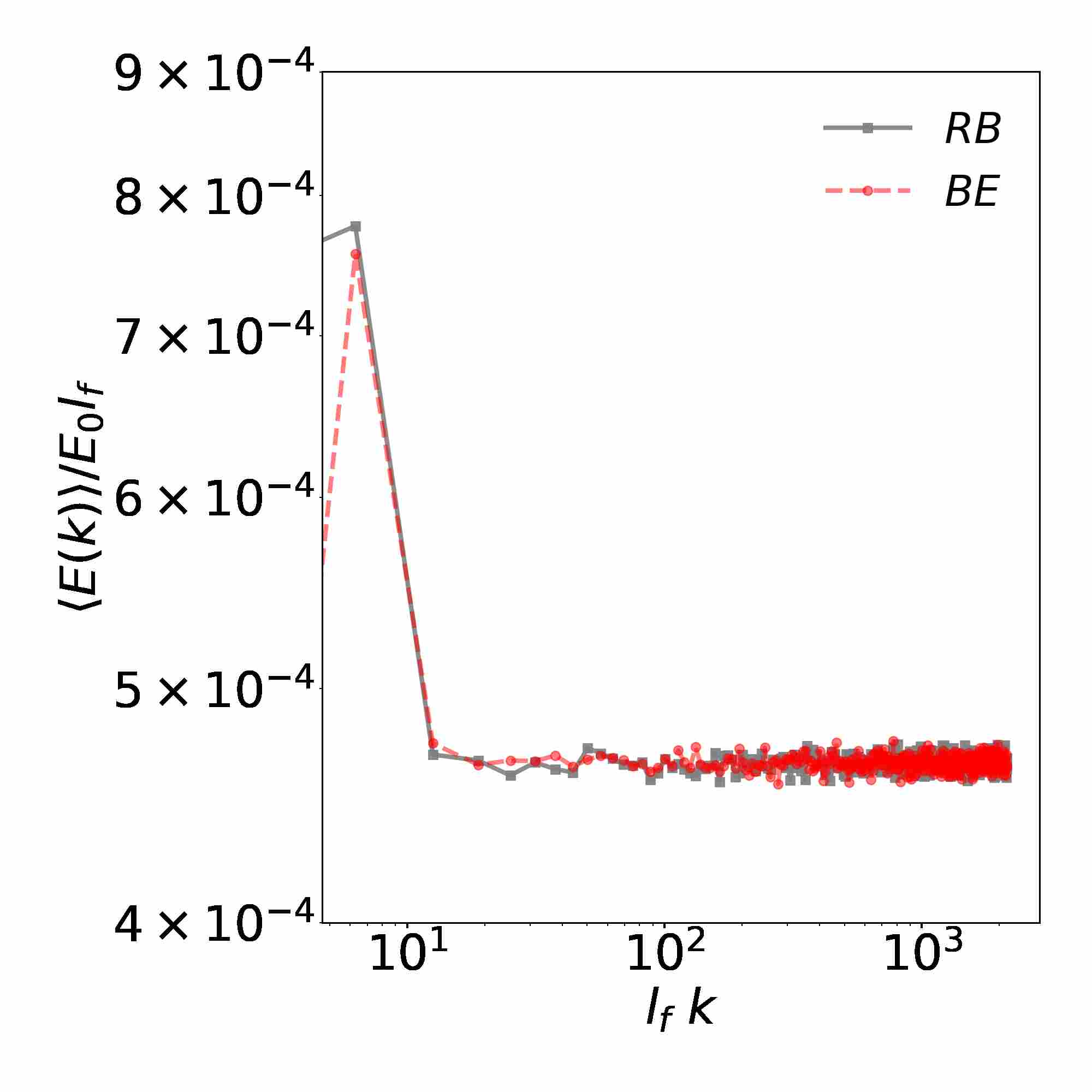}
	\put(-25,105){(c)}}
	\caption{Energy spectra showing the equivalence of RB and BE. (a) Regime A, $R=25.13$: Energy spectrum decays exponentially at large $R$. (b) Regime C, $R=0.31$: Oscillations pervade the spectrum. (c) Regime D, $R=0.28$: The thermalized regime. }
	\label{fig:equi_sp}
\end{figure*}

The characterization of the dynamical features so far, indicates that the RB system exhibits several features that are similar to those of the BE. But are the two systems statistically equivalent? Originally it was conjectured that the statistical features of the Navier-Stokes equation and the modified time reversible version are equivalent~\cite{gallavotti1995dynamical,gallavotti1996equivalence}. Now, if the conjecture holds for the present case, then we have two distinct microscopic ensembles wherein one is (formally) time-reversible (RB) and the other is not (BE), but both give rise to the same macroscopic behavior in a statistical sense. The above can be understood as follows. In RB, the energy $E_{RB}$ is held fixed and the viscosity $\nu_r$ is a fluctuating quantity, whereas in BE, the viscosity is fixed at $\nu_{BE}$ and the energy fluctuates. The equivalence, if established, allows us to write $\langle \nu_r \rangle_{RB} \approx \nu_{BE}$ and $E_{RB} \approx \langle E \rangle_{BE}$~\cite{margazoglou2022nonequilibrium,biferale2018equivalence,gallavotti2022navier} provided other parameters remain the same. This implies that the forcing chosen must also be the same for both systems.

To examine the conjecture for the Burgers system, we use the following numerical procedure. We perform the direct numerical simulations of the BE starting with initial condition $u(x,0)=u_0 \sin x$ and forcing $f(x)=\sin x$. The constant $u_0$ can be arbitrary, but the viscosity $\nu_{BE}$ is chosen such that the value of $R = f_0 \ell_f/\langle E_{BE} \rangle$ ($f_0=1, \ell_f=2\pi$) is in one of desired RB regime as discussed above; $\langle E_{BE} \rangle$ is the average energy of the BE system in the steady state. The BE simulations are run for sufficiently long durations, after a steady state has been achieved. Next, we use an arbitrary realization of the velocity field from the steady state, after an appropriate renormalization, to initiate the RB run. The initial velocity field for the RB run is set as $u_{RB}=\lambda u_{BE}$ with $\lambda=\sqrt[]{\langle E_{BE} \rangle / E_{BE}}$. This ensures the imposition of a global constraint of constant energy at $E_{RB}=\langle E_{BE} \rangle$.

Figures~\ref{fig:equivnu} (a)-(c) show the behavior of the RB system, starting from an appropriate BE steady state, in regimes A, C and D, respectively. These plots show that $\nu_r$ fluctuates around the mean value $\langle \nu_r \rangle = \nu_{BE}$. The insets depict the fluctuating energy $E_{BE}$ around the fixed energy $E_{RB}$ in each of the three regimes. Moreover, Fig.~\ref{fig:equi_sp} shows that the time-averaged energy spectrum in all the three regimes match very closely. In regime A (Fig.~\ref{fig:equi_sp} (a)), both the systems exhibit an exponentially decaying spectra; in regime C (Fig.~\ref{fig:equi_sp} (b)) both the systems display the presence of fast oscillations (same frequency) on the spectrum. Similarly, Fig.~\ref{fig:equi_sp} (c) indicates perfect agreement between the (time-averaged) spectra of the two systems. Furthermore, a comparison of the structure functions for the two systems shows an excellent agreement, for details we refer to Appendix~\ref{app:multifrac}.

The above observations clearly show that the statistical properties of the RB and BE are in a very good agreement, although the microscopic dynamics may differ considerably. The equivalence in regime A is very interesting, as this regime is devoid of any truncation effects and the system is described by a PDE. The rest of the regimes are dominated by the truncation effects, but even then the statistical equivalence holds strong. Thus, the conjecture seems to be true in general.

Also, the above demonstration of the equivalence conjecture motivates us to use the analytical machinery developed to study the dynamics of BE in gaining deeper understanding of the RB system; this is what we do in the next section.

\subsection{Influence of singularities on the dynamics}
{\label{section:pole}}
So far we have not provided a detailed justification for the observed behavior of energy spectrum in different regimes, except for suggesting that singularities of the solution may play an important role. We held off because to explain the behavior of the energy spectrum would require us to invoke a well known, powerful result for the viscous BE: a pole expansion solution~\cite{senouf1997dynamics}.

Given that the `\textit{equivalence conjecture}' holds across the different regimes of the RB system, the (statistically-averaged) behavior of the energy spectrum for the two systems are identical. We use this result in conjunction with the above analytical method to understand the various statistical regimes of the RB system. We first introduce the pole expansion, then use it to derive an analytic expression for the spectrum values and subsequently use the positions of the poles to justify the observations in the RB.

As such, a clear understanding of the existence of singularities in the BE and their dynamics are crucial to completely characterize the system. For our initial conditions, $u(x,0)=u_0\sin(x)$, it can be shown that in the inviscid and viscous BE a pole condensation occurs on the real axis at $x=\pi$ at $t_*=1/u_0$ (see Appendix~\ref{app:poleexpn}). Although we do not know the analytic equations describing the dynamics of the poles of the forced BE, we know their positions in a steady state, which we have derived in~\eqref{eq:sssBEpoles}. For the forced BE, we have also shown that we expect a pole condensation at $x=\pi$ as the viscosity becomes smaller.

Before delving further into the dynamics we develop an expression for the energy spectrum. Following~\cite{senouf1997dynamics}, a general meromorphic solution to the BE can be expressed using a Mittag-Leffler pole expansion as
\begin{equation}\label{eq:pole-fc}
	u = \frac{x}{t} - \sum_{n} \frac{2\nu }{x-a_{n}(t,\nu)},
\end{equation}\label{eqn:vel-pole}
where $a_{n} = a_{n}(t,\nu)$ are the time-dependant pole locations. The Fourier transform of the above equation gives
\begin{equation}
\hat{u}_{k} = \frac{\hat{x}_{k}}{t} - 2\nu \sum_{n} \int_{-\infty}^{\infty} \frac{e^{ikx}}{x-a_{n}} \,dx.
\end{equation}
Note that $\hat{x}_{k} =\int_{-\infty}^{\infty} x \, e^{ikx}\, dx$ does not converge. However, we can guess its behavior by considering the integration over the interval $[0,2\pi]$. This is justified since we are dealing with periodic boundary conditions. Thus, we find $\hat{x}_{k} = 2/ik$.

Now, depending on the location of poles (on the real line or complex plane), we investigate the following two cases.

$\textbf{Real poles}$: If $a_{n}\in \mathbb{R}$, then
\begin{equation}\nonumber
\int_{-\infty}^{\infty} \frac{e^{ikx}}{x-a_{n}}dx = \begin{cases} i\pi e^{ika_{n}}, & k>0\\
												    	-i\pi e^{ika_{n}}, & k<0
													\end{cases}
\end{equation}
This gives $ \hat{u}_{k} = 2/ikt - \sum_{n}2\nu i\pi e^{ika_{n}} \text{ for } k > 0.$ From this, we find
\begin{equation}\label{eqn:spectrum-real}
\begin{split}
	|u_{k}|^2 &= 4\Bigl[\frac{1}{k^2 t^2}+\frac{2\nu\pi}{kt}\sum_{n}\cos(ka_{n}) \\ &+\nu^{2}\pi^{2}\sum_{m,n} \cos[k(a_{n}-a_{m})]\Bigr].
\end{split}
\end{equation}

$\textbf{Complex poles}$: If $a_{n}\in \mathbb{C}$, then
\begin{equation}\nonumber
\int_{-\infty}^{\infty} \frac{e^{ikx}}{x-a_{n}}dx = \begin{cases} 2i\pi e^{ika_{n}}, & k\Im(a_{n})>0 \\
												    	-2i\pi e^{ika_{n}}, & k\Im(a_{n})<0
													\end{cases}
\end{equation}
This gives $\hat{u}_{k} = 2/ikt - \sum_{n}4\nu i\pi e^{ika_{n}} \text{ for } k > 0. $ From this, we find
\begin{equation}\label{eqn:spectrum-complex}
\begin{split}
	|u_{k}|^2 &= 4\Bigl[\frac{1}{k^2 t^2}+\frac{4\nu\pi}{kt}\sum_{n} e^{-k\Im(a_{n})}\cos{(k \Re(a_{n}))} \\
	&+4 \nu^{2}\pi^{2}\sum_{m,n} e^{-k\Im(a_{m}+a_{n})} \cos{[k \Re(a_{n}-a_{m})]}\Bigr]
\end{split}
\end{equation}

The sums are over positive $m,n, \text{ and } k$.

\begin{figure*}
	\centering
	\resizebox{\textwidth}{!}{
	\includegraphics[scale=0.2]{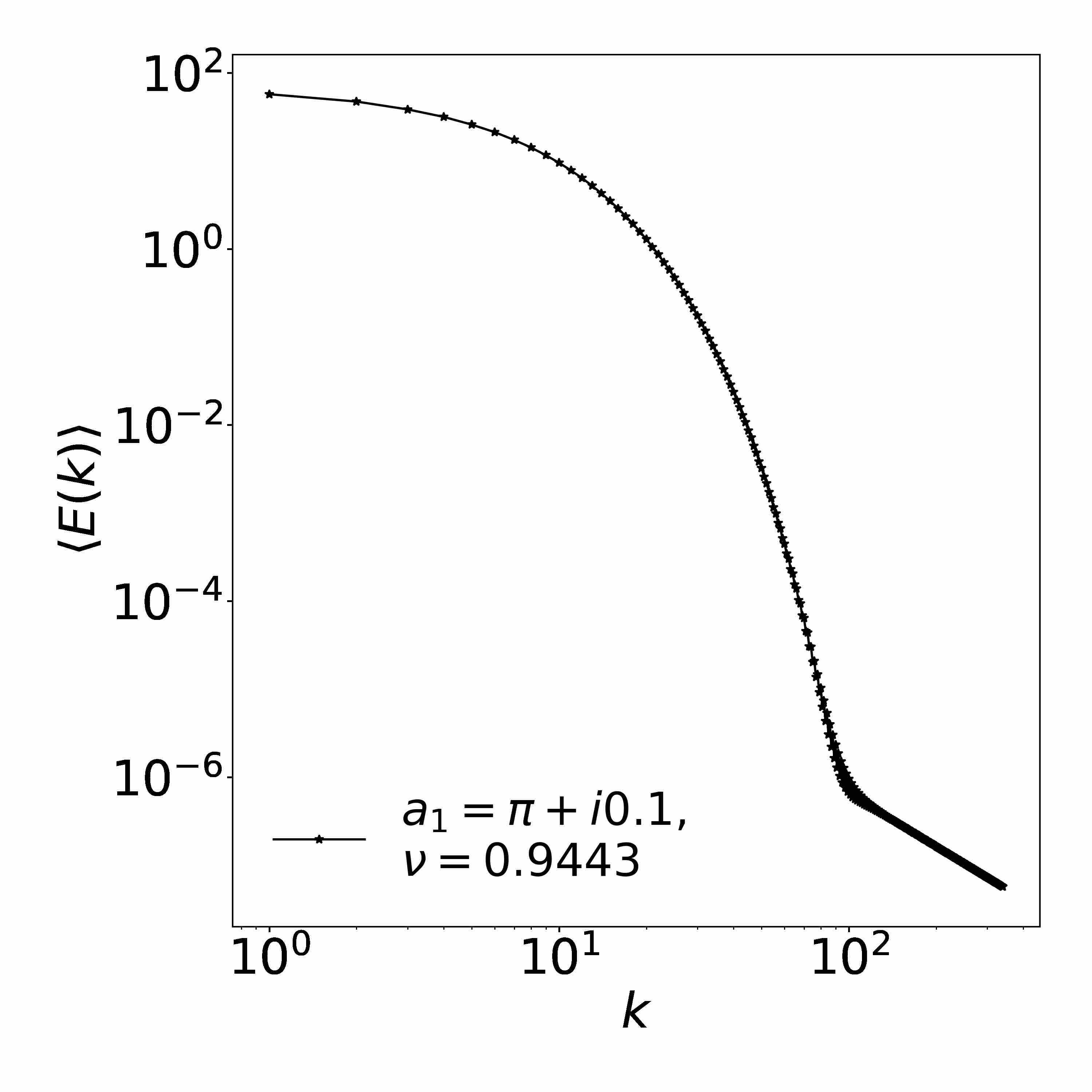}
	\put(-20,100){(a)}
	\includegraphics[scale=0.2]{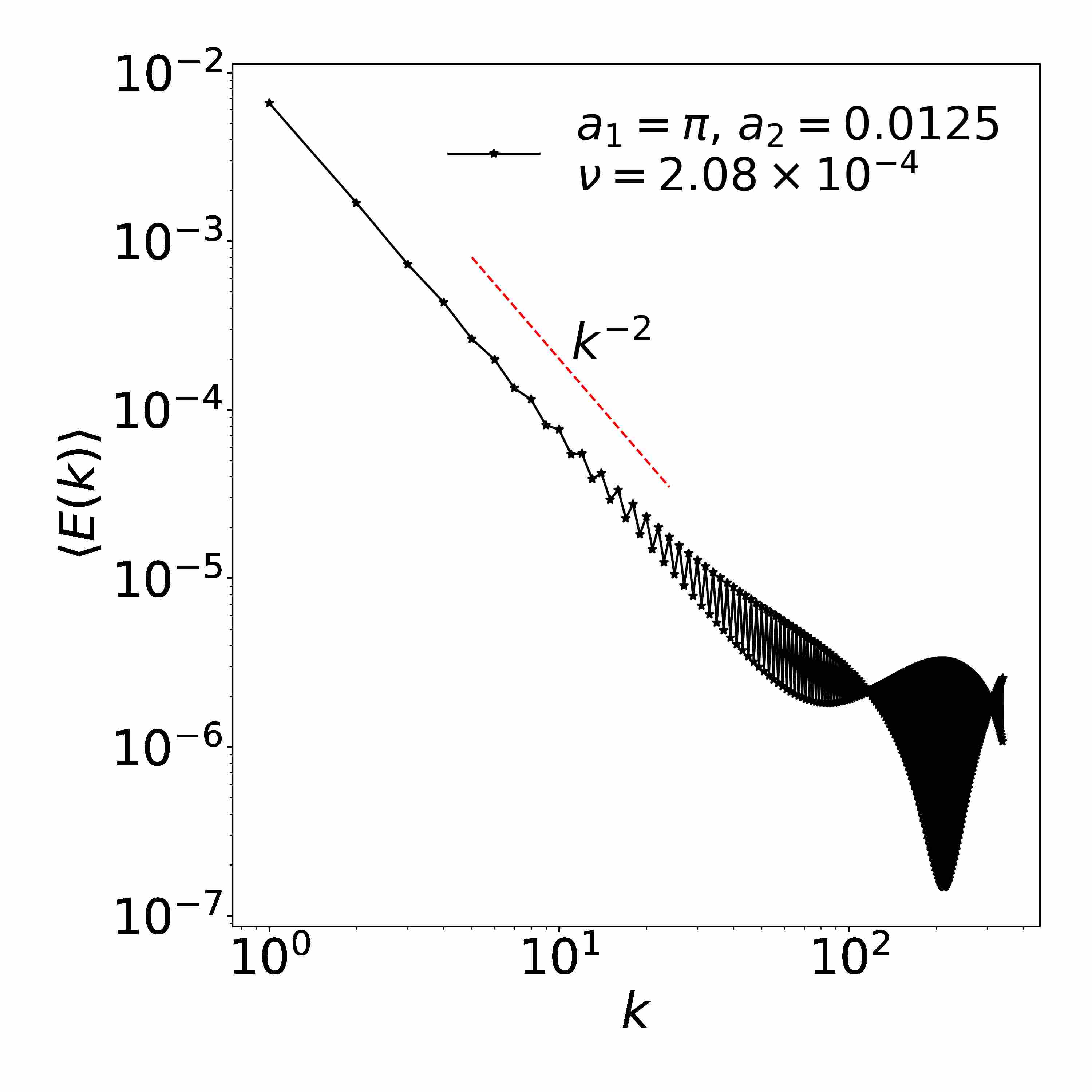}
	\put(-20,100){(b)}
	\includegraphics[scale=0.2]{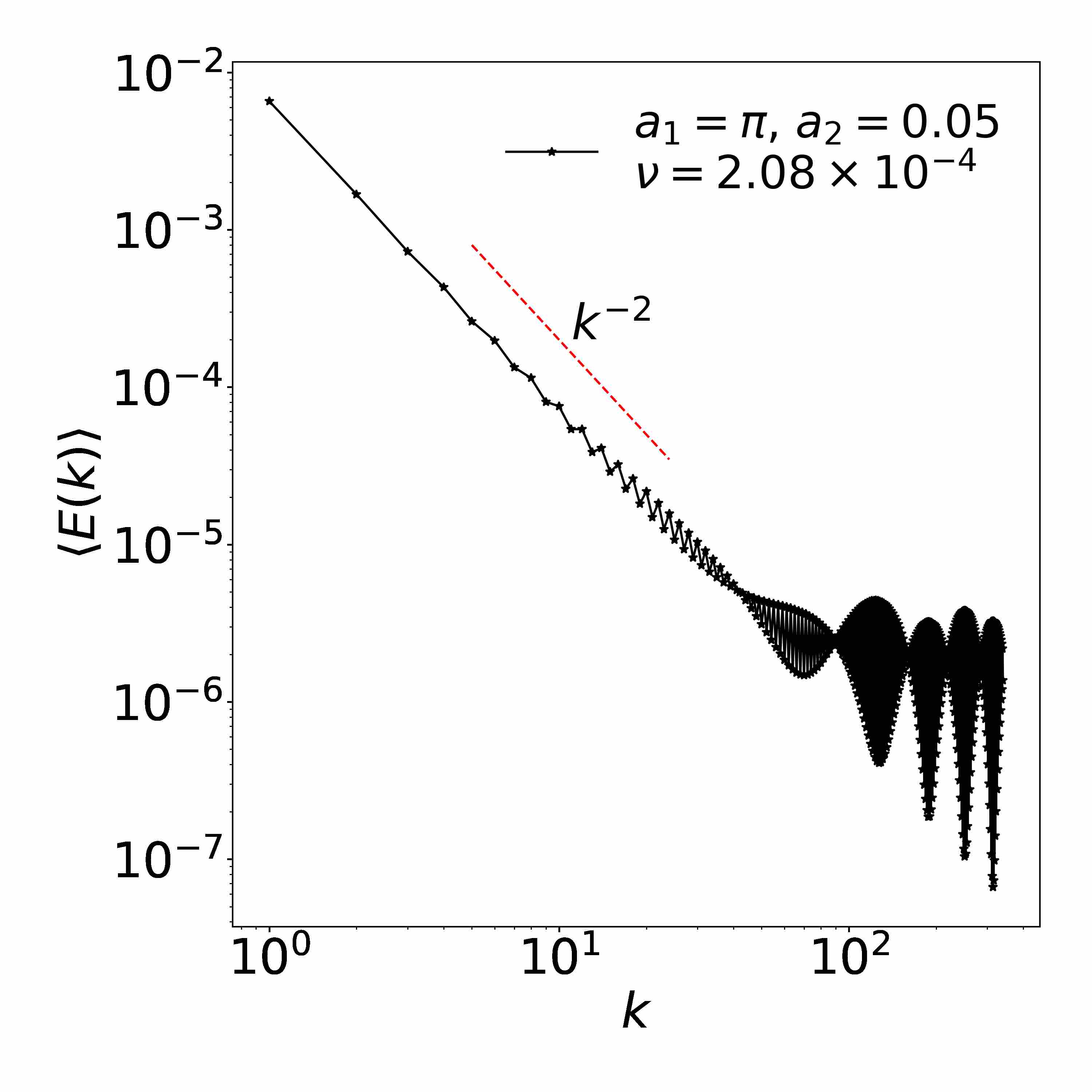}
	\put(-20,100){(c)}}
	\caption{Semi-analytically obtained energy spectrum. (a) Exponential behavior of the energy spectrum corresponding to a pole in the complex plane. (b), (c) Oscillatory behavior of the energy spectrum corresponding to the motion of one pole on the real axis and the other one fixed at $\pi$. }
	\label{fig:analyticc-spectrum}
\end{figure*}

Care must be taken while considering the sum in Eq.~\eqref{eq:pole-fc}. To have a consistent definition, the poles in the upper-half plane must contribute only to the positive wave-numbers, while the poles in the lower-half plane contribute only to the negative wave-numbers. Let us define $ \{ a_{n} : n > 0 \} $ as the poles in the upper half plane i.e $\Im(a_{n}) > 0$ for positive $n$. We also define $ \{ a_{-n} = \bar{a}_n : n > 0 \} $ as the poles in the lower half plane. Then,

\begin{flalign*}
 \hat{u}_{k} & \equiv 2/ikt - \sum_{n>0}4\nu i\pi e^{ika_{n}}\\
 \hat{u}_{-k} & \equiv -2/ikt + \sum_{n<0}4\nu i\pi e^{-ika_{n}}
\end{flalign*}
Thus, we can write
\begin{flalign*}
 \hat{u}_{-k}^{*} & = 2/ikt - \sum_{n<0}4\nu i\pi e^{ik\bar{a}_n}\\
 & = 2/ikt - \sum_{n>0}4\nu i\pi e^{ika_{n}}\\
 & = \hat{u}_{k}
\end{flalign*}
This justifies our reality condition, and also prevents the Fourier coefficients from blowing up when $k\Im{(a_n)}<0$.

Thus, we find a leading behavior of $k^{-2}$ in both cases of the energy spectrum, independent of whether the location of the pole is real or complex. However, the presence of the poles add a oscillating cosine term and a decaying exponential term, which causes drastic changes in the observed spectrum. In Fig.~\ref{fig:analyticc-spectrum} we plot the spectrum that is obtained in the regimes A and C.

In regime A, the truncation effects are absent and both the BE and RB exhibit an exponential fall-off in the spectrum. Figure~\ref{fig:analyticc-spectrum} (a) shows this behavior. The exponential terms in Eq.~\eqref{eqn:spectrum-complex} dominate and determine the behavior of the spectrum, with little to no oscillatory behavior. The exponential fall is a clear indication that the regime is without truncation effects, as the complex space singularities do not come within one Galerkin wavelength of the real axis. In this regime, the motion of the poles is arrested much before any condensation occurs and we are left with an exponential spectrum without any flattening or oscillations.

To understand the behavior of the spectrum in regimes B and C, we must remember that even though the BE and RB are equivalent in a statistical sense, with same macroscopic features, they have different microscopic dynamics. This is quite evident if one considers the pole dynamics. In the forced BE corresponding to regime B and C, a steady state is attained which also evidently indicates an arrest of the motion of the poles; however, corresponding to this in regimes B and C of theRB, the motion continues indefinitely albeit about the steady state of BE as fluctuations. For regime B, this behavior arises as a periodic motion in the spectrum, where the initial exponential spectrum becomes flat and returns back to an exponential state. This is the result of the poles, which were near $\pi$ in the complex plane (\textit{i.e.}, with non-zero imaginary parts), coming close to the real axis, at $x=\pi$, and then retracing their path away from $x=\pi$, due to the fluctuating viscosity [c.f. Fig.~\ref{fig:traceB}]. We must also take into consideration the effect of the Galerkin truncation. The presence of $e^{-k\Im(a_n)}$ terms results in the truncation error being exponentially small as long as the imaginary parts remain sufficiently large. Singularities approaching the real axis cause truncation effects which show up as tygers. This appears in the spectrum as oscillations (c.f. Fig.~\ref{fig:regimeB} (c)).
We can define the threshold of the appearance of truncation effects as the time $t_g$ such that $k_{\max}\Im[a_n(t_g)]=1$. Since in regime B, the poles move towards and away from the real axis in a cyclic manner, the tygers show a periodic nature with truncation effects disappearing and reappearing. Also, since the singularities are very close to $x=\pi$, oscillations start to develop in the spectrum with a $k^{-2}$ scaling. Such a scaling is a well known feature associated with the shock formation.

In regime C, the spectrum develops modulations on top of the oscillations and $k^{-2}$ scaling found in regime B. The modulations divide the spectrum into lobes. These are the result of poles appearing in the complex plane close to the origin, the position where the second discontinuity forms. These poles are in fact the reason for the formation of the secondary discontinuity. These new poles coupled with the poles at $x=\pi$ give rise to the structure in the velocity space. Furthermore, the closer the two groups of poles are, the more the number of lobes in the spectrum.

Figure~\ref{fig:analyticc-spectrum} (b) shows the energy spectrum, obtained semi-analytically by assuming the two poles to be at $a_1 = \pi$ and $a_2=0.0125$ (slightly off the origin). The viscosity is set to a very small value ($\nu \sim 10^{-4}$), so $R$ is in regime C. The contribution from the cross-terms can be neglected in Eq.~\eqref{eqn:spectrum-real}, as they come with weights proportional to $\nu^2$. The contribution responsible for the deviations from the base scaling is the second sum in Eq.~\eqref{eqn:spectrum-real}, which is $\sim \nu[\cos {k\pi}+\cos{ka_2}]/kt$. The total contribution from this term is responsible for the modulations and lobes in the spectrum. The width of the lobe is maximum, when this term is maximum and vice-versa. The pole at $a_1=\pi$ is responsible for the oscillations throughout the length of the spectrum, due to the $\cos(k\pi)$ term, which has the value $(-1)^{k}$. The second pole at $a_2$ is responsible for the modulations--it causes the sum to have zero contribution at points where
\begin{equation}
	\begin{split}
	\cos{ka_2} &=-\cos{k\pi}, \\
	k &= \begin{cases}
		(2n+1)\pi/a_2, & \text{$k$ is even and $n \in \mathbb{Z}$},\\
		2n\pi/a_2, & \text{$k$ is odd and $n \in \mathbb{Z}$}.
		\end{cases}
	\end{split}
\end{equation}
This reproduces the spectrum analytically.

Now, the larger the value of $a_2$, the more are the number of lobes. This is because the term $(\cos {k\pi}+\cos{ka_2})$ has values close to zero at larger number of $k$'s, as $k$ is smaller for larger $a_2$ and so more such $k$ values lie within $\kmax$. This also means that for larger $\kmax$, we should observe more modulations. Large $a_2$ also implies that the distance between the shock and the second discontinuity is less. As similar behavior (a large number of lobes appearing) is also encountered if the pole at $x=\pi$ moves, implying that a change in the distance between the shock and the second discontinuity results in more modulations. Hence, the motion of the shock is associated with these lobes.
Our numerical results are in agreement with this assessment, in Fig.~\ref{fig:microC} (c) $\langle E(k) \rangle$ for $R=1.79$ (red solid line) exhibits multiple lobes. In this case, the shock and secondary discontinuity in velocity move farthest in space (not shown). Also, the dependence of the modulations on $\kmax$ can be used to explain the appearance of the structure much earlier in regime B for the larger resolutions (greater than $1024$). For large $\kmax$, even a slight motion of the shock or a small $a_2$ can lead to the formation of a lobe in the spectrum, whereas at the smaller resolution this is not possible simply because those modes are not available. Since the motion of the shock depends on the truncation waves, which are more vigorous for large $\kmax$, the formation of the structure can be expected at larger $R$ for large resolutions.

Thus, the use of the pole expansion method, combined with some observations from our numerics, allows us to predict the behavior of the spectrum, at least qualitatively, and also identify the mathematical features that are responsible for the structure that appears in our simulations. Also, we have been able to explain how such a structure forms at the higher resolution for larger $R$. However, some clarifications must be made. First, we do not explicitly know the full analytic form of the solution--we only know a stationary state approximation to the entire solution. Thus, the complete dynamics of the singularities is not known explicitly. Nevertheless, the singularities of the steady state solution allow us to make some assumptions regarding the poles. For example, we have assumed simple poles. Second, in our semi-analytical analysis we have only assumed two poles, but in reality there may be an infinite number of them. And finally, we have neglected any dynamics of the poles, when calculating the time-averaged spectra shown in Fig.~\ref{fig:analyticc-spectrum}, whereas in reality the pole locations are time-dependent. Despite these shortcomings, it is nonetheless remarkable that we have been able to capture all the broad features of the system using this crude approach.

\subsection{Phase transitions: Continuous and discontinuous}

\begin{figure*}
	\centering
	\resizebox{\textwidth}{!}{
	\includegraphics[scale=0.3]{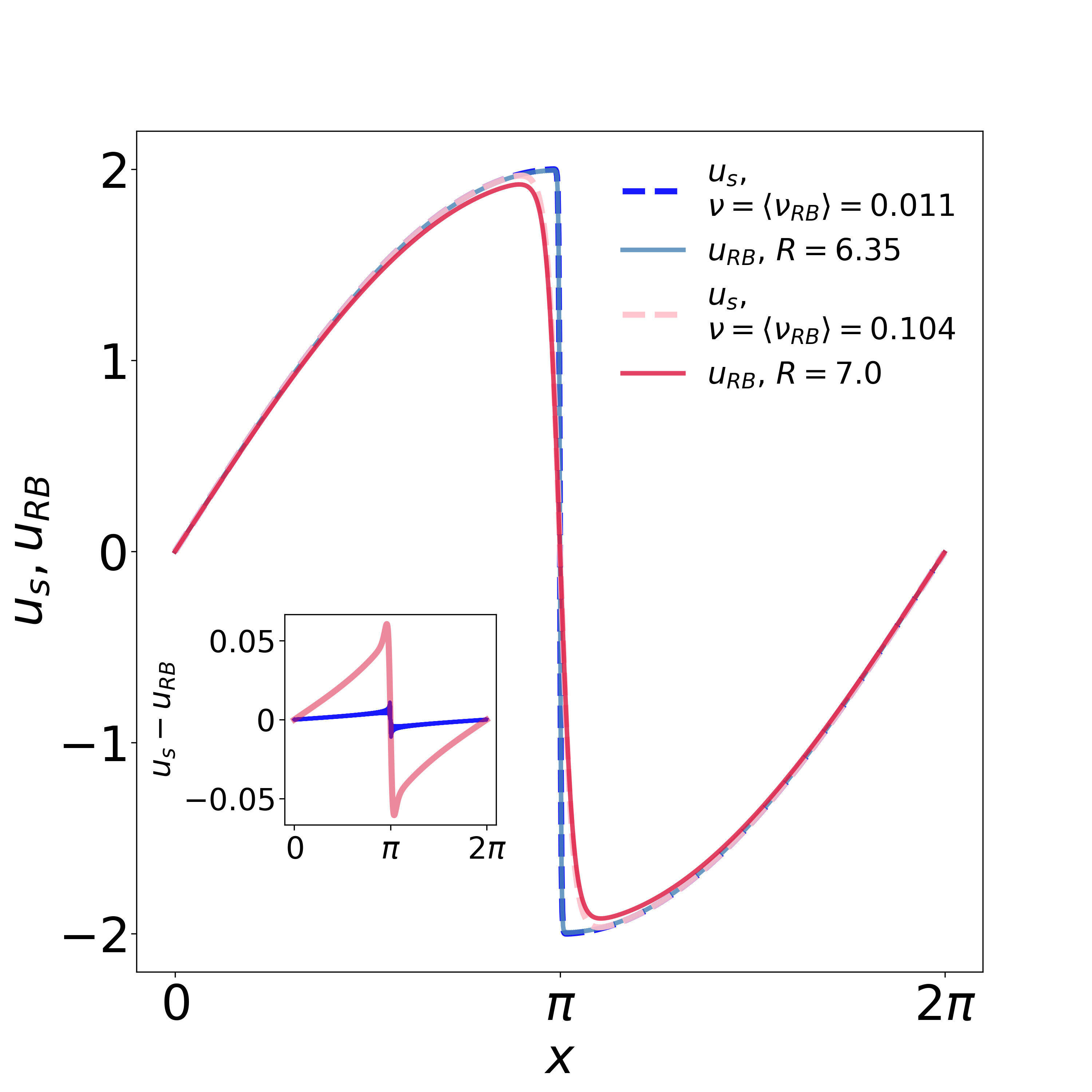}
	\put(-180,160){(a)}
	\includegraphics[scale=0.3]{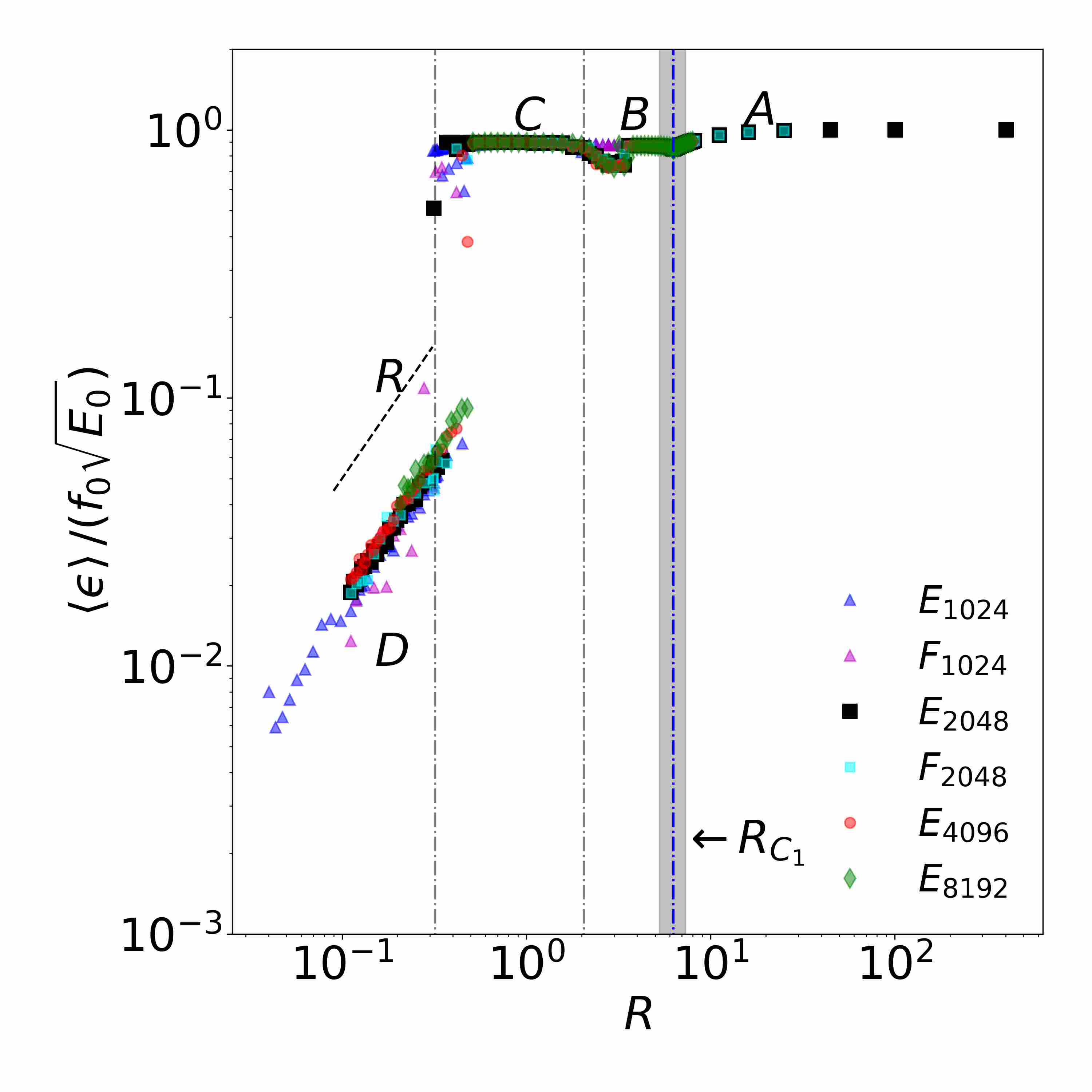}
	\put(-165,170){(b)}}
	\caption{(a) Comparison of the asymptotic steady-state velocity of the BE, $u_s$, and  RB, $u_{RB}$. The inset shows the difference. (b) Plot of $\langle \epsilon \rangle$ vs $R$. $\langle \epsilon \rangle$ is a clear indicator of the first order transition associated with a discontinuous jump with a change of scaling from constant $R^0$ to $R$.}\label{fig:phasetrans}
\end{figure*}

\subsubsection{The continuous phase transition}
\label{sec:cpt}

The first phase transition $\mathcal{T}_{AB}$ is essentially an indicator of the birth of truncation effects and it precedes the birth of tyger phenomenon. In our DNSs, the critical point of this transition is observed to be at $R_{c_1}=6.3$. Most interestingly, after this transition tygers are prevalent in the RB, and the behavior of $\Omega$ and $\nu_r$ for $R \lessapprox R_{c_1}$ is a consequence of the tyger phenomena.

It is known that in the case of an inviscid BE, a pole condensation takes place on the real axis with the formation of a shock. The tygers then arise when the singularities causing the shock come within one Galerkin wavelength of the real axis. The story is similar for the RB: tygers and all the truncation phenomena result from the shock and the singularities that give rise to it. However, the RB is a forced-dissipated system in which forcing plays a major role and dictates the shock formation. Therefore, to understand this transition, we need to have a clear picture of how forcing affects the nature of the shocks.

Let us consider the steady state solution, Eq.~\eqref{eq:sssBEmatched}, of the forced BE, which we reproduce below:
\begin{equation}\nonumber
\begin{split}
u_s = &2\sgn(x-\pi)\sqrt{f_0}(\sin(x/2)-1) \\
&+2\sqrt{f_0}\tanh(\sqrt{f_0}(\pi-x)/\nu).
\end{split}
\end{equation}
The outer solution $u_{outer}=2\sgn(x-\pi)\sqrt{f_0}\,\sin(x/2)$ acts as the envelope of the solution and dictates the energy content of the steady state when a shock has formed. The energy calculated for the steady state using the outer solution is $E_s=\langle u^2_{outer}/2 \rangle=f_0$. Thus, the dimensionless parameter $R=2\pi$. Note that it is surprisingly close to the numerically observed $R_{c_1}$. Now we make use of the `equivalence conjecture' and the fact that the transition $\mathcal{T}_{AB}$ is accompanied by the appearance of a shock and $\nu_r$ drops to a small value to appreciate this better. For the RB system the energy $E_0$ is held fixed, the shock forms when $E_0$ is close to the value $E_s$. For protocol E, it gives $E_0=E_s=f_0=1$, whereas for the protocol F, this occurs at $f_0=E_s=E_0=0.25$. As a result, the value of $R_{c_1}$ can be predicted by using $E_0=f_0$ in the RB for which the mean $\langle \nu_r \rangle$ is sufficiently small. This prediction relating the energy content of the outer solution to the transition point $R_{c_1}$ clearly depicts that the first phase transition is a result of the shock formation; thus, a pole-condensation.

A comparison of the asymptotic steady state solution $u_s$ and the numerically determined solution $u_{RB}$ in Fig.~\ref{fig:phasetrans} (a) shows that the agreement between the two becomes better as $R$ approaches $R_{c_1}$; the difference $u_s-u_{RB}$ goes to zero [see the inset of Fig.~\ref{fig:phasetrans} (a)]. In fact, at $R=R_{c_1}$ they lie on top of each other. The reason $R_{c_1}$ is not exactly $2\pi$ is associated with the fact that the truncation effects are known to be felt a bit earlier than the predicted shock formation, for the RB this means that $E_0$ is slightly less than $f_0$ (evaluated earlier). Hence, $R_{c_1}$ is slightly greater than $2\pi$.

\subsubsection{The discontinuous phase transition}
\label{sec:dpt}

The discontinuous jump $\mathcal{T}_{CD}$ is the other striking feature of the RB dynamics. The jump is reflected in all the global quantities and depends on the truncation wave number $\kmax$. The latter is a clear indication of thermalization where the quantities, such as, $\Omega$, etc, reach their absolute equilibrium values.

In the forced-dissipated BE dynamics, the building up of a shock is accompanied by the piling up of energy at the location of the shock. Eventually, this pile-up leads to a catastrophe when a shock discontinuity forms and the energy from the shock is lost to the viscous dissipation; this continues until a steady flux of energy is established. This is indeed what happens in a well-resolved DNS. However, once the truncation effects become pronounced, the above mechanism is altered. For example, in the case of inviscid BE, the energy piled up in the shock shows up as truncation waves at particular resonance points and we witness the birth of tygers. The transfer of energy from the shock to the tygers continues until the latter attain similar amplitudes as the former; subsequently, the white noise born from the collapse of the tyger envelope leads to complete thermalization. Now coming back to the RB, apart from regime A, the scales are not well resolved and truncation effects play a major role. The energy transfer from the shock to the tygers sets in, instead of being dissipated by the viscosity. However, unlike the inviscid BE, in which the viscous dissipation is absent, in the RB system the state-dependent viscosity dissipates the energy associated with the tygers and a steady state is reached.

We use the above general considerations to advance our understanding of the observed discontinuous transition. In regime C of the RB, a collapse of the tyger envelope is observed and subsequently, a secondary discontinuity is seen to form at the origin, buffeted by wings of white noise. We have already mentioned in a previous section that the amplitude of this secondary discontinuity increases as $R$ decreases; the height of this discontinuity and the amplitude of the white noise determines whether the system will eventually thermalize or not. When the height is less than the height of the shock, the breakdown of the tyger envelope is not extremely severe and the (spatial) structure shown in Fig.~\ref{fig:RC_minishock}(a) is formed, which is essentially a result of regularization occurring due to the finite viscosity post tyger collapse. This regularization process effectively holds up the thermalization and only a partially thermalized state is observed. In other words, whether the RB system will thermalize or not, depends on the stage of the energy transfer mechanism at which the effect of the viscosity is felt. In regime C, the viscosity acts before the white noise can attain amplitudes as large as the shock, whereas in regime D the viscosity is so low that it fails to prevent the white noise from catching up with the shock amplitude. The viscosity does act to maintain the steady state, but by then thermalization has taken hold. With the appearance of thermal states, the energy that had been pouring into the tyger is now equally distributed among all the length scales and the energy spectrum $E(k)\sim k^0$. Thus, when the RB system transits from regime C to D, the above signature energy pile-up mechanism disappears and the global quantities exhibit a truncation wave number ($\kmax$)-dependent jump. This $\kmax$ dependence derives from the truncation dependence of the tyger; see Appendix~\ref{app:tygersiBE}. This can be seen from truncation waves $\sin(k_{max}+1/2)x / \sin(x/2)$, that give rise to the tygers. The truncation waves and subsequently the thermal white noise clearly depend on the Galerkin truncation. The breakdown is much more vigorous at larger $k_{max}$ since there are much more truncation waves and subsequently much more white noise disturbances. As a result, smaller $E_0$ is required for the white noise to match the shocks. This explains the resolution dependence of the second critical point $R_{c_2}$.

We finally examine the behavior of the energy injection rate $\widetilde{\epsilon} \coloneqq \epsilon/f_0\sqrt{E_0}$, as $R$ is varied across different regimes of the RB; see Fig.~\ref{fig:phasetrans} (b). The plots from all the set of runs show that $\widetilde{\epsilon}$ is nearly constant across regimes A, B, and C; however, it shows a sudden discontinuity at $R_{c_2}$ and its value goes down by an order of magnitude. Thereafter, in regime D, we have $\widetilde{\epsilon} \sim R$. We can understand this scaling behavior as follows. Note that in the thermalized regime $\epsilon \sim f^2_0\tau_{eq}$, where $\tau_{eq}=\ell_f/\sqrt{E_0}$ is the equilibrium correlation time. Hence, $\widetilde{\epsilon}=f^2_0\tau_{eq}/f_0\sqrt{E_0}=R$.
Recalling the unit of time that we had introduced earlier, $\tau=(\ell^2/\langle \epsilon \rangle)^{1/3}=(\ell_f\sqrt{E_0}/f^2_0)^{1/3}$, the viscosity in regime D can be evaluated as
\begin{equation}
\frac{\langle \nu_r\rangle\tau}{\ell^2_f}=\frac{\langle \epsilon \rangle \tau}{\Omega_{eq}\ell^2_f} \sim \frac{R f_0 \sqrt{E_0}\tau}{E_0 (\kmax\ell_f)^2} \sim \frac{R^{4/3}}{(\kmax\ell_f)^2}.
\end{equation}
Similarly, we can estimate $\langle \nu_r \rangle$ by observing that $\widetilde{\epsilon}$ is roughly constant and $\Omega \propto \Omega_{eq}$ in regime C:
\begin{equation}
	\frac{\langle \nu_r\rangle\tau}{\ell^2_f} \sim \frac{{\langle \epsilon \rangle}^{2/3}\ell^{2/3}_f}{E_0(\kmax\ell_f)^2} \sim \frac{R^{2/3}}{(\kmax\ell_f)^2}.
\end{equation}

\subsubsection{Effective theory}

We have established that the energy spectrum has two broad features, neglecting the oscillations and modulations. It is either flat or has a $k^{-2}$ scaling. We use these properties to develop an effective theory to explain the jump observed in the global quantities. To do so we approximate the energy spectrum as follows:

\begin{equation}\nonumber
	E(k)=\begin{cases}
			\alpha k^{-2}, & k_0 \leq k \leq k_{p} \\
        	\beta , & k_{p} < k \leq k_{max},
    	\end{cases}
\end{equation}
where $k_0(=1)$ and $k_{p}$ are the smallest wave numbers in our numerics and the cut-off wave number of $k^{-2}$ behavior, respectively. Note that $\alpha$ has dimensions of energy by length and $\beta$ has dimensions of energy $\times$ length. Also, we can write the total energy $E_{0} = E_{1}+E_{2}$, where $E_{1}$ is the energy content in the $k^{-2}$ modes and $E_{2}$ is the energy in the thermalized modes. Therefore, we have, assuming continuously varying $k$:
\begin{equation*}
    E_{1} = \alpha \int_{k_0}^{k_{p}} k^{-2} dk
    \implies \alpha = \frac{E_{1}}{(1/k_0-1/k_{p})};
\end{equation*}

\begin{equation*}
    E_{2} = \beta \int_{k_{p}}^{\kmax} dk
    \implies \beta = \frac{E_{2}}{\kmax-k_{p}}.
\end{equation*}
This allows us to express $\Omega$ using $E_1$ and $E_2$ as follows:
\begin{align*}
	\Omega &= 2\int_{k_0}^{\kmax} k^{2} E(k)dk \\
	&=2 E_{1}k_{p}k_{0} + \frac{2}{3}E_{2}(k_{p}^{2}+k_{p}\kmax+\kmax^{2}).
\end{align*}

If we consider $E_{1} = \lambda_{E} E_{0}$ and $k_{p} = \lambda_{k}\kmax$ for $0\leq \lambda_{E},\lambda_{k} \leq 1$, this yields
\begin{equation}\label{effective2}
\frac{\Omega}{E_{0}} = 2 \lambda_{E} \lambda_{k} k_0 \kmax+ \frac{2}{3}(1-\lambda_{E})(1+\lambda_{k}+\lambda_{k}^{2})\kmax^2
\end{equation}

We now consider the two limiting cases. (1) When the first transition $\mathcal{T}_{AB}$ starts to set in, we have $k_{p} = k_{max}$; thus, $\lambda_{E} = 1$ and $\lambda_{k} = 1$. Equation~\eqref{effective2} gives $\Omega = 2E_{0}k_{max}$, which is in agreement with numerical results at $R\approx R_{c_1}$. (2) When the transition $\mathcal{T}_{CD}$ starts to set in, we have $k_{p}\approx k_0$, which implies $\lambda_{E} = 0$ and $\lambda_{k} = k_0/k_{max}$. We obtain  $\Omega = (2/3) E_{0}\kmax^{2}$.

\begin{figure}
	\centering
	\resizebox{\columnwidth}{!}{
	\includegraphics[scale=0.2]{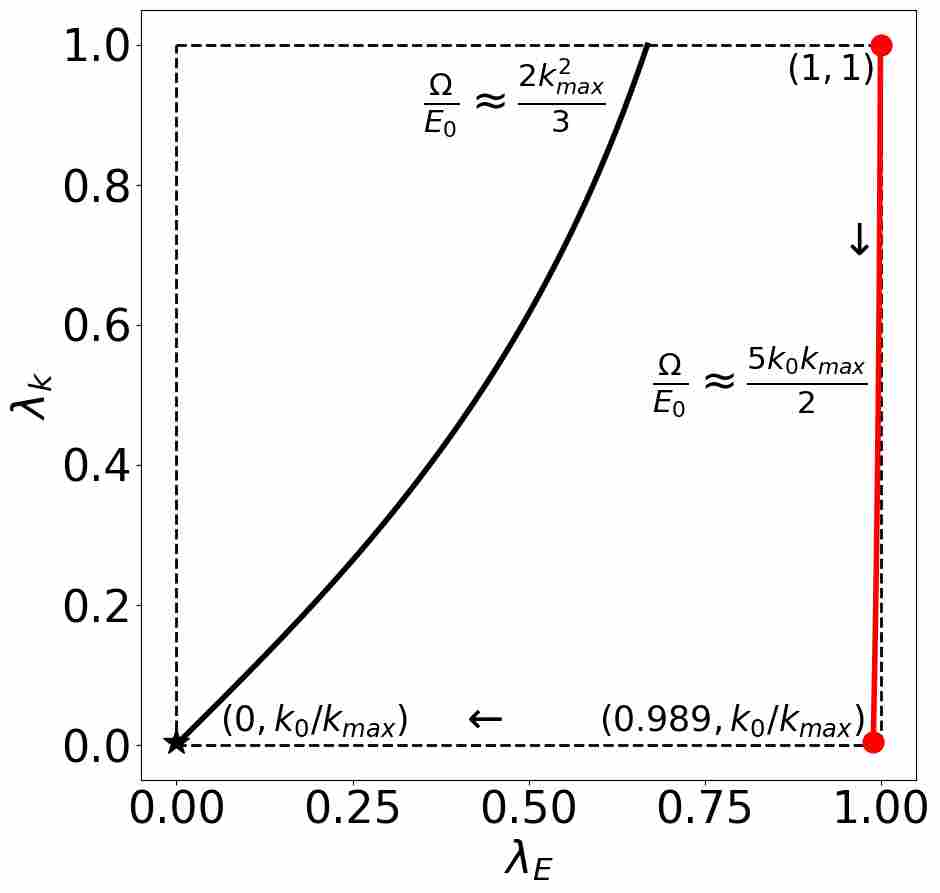}}
	\caption{Plot depicting the values of $\lambda_E$ and $\lambda_k$ for the two possible forms of $\Omega/E_0$, linear (red) and quadratic (black) in $k_{max}$. Here, $N_c=1024$ and $k_{max}=341$.}
	\label{fig:eff-jump}
\end{figure}

DNS results show that in regime C, $\Omega/E_{0} \propto \kmax$, the prefactor is $5/2$. For such a scaling to hold, Eq.~\eqref{effective2} suggests that $(1-\lambda_{E})(1+\lambda_{k}+\lambda_{k}^{2}) \propto k_0/\kmax$. In Fig.~\ref{fig:eff-jump}, we show a plot of the right-hand side of Eq.~\eqref{effective2} for $0\leq \lambda_{E},\lambda_{k} \leq 1$, and we observe that such a scaling survives only when $\lambda_E$ is very close to $1$.

Moreover, the expression for the energy spectrum can be rewritten as
\begin{equation}\nonumber
	\frac{E(k)}{E_0}=\begin{cases}
		\frac{\lambda_E \lambda_k k_0\kmax}{\lambda_k\kmax-k_0}k^{-2}, & k_0 \leq k \leq k_{p} \\
		\\
		\frac{(1-\lambda_E)}{(1-\lambda_k)\kmax}, & k_{p} < k \leq \kmax.
	\end{cases}
\end{equation}

Now, as $R$ decreases, $\lambda_{k}$ and $\lambda_{E}$ adjust their values until $\lambda_{k} \gtrapprox k_0/k_{max}$ and $\lambda_{E}$ given by the Eq.~\eqref{effective2}. For $k_{max} = 341$, $\lambda_{E}\approx 0.989$. The behavior of the system can be understood using Fig.~\ref{fig:eff-jump}. At the very onset of the transition $\mathcal{T}_{AB}$, we have $(\lambda_E,\lambda_k)=(1,1)$; hereafter, as we decrease $R$, the system moves along the red line, while retaining the $\Omega/E_0 \approx 5k_0\kmax/2$ behavior. This continues until $(\lambda_E,\lambda_k) \approx (0.989,k_0/\kmax)$ with $\lambda_k > k_0/\kmax$. $\lambda_E$ is very close to one, which suggests that the energy contained in the thermalized modes is very small, even though $\lambda_k$ is very close to, but greater than $k_0/\kmax$. This signifies that the truncation effects that tend to flatten the high wave number modes, have reached the smaller wave number modes. Note that when $\lambda_k = k_0/\kmax$, the only way $E(k)$ can have the above form is if $\lambda_E=0$, which corresponds to a complete thermalization. Thus, in Fig.~\ref{fig:eff-jump} the system jumps from the point $(0.989,k_0/\kmax)$ on the red line to $(0,k_0/\kmax)$ on the black line. This jump is associated with the discontinuous transition $\mathcal{T}_{CD}$ for the RB system.

\section{Concluding Remarks}
\label{sec:conclusions}

We have constructed a time-reversible formulation of the Burgers equation in the presence of a forcing and dissipation. We achieved this by imposing a global constraint on the Burgers dynamics in the form of a conservation of its total energy. This, in turn, modifies the viscous dissipation process and the fluid is no longer subjected to a constant viscosity, but has a dissipation coefficient that, in principle, can fluctuate in time and depends on the state of the system. The key questions that we address are whether this time-reversible Burgers equation reproduces the statistical properties of the BE and what are its different dynamical regimes.

Following Refs.~\cite{shukla2019phase,seshasayanan2021equivalence}, we introduced a dimensionless parameter $R$ to identify the different statistical regimes of the RB. We find that in between the hydrodynamic regime ($R \to \infty$) and the thermalized regime ($R \to 0$), the RB system goes through two major transitions or bifurcations: (i) a continuous transition with a critical point at $R_{c_1}$ and (ii) a discontinuous transition with a critical point $R_{c_2} \ll R_{c_1}$.

The continuous transition is associated with the shock formation and subsequently with the birth of truncation effects. The critical point $R_{c_1}$ is independent of the Galerkin truncation wave number. Interestingly for $R < R_{c_1}$, the truncation effects lead to the formation of spatial structures, which are similar to \textit{tygers}, originally observed in the inviscid BE~\cite{ray2011resonance}. However, the inherent difference between the inviscid BE and the RB makes these structures different from their inviscid counterparts. Moreover, the RB tygers eventually give way to an additional structure consisting of a smaller secondary discontinuity at the resonance point. The arguments based on the steady-state solution, obtained using the matched asymptotic expansion, allowed us to predict the value of the first critical point $R_{c_1}$ as $2\pi$, which agrees very well with numerically observed value of $\sim 6.3$.

At the final transition, global quantities exhibit jumps and attain values that are consistent with what is expected based on the establishment of a quasi-equilibrium state; the energy spectrum shows $k^0$ scaling and the thermalization has taken hold in the system. Also, we provide an effective phenomenological theory, with inputs from the numerical observations of global quantities and the behavior of partially thermalized spectrum, to describe the discontinuous transition at $R_{c_2}$. Hence, both transitions are strongly associated with truncation effects.

Our results show that the \textit{equivalence conjecture} holds for all regimes of the RB, and not just for the well-resolved DNSs in the hydrodynamic regime. Also, we have checked that these results extend to a randomly forced system as well. This suggests that the conjecture holds in general, irrespective of the proliferation of truncation effects because of limited numerical resolution. Furthermore, after establishing the conjecture, we used the analytical machinery of pole-expansion solutions developed for the BE to explain the statistical features of the time-reversible system. We believe that the results presented here could help shed light on the dynamical behavior of 2D and 3D RNS systems.

\textbf{Acknowlegments} V.S. would like to thank B. Dubrulle for several interesting discussions. The authors acknowledge National Supercomputing Mission (NSM) for providing computing resources of ‘PARAM Shakti’ at IIT Kharagpur, which is implemented by C-DAC and supported by the Ministry of Electronics and Information Technology (MeitY) and Department of Science and Technology (DST), along with the NSM Grant DST/NSM/R\&D HPC Applications/2021/03.21.  VS would like to acknowledge support from the Institute Scheme for Innovative Research and Development (ISIRD), IIT Kharagpur, Grant No. IIT/SRIC/ISIRD/2021-2022/03.

\appendix

\section{Error in the conservation of energy}
\label{app:errconsE}

The table below shows the error in the conservation of energy for the various regimes at different resolutions. The maximum error is found to be of the order of $10^{-5} \%$.

\begin{table}
	\centering
	\begin{tabular}{ |c|c|c|c| }
		\hline
		$N_c$ & R & $E_0$ & $100 \times | 1-E/E_0|$ \\
		\hline\hline
		{1024} & 11.17 & 0.5625 & 8$\times $$10^{-8}$\\
		& 6.28 & 1 & 2.5$\times $$10^{-7}$ \\
		& 1.01 & 6.25 & 3.6$\times $$10^{-6}$\\
		& 0.25 & 25 & 7.1$\times $$10^{-6}$\\
		\hline
		{2048} & 11.17 & 0.5625 & 1.59$\times $$10^{-7}$\\
		& 6.28 & 1 & 6.7$\times $$10^{-7}$ \\
		& 1.01 & 6.25 & 7.5$\times $$10^{-6}$\\
		& 0.25 & 25 & 6.7$\times $$10^{-6}$\\
		\hline
		{4096} & 6.28 & 1 & 3.4$\times $$10^{-6}$ \\
		& 1.01 & 6.25 & 6.2$\times $$10^{-6}$\\
		& 0.25 & 25 & 6.6$\times $$10^{-6}$\\
		\hline
	\end{tabular}
	\caption{List of errors at different values of $R$.}
	\label{tab:enconsvserr}
\end{table}

\section{Absolute equilibrium}
\label{app:abseq}

Solutions of the Galerkin truncated inviscid BE are known to thermalize. In the wave number space, the inviscid BE is written as
\begin{equation}
\frac{ \partial \hat{u}_k}{ \partial t} + \frac{1}{2}\sum_{k^\prime=-k_{max}}^{k_{max}}i\,k\, \hat{u}_{k^\prime} \hat{u}_{k-k^\prime}= 0.
\end{equation}
The introduction of the truncation in the wave numbers at $k_{max}$ reduces the PDE to a set of finite number of ODEs, which still retain conserved quadratic quantities like the energy, and also satisfy Liouville's theorem~\cite{majda2000remarkable,abramovmajda2003,majdatomofeyev2002}. As a result, the solutions converge to a thermal state with Gibbs statistics. These are called absolute equilibria. Now, from the equipartition of energy, one can predict the energy spectrum in this state as
\begin{equation}
E_{eq}(k)=E_0/\kmax.
\end{equation}
Assuming a continuous distribution of wave numbers one can predict the value of the strain-square term as
\begin{equation}
\Omega_{eq} = 2\int_1^{\kmax} k^2E(k) \sim \frac{2}{3} E_0 \kmax^2.
\end{equation}

\section{Tygers in the inviscid BE}
\label{app:tygersiBE}

\begin{figure*}
	\begin{subfigure}{.497\textwidth}
		\centering
		\includegraphics[width=\textwidth]{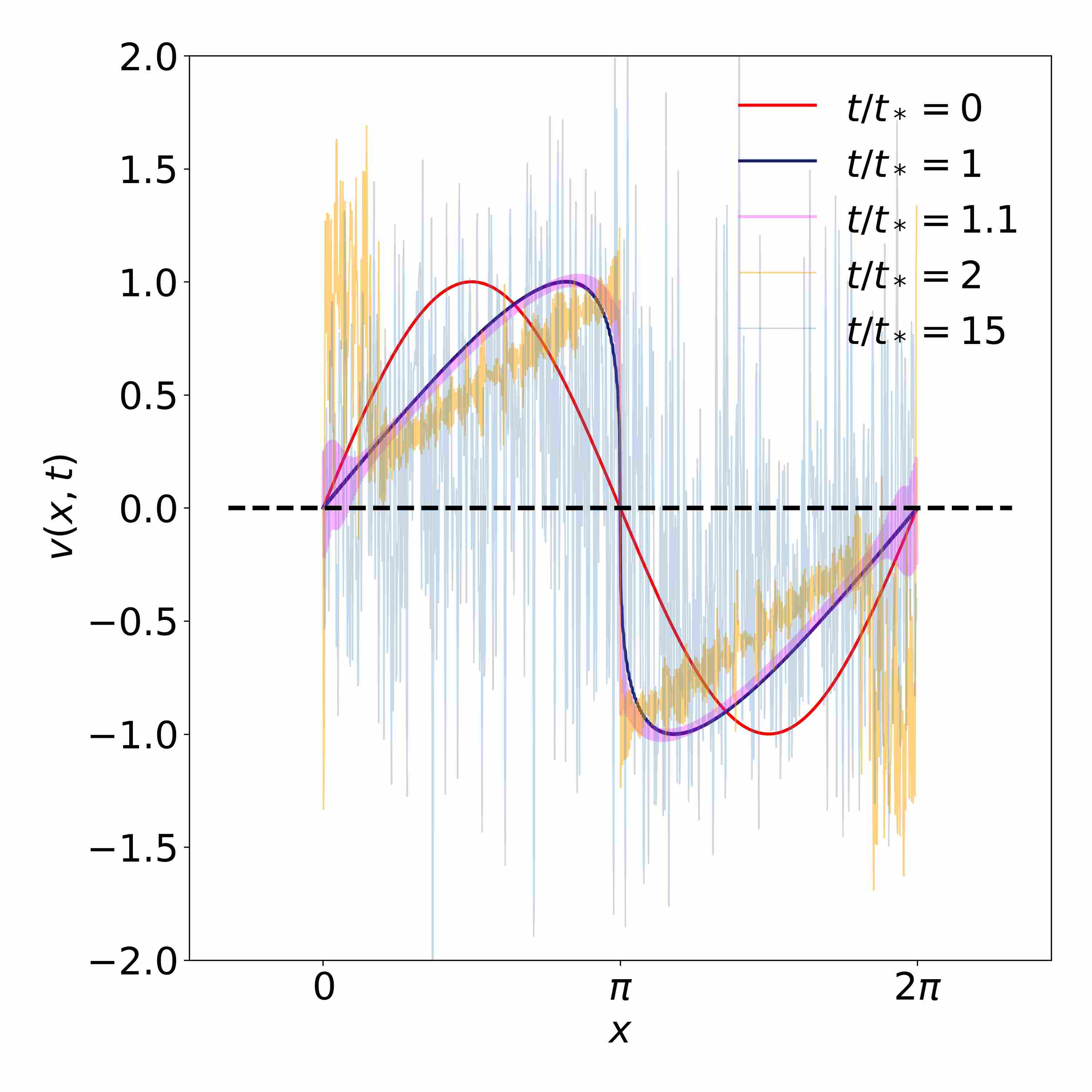}
		\put(-205,230){(a)}
		\phantomsubcaption\label{fig:tyg-vel}
	\end{subfigure}%
	\begin{subfigure}{.497\textwidth}
		\centering
		\includegraphics[width=\textwidth]{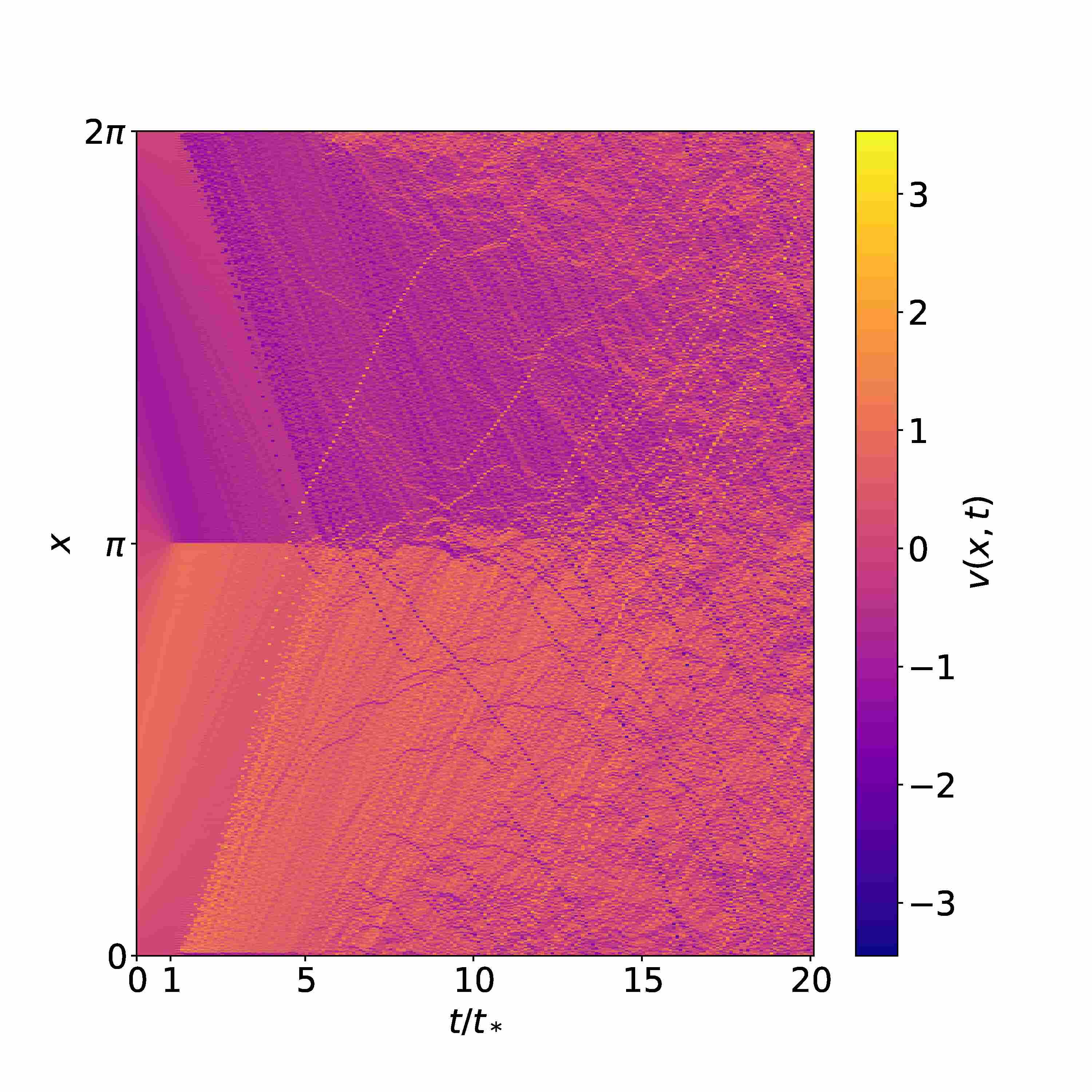}
		\put(-230,230){(b)}
		\phantomsubcaption\label{fig:tyg-sptemp}
	\end{subfigure}
	\caption{\textbf{Inviscid BE.-}(a) Velocity fields $u(x,t)$ at time instants $t/t_*=0$, $1$, $1.1$, $2$, and $15$. (b) Spatio-temporal plot of the velocity field. A shock forms at $x=\pi$ at time $t/t_*=1$; corresponding to this, tygers appear at $x=0$ and begin to spread along both its left and right sides until the entire space has been engulfed. At later stages, the system has completely thermalized.}
\end{figure*}

The Galerkin truncated inviscid BE is
\begin{equation}
\frac{\partial v}{\partial t} + Pv\frac{\partial v}{\partial x} = 0, 
\end{equation}
where $P$ is the Galerkin projection operator and
\begin{equation}
v(x,t)=P\,u(x,t)=\sum_{k=-\kmax}^{\kmax}\hat{u}_k \, e^{ikx}.
\end{equation}
Pseudo-spectral simulations of the above equation have shown that solutions transition to a thermal state. This transition begins with the formation of tygers, which are the precursors of thermalization~\cite{ray2011resonance,venkataraman2017onset,feng2017thermalized}.

For a clearer picture, let us consider the inviscid BE with an initial condition given by
\begin{equation}
u(x,0)=\sin{x}.
\end{equation}
It is well known that such an initial condition has an implicit solution given by
\begin{equation}
u(x,t)=\sin{(x-u(x,t)t)},
\end{equation}
which remains valid till the formation of a pre-shock singularity at time $t_*=1$ [blue solid line in Fig.~15 (a)]. A stationary shock is known to form at $x_*=\pi$ for such an initial condition. A derivation of these results are given in Appendix~\ref{app:poleexpn}. Now, the Galerkin projection operator in the above equation renders it non-local in real space. Localized non-linearity like a shock or preshock acts as a source of \textit{truncation waves}. More specifically, the non-linear term involves a convolution in real space with
\begin{equation}
g(x) \equiv \sum_{k=-\kmax}^{\kmax}e^{ikx} = \frac{\sin(\kmax+1/2)x}{\sin(x/2)}.
\end{equation}
Such a truncation wave is like a plane wave with a wave number close to $\kmax$, when it is away from the source. Near the source, the wave resembles a Dirac measure for large $\kmax$. The truncation waves also have velocity equal to that of the source, \textit{i.e.} the shock. For the chosen initial condition $u = \sin x$, the shock is stationary and as a result the truncation wave velocity is zero.

Furthermore, a Lagrangian particle description of the inviscid BE is possible. At positions away from the shock, the fluid particles move with constant velocity. Particles with velocity close to that of the truncation waves react resonantly with them. Now, for the $\sin x$ initial condition, the particles with velocity zero are located at positions $x=0$ and $x=\pi$. Also, the resonance is possible only when the strain is positive. At locations of negative strain, the truncation wave is squeezed, acquiring wave numbers larger than the Galerkin truncation and subsequently disappears. The strain is initially given by $u_x=\cos x$. It is maximum ($u_x=1$) at $x=0$ and minimum at $x=\pi$ ($u_x=-1$). Thus, one can predict the formation of a tyger at the origin. In Fig~15(a), a shock forms at $t/t_*=1$ and at $t/t_*=1.1$ bulges from truncation waves forming around $x=0$ as stated (the pink solid line). These bulges are the tygers, which grow and spread the thermalization. The effect of the negative strain can be seen on the shock, as a thickening at the edges.

Initially, the tygers appear as symmetric bulges around the point of resonance. However, as time progresses the symmetry is lost as the bulge of the tyger increases in amplitude due to the truncation wave interactions. Eventually, the bulge collapses, giving way to two fronts of white noise (velocity field at $t/t_*=2$, the yellow solid line). The evolution of the tyger is also clearly evident in Fig.~15(a), where the two parts of the tyger can be seen as lines emanating from $x=0$. The tygers initially remain confined by the shock and a partially thermalized state is observed. Eventually, as the tyger amplitude becomes as large as the shock amplitude, the velocity field appears globally thermalized.

\section{Tygers and the lack of thermalization in the forced inviscid BE}
{\label{app:tygersfiBE}}

\begin{figure*}
	\begin{subfigure}{.497\textwidth}
		\centering
		\includegraphics[width=\textwidth]{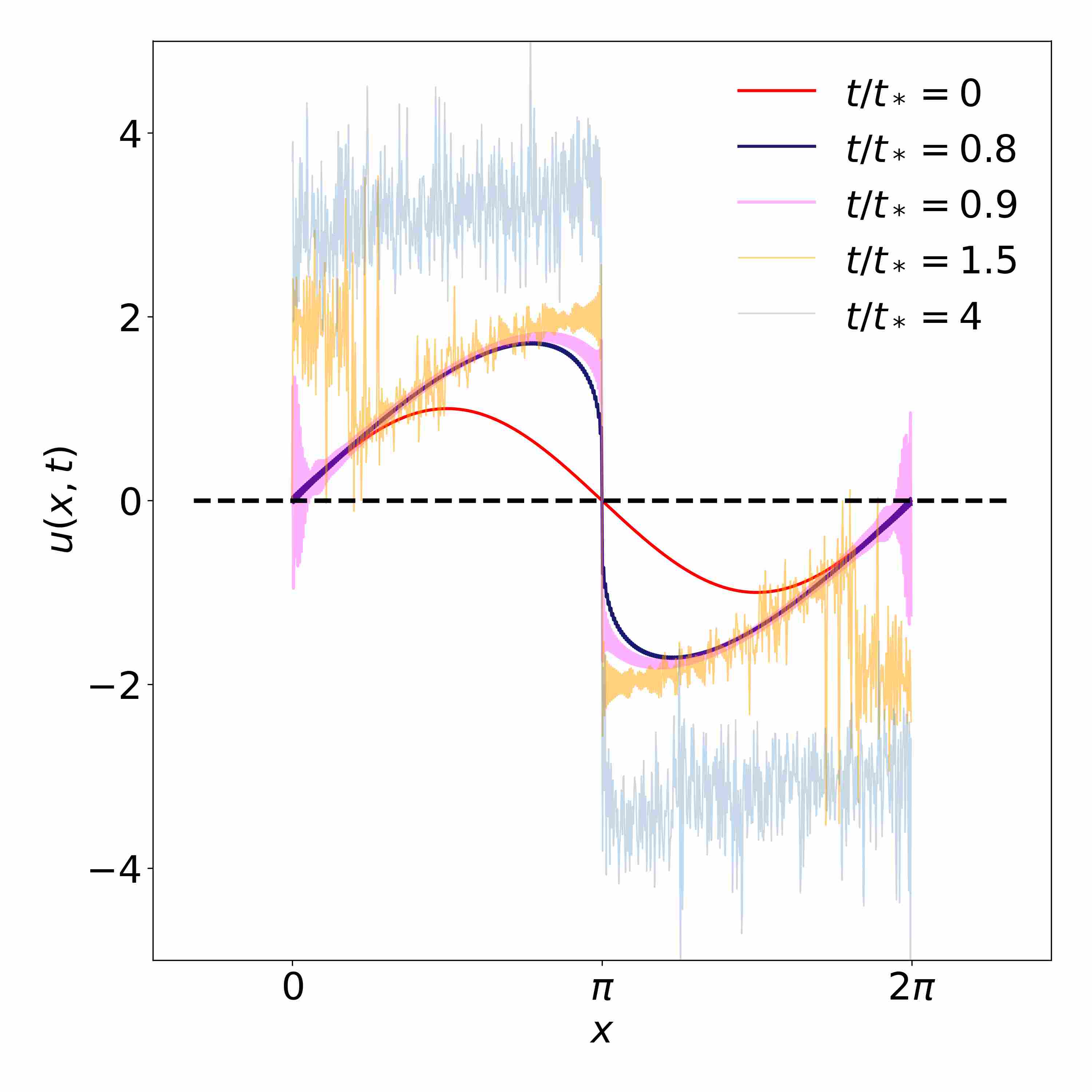}
		\put(-205,230){(a)}
		\phantomsubcaption\label{fig:fbetyg-vel}
	\end{subfigure}%
	\begin{subfigure}{.497\textwidth}
		\centering
		\includegraphics[width=\textwidth]{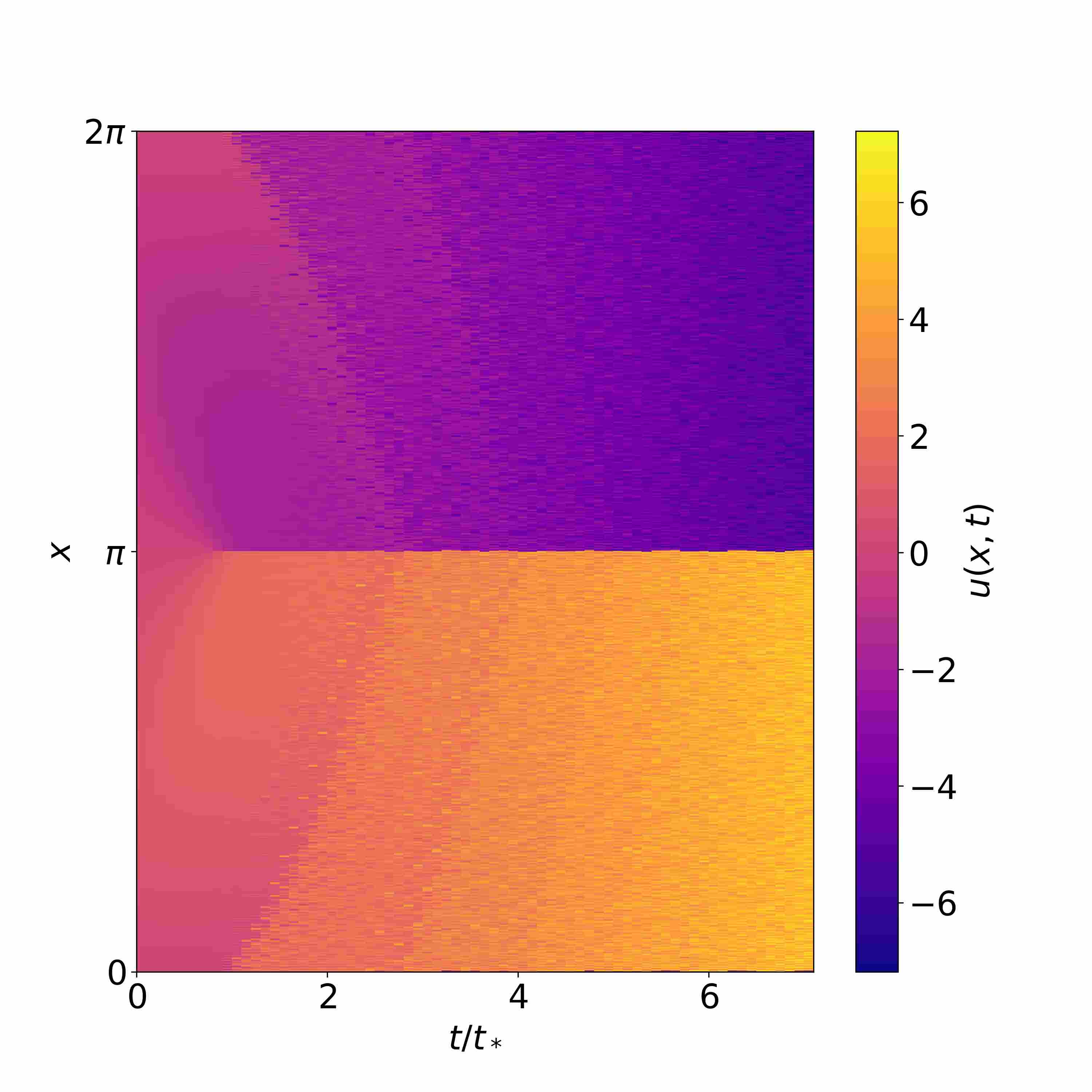}
		\put(-230,230){(b)}
		\phantomsubcaption\label{fig:fbetyg-sptemp}
	\end{subfigure}
	\caption{\textbf{Forced inviscid BE.} (a) Velocity fields $u(x,t)$ at time instants $t/t_*=0$, $0.8$, $0.9$, $1.5$, and $4$. (b) Spatiotemporal plot of the velocity field. A shock forms at $x=\pi$ around $t\approx 0.8$; corresponding to this, tygers appear at $x=0$ and begin to spread along both its left and right sides. $t_*=1$ is the time of shock formation in the inviscid BE for $u(x,0)=\sin x$.}
	\label{fig:fbetyg}
\end{figure*}

Like its unforced counterpart, the sinusoidally forced, inviscid BE
\begin{equation}
\frac{\partial u}{\partial t} + u\frac{\partial u}{\partial x} = \sin x
\end{equation}
also form tygers. What is interesting is that in the forced inviscid BE, tygers do not cause thermalization, for the chosen initial condition. Instead, the tygers lead to a secondary discontinuity at $x=0$, where most of the energy is pumped. As a result, a square wave forms where the ends are partially thermalized due to the wings of white noise arising from the tygers. The total system, however, remains partitioned into two parts, with energy going into the shock and the tyger-born discontinuity. In the inviscid BE, the energy lost at the shock passes to the tygers. However, in this case the forcing causes a pile-up of energy as there is no dissipation mechanism. The energy pumped into the system is forced into the shock which subsequently passes into the tyger discontinuity. This pile-up at the two discontinuous forms an insurmountable energy barrier, which the white noise born from the tyger collapse is not able to traverse. This causes the partitioning of the system and prevents the attainment of complete thermalization. This behavior is inevitably the result of the symmetry caused by choosing both the forcing and initial velocity profile as sinusoidal. If this symmetry is broken, say by using the forcing $\sin (x-\pi/2)$, then the position of the shock changes and thermalization does occur (even though the energy keeps increasing) with energy equipartition among the available modes. Thus, the symmetry of the chosen forcing and initial conditions are crucial. See Fig.~\ref{fig:fbetyg} for velocity profiles and spatiotemporal plot of the forced inviscid BE.

\section{Fate of the tygers in regime B}
{\label{app:tygersBEregB}}

\begin{figure*}
	\begin{subfigure}{.497\textwidth}
		\centering
		\includegraphics[width=\textwidth]{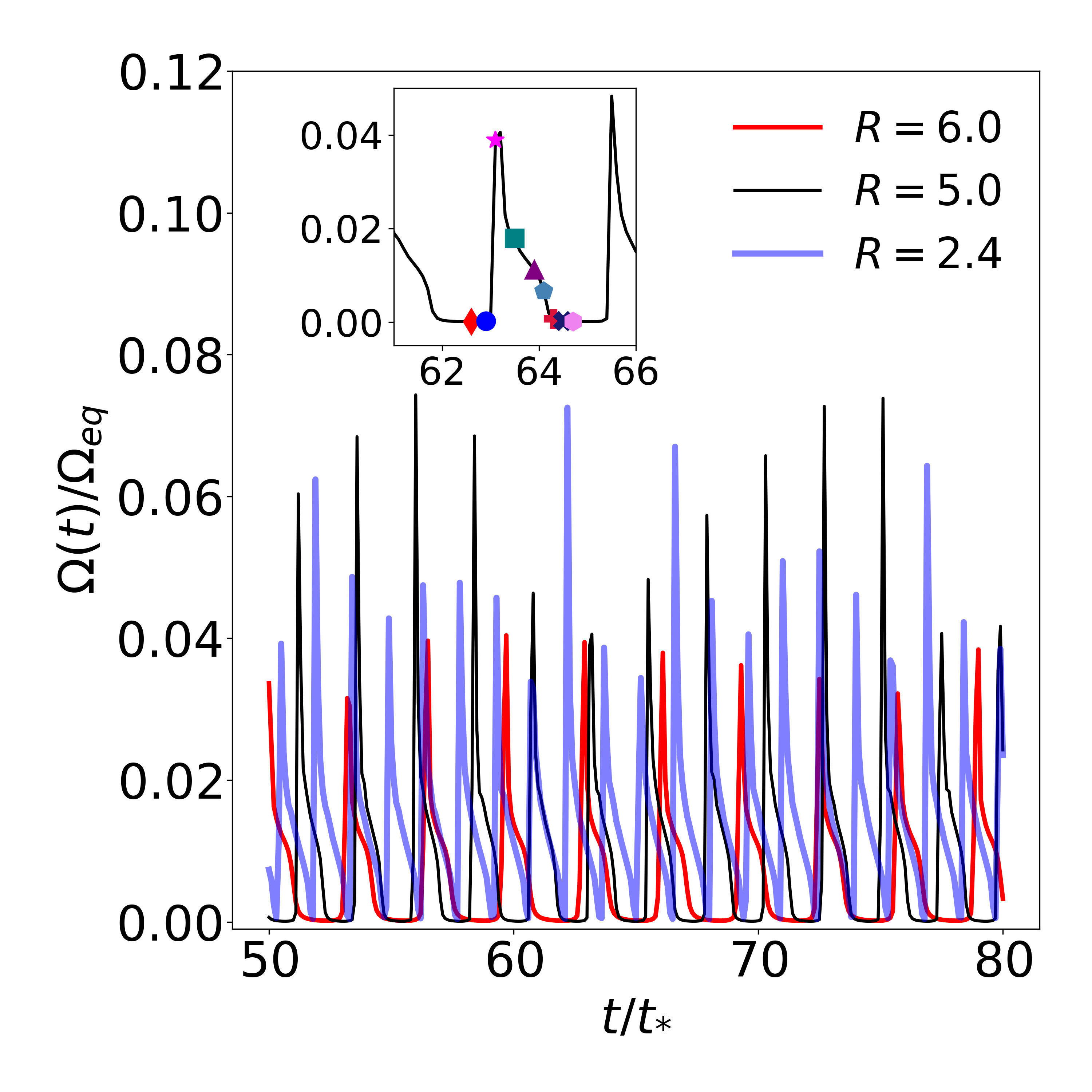}
		\put(-30,150){(a)}
		\phantomsubcaption\label{fig:ens tracing}
	\end{subfigure}
	\begin{subfigure}{.497\textwidth}
		\centering
		\includegraphics[width=\textwidth]{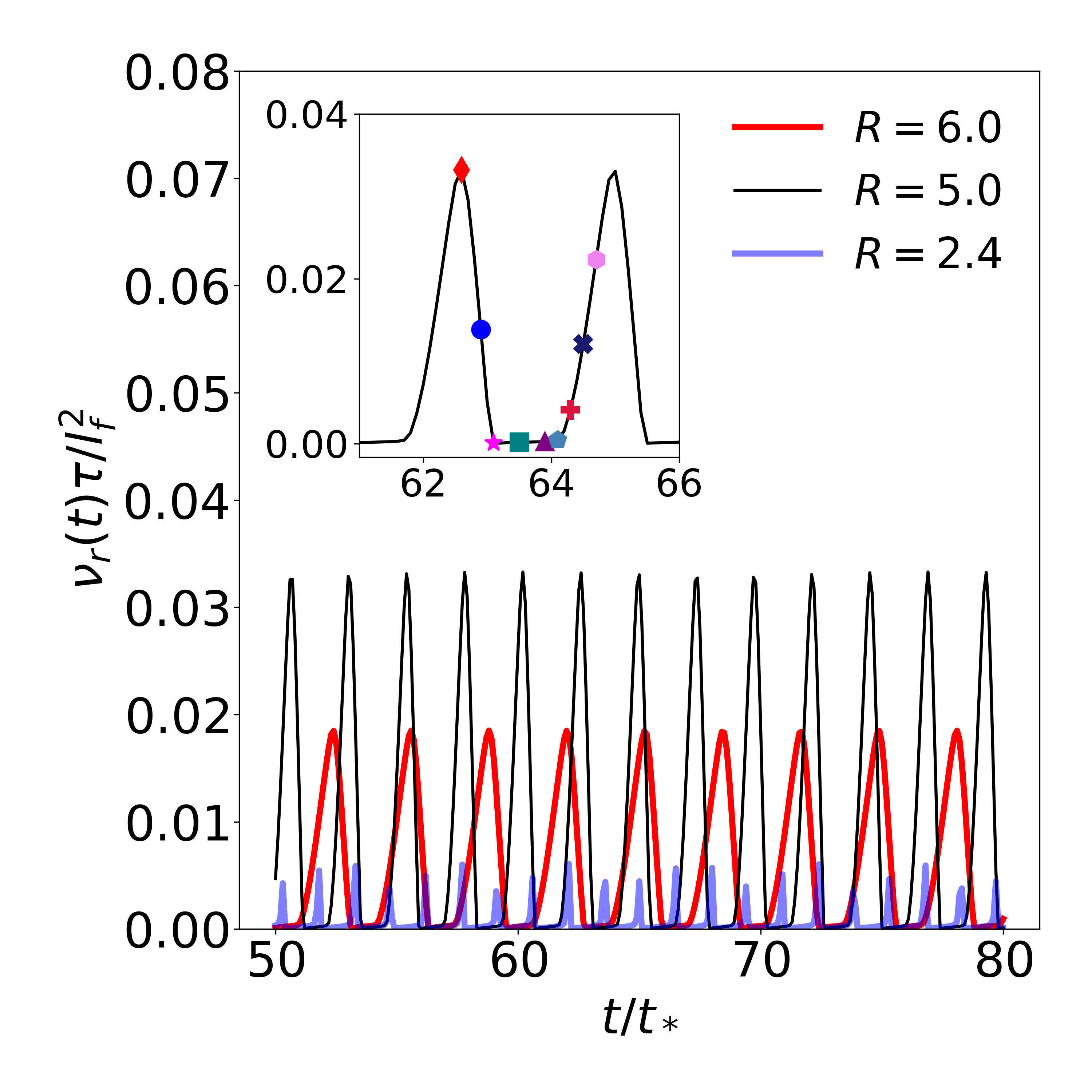}
		\put(-30,150){(b)}
		\phantomsubcaption\label{fig:viscosity tracing}
	\end{subfigure}
	\caption{ \textbf{Oscillations in regime B.} Time series of: (a) $\Omega/\Omega_{eq}$ and (b) $\nu_r\tau/\ell^2_f$ exhibit oscillations. The insets show two peaks for $R = 5$, wherein points with different markers correspond to different instants of time and have been used to demonstrate the appearance and disappearance of tygers in Fig.~\ref{fig:fulltygers}. $\Omega_{eq}=(2/3)E_0\kmax^2$, $\tau=(\ell^2_f/\langle \epsilon \rangle)^{1/3}$, where $\langle \epsilon \rangle$ is the energy injection rate, and $\ell_f$ is the forcing length scale. $t_*=1$ is the time of shock formation in the inviscid BE for $u(x,0)=\sin x$.}
	\label{fig:tracingtygers}
\end{figure*}

\begin{figure*}
	\centering
	\includegraphics[scale=0.45]{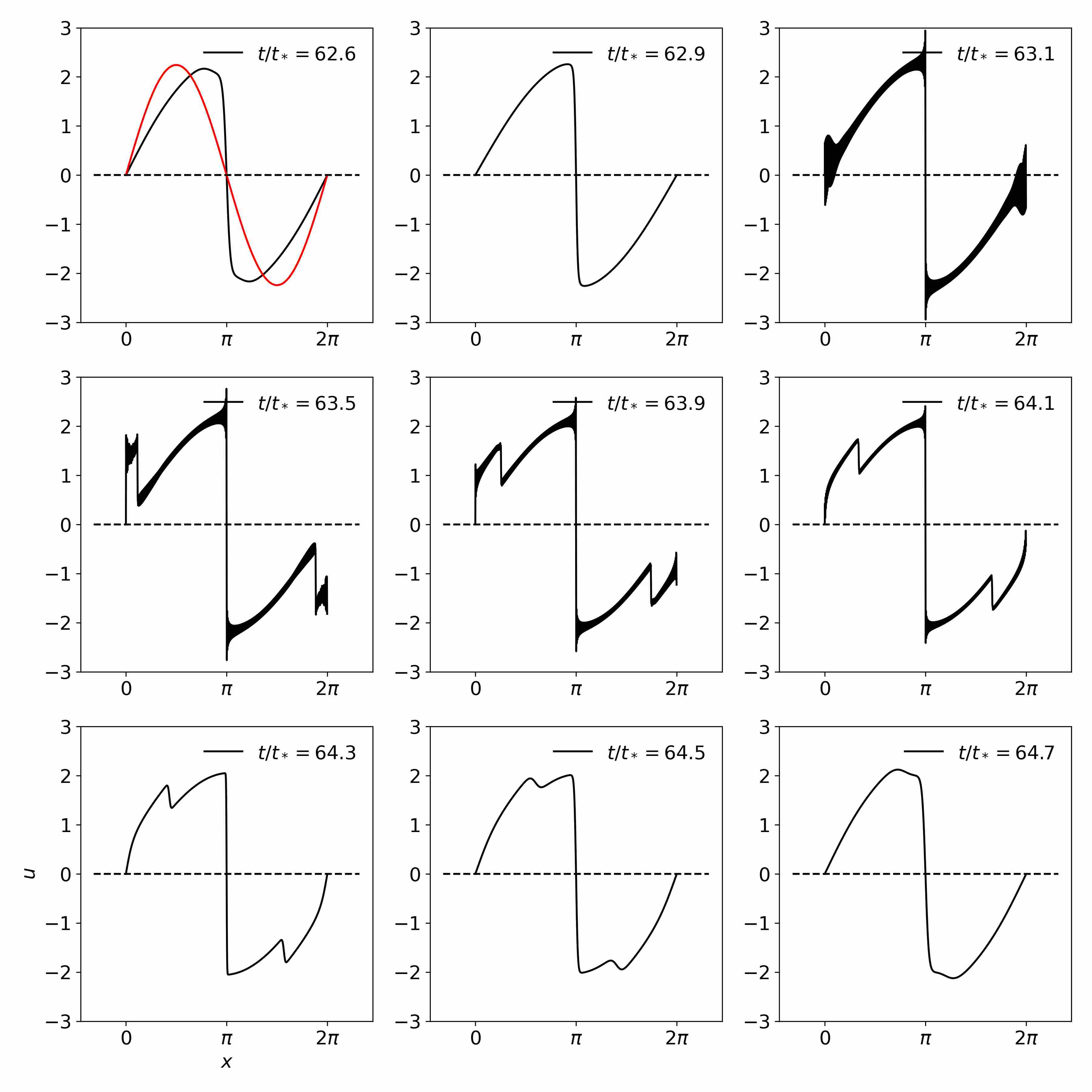}
	\caption{Plot of the velocity profile $u(x,t)$ in regime B, showing one cycle of birth and subsequent evolution of tygers at different instances. $t_*=1$ is the time of shock formation in the inviscid BE for $u(x,0)=\sin x$.}
	\label{fig:fulltygers}
\end{figure*}

We illustrate the full dynamics of tygers using Figs.~\ref{fig:tracingtygers} and~\ref{fig:fulltygers}, as it occurs in regime B (see main text Sec.~\ref{sec:regimeB}). To elucidate the dynamics, we carefully choose the time instants from the oscillation profile (part of the time series) of $\Omega(t)$ and $\nu_{r}(t)$ shown in the insets of Figs.~\ref{fig:fulltygers} (a) and (b), respectively.

\section{Structure functions and Multifractal spectrum}\label{app:multifrac}

In this appendix, we use the stationary state asymptotic solution of the BE (see Sec.~\ref{sec:masymp}) in the inviscid limit ($\nu \rightarrow 0)$ and derive an analytical form for the $p$th order structure function, which is given by
\begin{equation*}
	S_{p}(l) = \langle [\delta_{l} u(x)]^p \rangle
\end{equation*}
where $\delta_{l} u(x)$ = $u(x+l)-u(x)$ and $\langle \cdot \rangle$ denotes spatial averaging.

From the stationary state asymptotic solution, we get in the inviscid limit:
\begin{equation}
	f(x) =\begin{cases}
		2 \sin(x/2) & 0 \leq x < \pi\\
		-2 \sin(x/2) & \pi < x \leq 2 \pi.
	\end{cases}
\end{equation}
For such a form of the velocity, $S_{p}(l)$ is given by
\begin{equation}\label{eq:ssasf}
	\begin{split}
		S_{p}(l) = &\frac{4^p}{\pi} \bigg[ \sin^{p}(l/4)F_{p}(l) + (-1)^{p} \cos^{p}(l/4)G_{p}(l) \\
		& + (-1)^{p} \sin^{p}(l/4)H_{p}(l) \bigg],
	\end{split}
\end{equation}
where
\begin{flalign*}\nonumber
	F_{p}(l) & = \int_{l/4}^{\pi/2 - l/4} \cos^{p}(z) dz \\
	G_{p}(l) & = \int_{\pi/2 - l/4}^{\pi/2 + l/4} \sin^{p}(z) dz\\
	H_{p}(l) & = \int_{\pi/2 + l/4}^{\pi + l/4} \cos^{p}(z) dz
\end{flalign*}

\begin{figure*}
	\centering
	\resizebox{\textwidth}{!}{
		\includegraphics[scale=0.45]{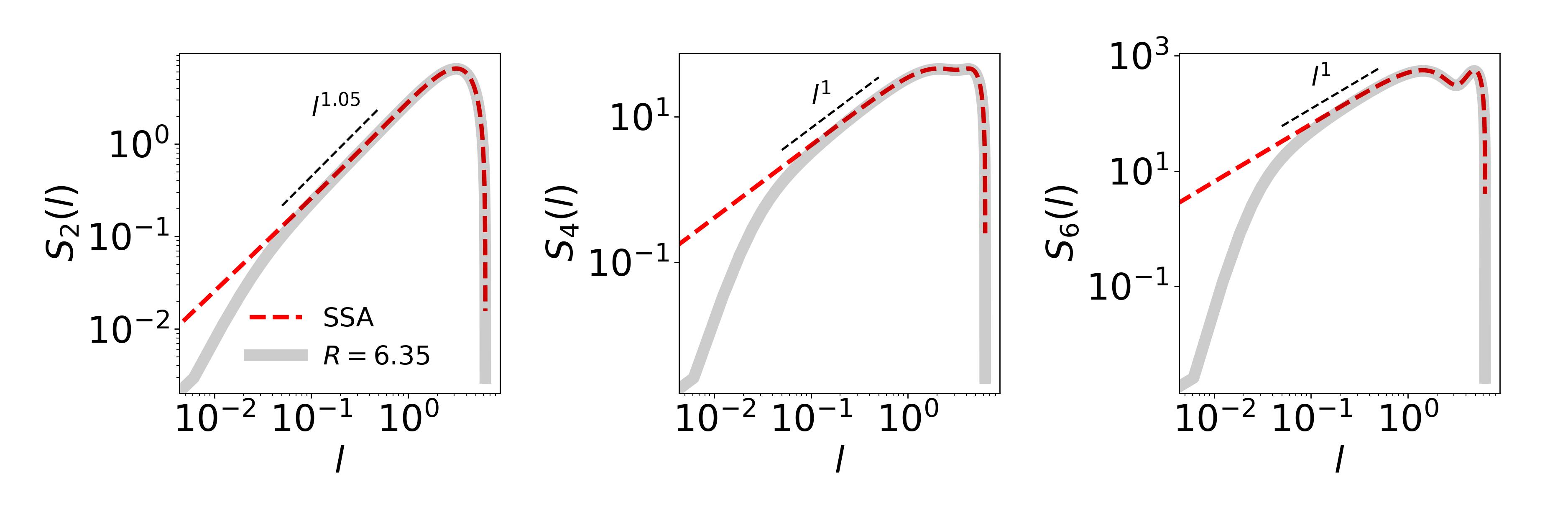}
		\put(-50,90){(c)}
		\put(-220,90){(b)}
		\put(-420,90){(a)}}
	\caption{Comparison of structure functions for the steady state asymptotic (SSA) solution in (~\ref{eq:ssasf}) with the structure functions for the RB at $R=6.35$ for (a) $p=2$,  (b) $p=4$, (c) $p=6$. Structure functions computed using the SSA solution and RB data are indicated by dashed red lines and solid grey lines, respectively.}
	\label{fig: multifracc}
\end{figure*}

We compare the above structure function for $p = 2,4,6$ in Fig.~\ref{fig: multifracc} with the ones from RB at $R = 6.35$ and observe that they are in a very good agreement. In each case, we find that $S_{p}\sim l$.

This can be verified exactly for each $p$. We show this for $p = 2$, where for small $l$ one can find that
\begin{equation}\nonumber
	\frac{d}{d \big[\ln l \big]} \big[ \ln {S_{2}(l)} \big] \sim 1 + \frac{\pi l}{16} + O(l^{2}).
\end{equation}

\begin{figure*}
	\centering
	\resizebox{\textwidth}{!}{
		\includegraphics[scale=0.2]{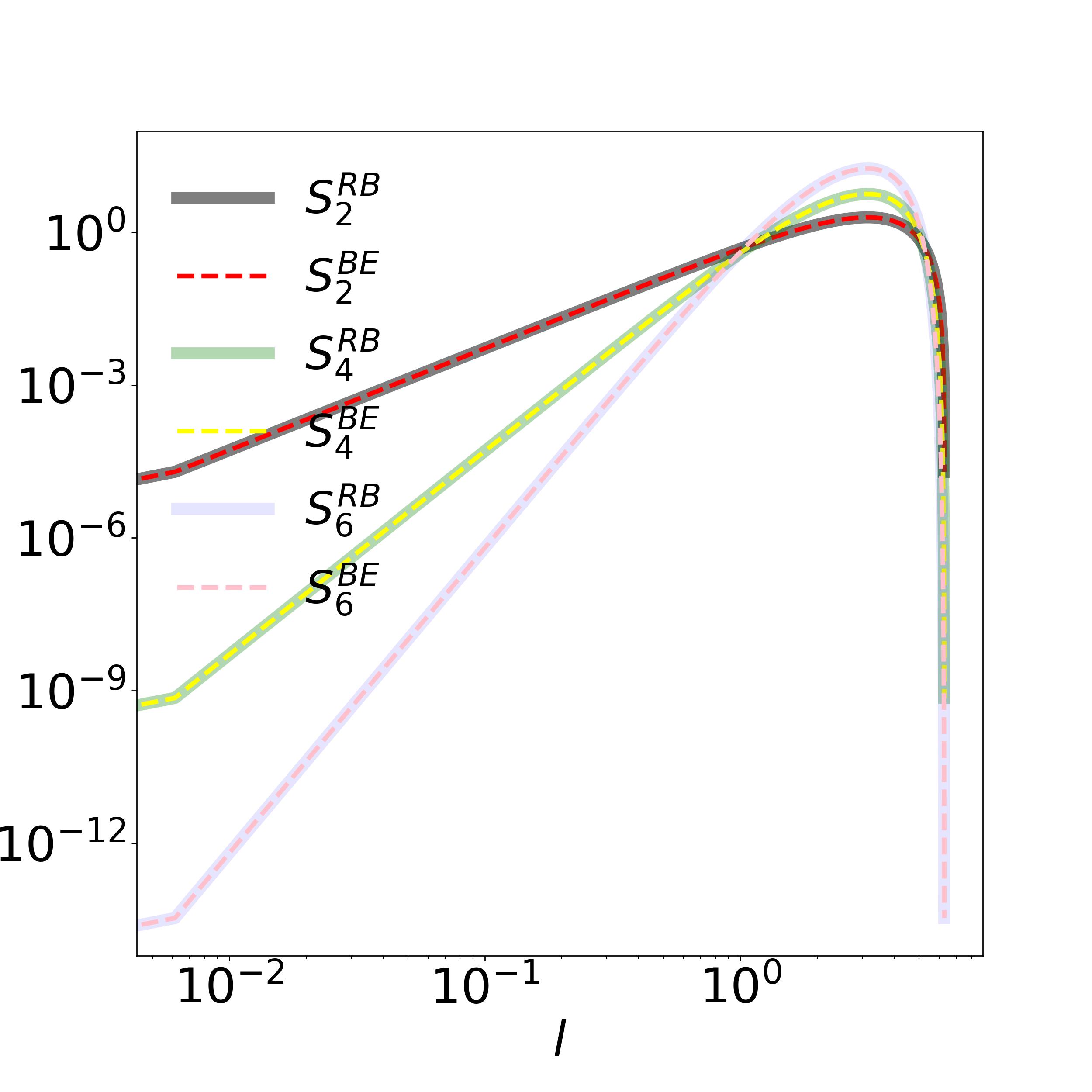}
		\put(-40,100){(a)}
		\includegraphics[scale=0.2]{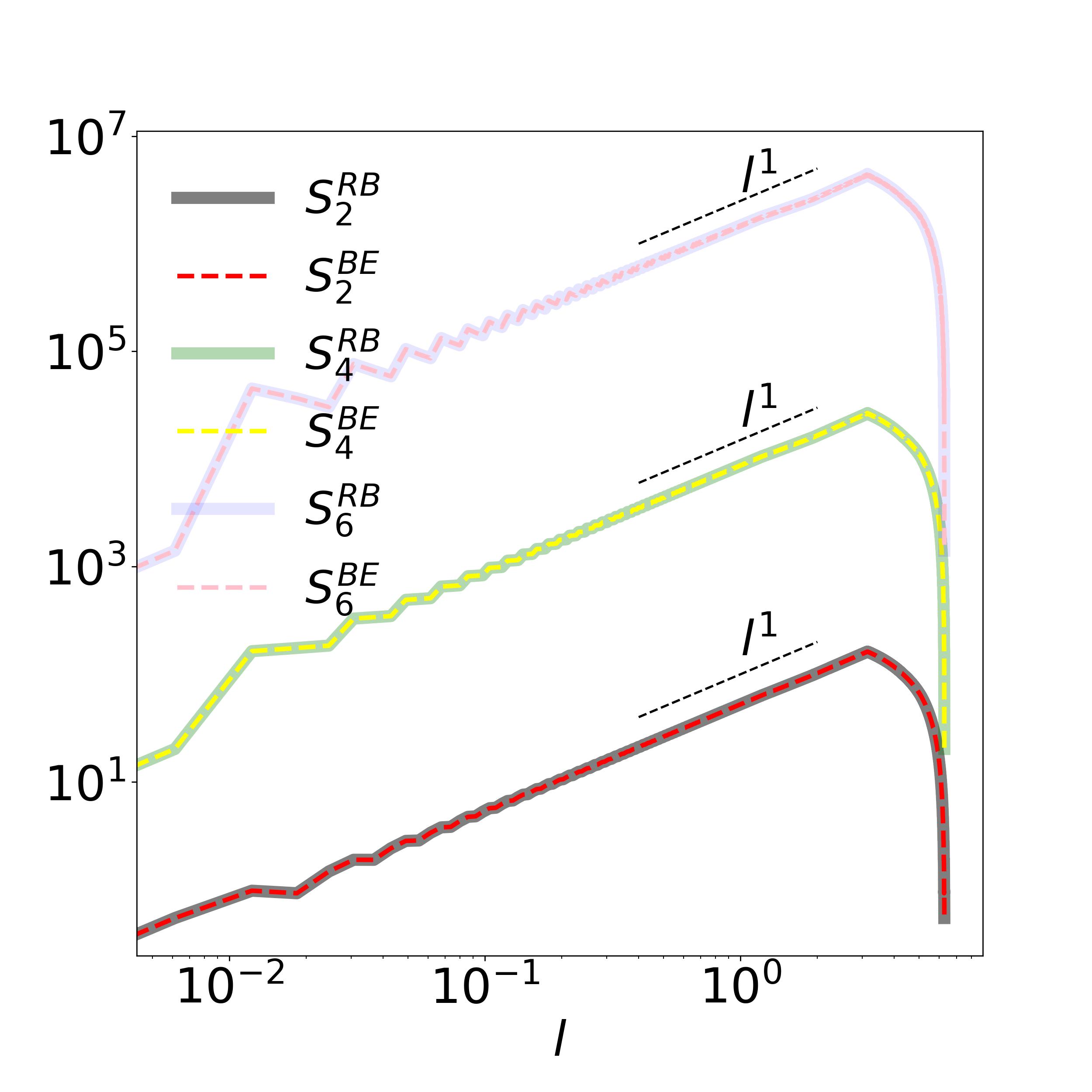}
		\put(-40,100){(b)}
		\includegraphics[scale=0.2]{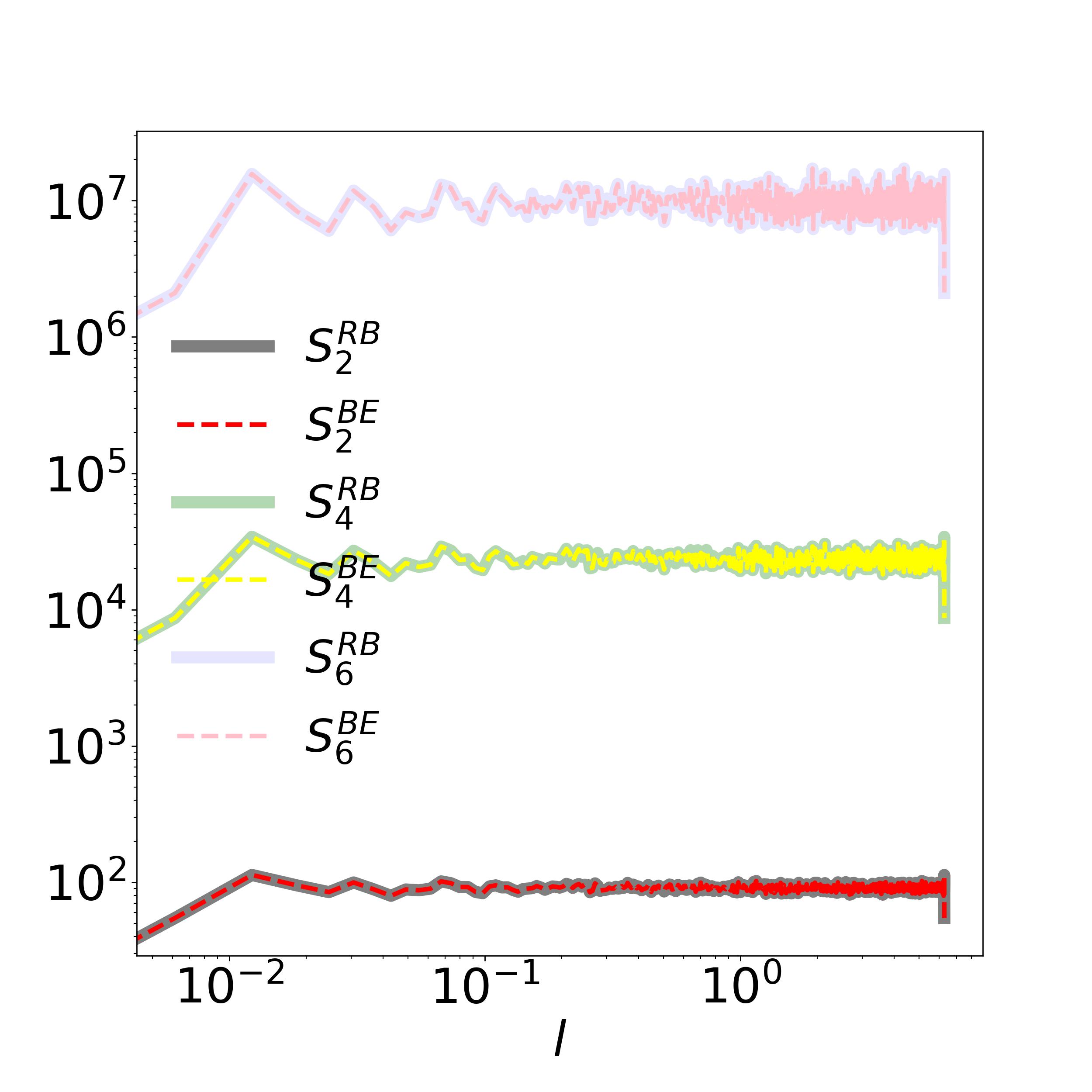}
		\put(-40,100){(c)}}
	\caption{Comparison of structure functions for the RB and BE in the various regimes: (a) Regime A, $R=25.13$; (b) regime C, $R=0.31$; and (c) regime D, $R=0.28$.}
	\label{fig:stf-eq}
\end{figure*}

Similar structure functions and scaling exponents have been observed in~\cite{xiesptempstrucfunc}, where a similar sinusoidal forcing with random coefficients has been used. Furthermore, in Fig.~\ref{fig:stf-eq} we compare the structure functions for the BE and RB in various regimes and find that they are always in a perfect agreement; thereby, further strengthening the equivalence conjecture.

\section{Pole expansion solution of Burgers equation}
\label{app:poleexpn}

This appendix provides a description of the Mittag-Leffler pole expansion solution of the viscous BE as documented in~\cite{senouf1997dynamics}. We start by considering the solution of the BE given by the Cole-Hopf transformation $u = -2\nu \partial_{x}\log(\phi)$, which transforms the BE into the heat equation $\partial_{t}\phi = \partial_{xx} \phi$  with the initial condition given by $\phi(x,0) = \exp \big[-\frac{1}{2\nu}\int_{0}^{x}u(y,0)dy \bigl]$. Solving the heat equation is a trivial task (using Fourier transformations or Green's equation). This gives us $\phi(x,t)  = K(x,t) E(x,t).$
Here,
\begin{flalign*}
K(x,t) & =  \frac{1}{\sqrt{4\pi \nu t}} \exp[-x^{2}/{4\nu t}]\text{ and}\\
E(x,t) & = \int_{-\infty}^{\infty}\exp\bigg[\frac{1}{2\nu}\int_{0}^{y}\big(\frac{x}{t}-\frac{\eta}{t}-u(\eta,0)\big)d\eta \bigg]dy\text{.}
\end{flalign*}
 \vspace{2mm}\newline
Plugging $K$ into the original transformation gives $-2\nu \partial_{x}(\log K) = x/t$ and therefore,
\begin{equation}\label{equ:cole-hopf}
	u(x,t) = \frac{x}{t}- 2\nu \partial_{x} \log(E(x,t)).
\end{equation}\vspace{2mm}\newline
For the initial condition, $u(x,0) = \sin x$,
\begin{equation}
	E(x,t) = \int_{-\infty}^{\infty}\exp \bigg[\frac{1}{2\nu} \big(\frac{x}{t}y-\frac{y^{2}}{2t}+ \cos y - 1 \big)\bigg]dy.
\end{equation}\vspace{2mm}\newline
For the paradigmatic initial condition $u(x,0) = 4x^{3}-x/t_{*}$,
\begin{equation}
	E(x,t) = \int_{-\infty}^{\infty}\exp \bigg[ \frac{1}{2\nu} \big(\frac{x}{t}y+\alpha y^{2} + y^{4}\big) \bigg]dy,
\end{equation}\vspace{2mm}\newline
where $\alpha = (t-t_{*})/2 t t_{*}$, $t_{*}$ being the time of appearance of a shock. Both $E$'s are even, analytic functions of $x$.\vspace{2mm}\newline\
It can be shown that the order $\lambda$  of $E(x,t)$ for  $u(x,0) = 4x^{3}-x/t_{*}$ is fractional, and hence $E(x,t)$ has infinite number of zeros. Using a canonical product expansion over the zeros $a_{n}$, we get \[E(x,t) = Cx^{m}e^{g(x)}\prod_{n}(1-\frac{x}{a_{n}}) e^{x/a_{n}+\frac{1}{2}(x/a_{n})^{2}+ \cdots \frac{1}{p}(x/a_{n})^{p}},\] where $p$ is an integer and $g(x)$ is a polynomial of degree $q$. The genus of the expansion is given by $h = \max(p,q)$. Also, $h\leq \lambda\leq h+1 \implies h = 1 \implies p,q \leq 1$. Again as  $E(x,t)$ is even, $g(x)$ cannot have degree 1. Therefore, $q = 0 \implies p = 1$. Also, $C\neq 0$ can be shown, hence $m = 0$ (as there are no roots at $x = 0$). This implies \[E = C(t)\prod_{n}(1-\frac{x}{a_{n}}) e^{x/a_n}.\]
Again, as $E$ is even, zeros occur in opposite pairs which gives $E = C\prod_{n}(1-\frac{x^2}{a_{n}^2})$ and $\sum \frac{1}{|{a_{n}}|} < \infty$,  $\sum \frac{1}{|{a_{n}}^2|} < \infty$ must hold where $C(t)$ can be found. From the form of the relation between $u$ and $E$, we know that the zeros of $E$ are poles of $u$. Finally plugging it in (\ref{equ:cole-hopf}), we get \[u = \frac{x}{t} - \sum_{n=1}^{\infty} \frac{4\nu x}{x^2+a_{n}^2} = \frac{x}{t} - \sum_{n=-\infty, \\ n\neq 0}^{\infty} \frac{2\nu }{x-a_{n}},\] where $a_{n} = a_{n}(t,\nu)$ are time-dependant pole locations.\vspace{2mm}\newline
In general, a meromorphic solution to the BE can be expressed as
\begin{equation}
	u = \frac{x}{t} - \sum_{n} \frac{2\nu }{x-a_{n}(t,\nu)}.
\end{equation}
Now, we apply the saddle-point method to determine the asymptotic behavior of $E$, as $\nu \rightarrow 0+$. In our system, $u(x,0) = u_{0} \sin x$. We write
\begin{flalign*}
	w(z,x) & = \int_{0}^{z}(\frac{x}{t}-\frac{\eta}{t}-u(\eta,0))d\eta \\
	& = \int_{0}^{z}(\frac{x}{t}-\frac{\eta}{t}-\cos \eta)d\eta \\
	& = \frac{xz}{t} - \frac{z^2}{2t} + u_{0} (\cos z - 1)
\end{flalign*}
Here $w(z,x)$ is the phase function of the integrand in the definition of $E$. The following system of equations:
\begin{flalign}
	w_{z} & = \frac{x}{t} - \frac{z}{t} - u_{0} \sin z = 0, \\
	w_{zz} & = -\frac{1}{t} - u_{0} \cos z = 0,
\end{flalign}
determine the envelope of the characteristics of the inviscid BE, given by $x_s(t)$. From these two equations, we find
\begin{flalign*}
	z_{s} & = \cos^{-1}(-\frac{1}{tu_0}),\\
	x_{s} & = u_{0} t \sqrt{1-\frac{1}{t^2 u_0^2}} + \cos^{-1}(-\frac{1}{tu_0}).
\end{flalign*}
The condition for the coalescence of singularities on the real axis can be determined by setting the term under the square root in the second equation to zero. This gives the time and position of coalescence as $t_{*} = 1/u_{0}$ and $x_{*} = \pi$, respectively. Thus, these two conditions give the shock time and position, respectively.



\bibliography{reference}

\end{document}